\documentclass[prd,twocolumn,nofootinbib,longbibliography,aps,superscriptaddress
]{revtex4-1}

\usepackage{amsmath,amssymb,ifpdf,url,mathrsfs,slashed,multirow,tabularx,type1cm}
\usepackage{subfigure}
\usepackage[dvipdfmx]{graphicx}
\usepackage{color}
\usepackage{braket}
\usepackage{bm}
\usepackage[toc,page]{appendix}
\usepackage{comment} 
\usepackage{ulem}

\allowdisplaybreaks
\definecolor{BlueViolet}{rgb}{0.2, 0.00, 0.7}
\definecolor{Blue}{rgb}{0.15, 0.00, 0.9}
\usepackage[
colorlinks=true,linkcolor=Blue,citecolor=Blue,
urlcolor=BlueViolet,breaklinks=true]{hyperref}

\newcommand{\Slash}[1]{{\ooalign{\hfil \hspace*{-5pt}~#1\hfil\crcr\raise.167ex\hbox{/}}}}

\def\({\left(}
\def\){\right)}
\def\<{\langle}
\def\>{\rangle}

\newcommand{\matl}{\left( \begin{array}}
\newcommand{\matr}{\end{array} \right)}

\def\beq#1\eeq{\begin{align}#1\end{align}}

\newcommand{\du}[1]{\textcolor{red}{#1}}

\begin{document}

\title{
Entropy constraints on effective field theory
}

\author{Qing-Hong Cao}
\email{qinghongcao@pku.edu.cn}
\affiliation{Department of Physics and State Key Laboratory of Nuclear Physics and Technology, Peking University, Beijing 100871, China}
\affiliation{Collaborative Innovation Center of Quantum Matter, Beijing 100871, China}
\affiliation{Center for High Energy Physics, Peking University, Beijing 100871, China}

\author{Daiki Ueda} \email{ueda@pku.edu.cn} 
\affiliation{Center for High Energy Physics, Peking University, Beijing 100871, China}

\begin{abstract}
\noindent
In effective field theory, the positivity bounds of higher derivative operators are derived from analyticity, causality, and unitarity. We show that the positivity bounds on some operators of the effective field theory, e.g., dimension-eight term of a single massless scalar field, the Standard Model Effective Field Theory dimension-eight $SU(N)$ gauge bosonic operators, and higher-derivative operators in the Einstein-Maxwell theory, generated by interactions between heavy and light degrees of freedom can be derived by the non-negativity of relative entropy.
For such effective field theories, we prove that the interactions increase thermodynamic entropy at a fixed charge and an extremal point of energy, which is intimately connected with the extremality relations of black holes exhibiting Weak-Gravity-Conjecture.
These arguments are applicable when corrections from the interactions involving higher-derivative operators of light fields are not dominant in the effective field theories.
The entropy constraint is a consequence of the Hermiticity of Hamiltonian, and any theory violating the non-negativity of entropy would not respect the second law of thermodynamics.
\end{abstract}

\maketitle 

\renewcommand{\thefootnote}{\#\arabic{footnote}}
\setcounter{footnote}{0}

\section{Introduction}
Relative entropy~\cite{10.1214/aoms/1177729694,10.2996/kmj/1138844604,RevModPhys.50.221} is a fundamental quantity in probability theory and information theory. 
The relative entropy, which is non-negative, depicts a {\it distance} between two probability distributions and plays important roles in statistical mechanics~\cite{2000cond.mat..9244T,2010,2012} and quantum information theory~\cite{1989RpMP...27...19O,doi:10.1080/00107514.2011.587535,RevModPhys.74.197}. 
In the context of information-thermodynamics, the distance between two probability distributions is an essential concept to derive a non-negativity of difference in von-Neumann entropy between initial and final states~\cite{2000cond.mat..9244T,2010,HASEGAWA20101001}, so-called second law of thermodynamics.  

Recently, thermodynamics of black hole~\cite{
Kats:2006xp,Cheung:2018cwt,Cheung:2019cwi,Loges:2019jzs,Reall:2019sah,Goon:2019faz} have been studied in the context of Weak Gravity Conjecture (WGC)~\cite{Arkani-Hamed:2006emk},  
which is motivated to distinguish the landscape from the swampland~\cite{Vafa:2005ui}.
The WGC states that the $U(1)$ charge-to-mass ratio of extremally charged black holes is larger than unity in any gravitational effective field theory (EFT) that admits a consistent UV completion~\cite{Goon:2019faz}.
Some proofs for this statement have been made using black holes and entropy consideration~\cite{Cheung:2018cwt, Cheung:2019cwi}, or positivity bounds from unitarity and causality~\cite{Bellazzini:2019xts,Hamada:2018dde}.
In particular, Refs.~\cite{Cheung:2018cwt, Cheung:2019cwi} are based on a positivity of entropy difference between 
Einstein-Maxwell theories with and without perturbative corrections that are described by higher-derivative operators.

The crucial role of relative entropy in information-thermodynamics suggests that the positivity of entropy difference in the WGC is intimately connected to the distance between two theories, which has been studied in different contexts~\cite{Gaite:1995yg,Calmet:2011zz,Balasubramanian:2014bfa,Casini:2016udt}.
That inspires us to establish a connection between relative entropy and positivity bounds in the EFTs. 
In the work, we provide lower and upper bounds on perturbative corrections from interactions between heavy and light degrees of freedom to the Euclidean effective action.
From the upper bound, we obtain the same bounds on some operators of EFTs, e.g., dimension-eight term of a single massless scalar field, the Standard Model EFT (SMEFT) dimension-eight $SU(N)$ gauge bosonic operators, and higher-derivative operators in the Einstein-Maxwell theory, as those positivity bounds achieved in conventional EFT studies~\cite{Arkani-Hamed:2006emk,Adams:2006sv,Remmen:2019cyz} when the  higher-derivative operators are generated by the  interactions between heavy and light fields.
The constraints on such EFTs are applicable when the perturbative corrections from the interactions involving higher-derivative operators of the light fields are not dominant in the EFTs.

Ref.~\cite{Goon:2019faz} implies a possibility that the WGC-like behavior in the perturbative correction to extremality relations of black hole~\cite{Cheung:2018cwt} can be generalized to a broad class of thermodynamic systems on the condition that the correction to entropy is non-negative.
We prove that the corrections to the entropy at a fixed charge and an extremal point of energy from the operators, such as the dimension-eight term of a single massless scalar field, the SMEFT dimension-eight $SU(N)$ gauge bosonic operators and higher-derivative operators in the Einstein-Maxwell theory, are non-negative when the corrections from the interactions involving higher-derivative operators of the light fields are not dominant in the EFTs.
%


\section{Distance between two theories}
\label{sec:2}
\begin{figure}[t]
\begin{center}
\includegraphics[width=6.5cm]{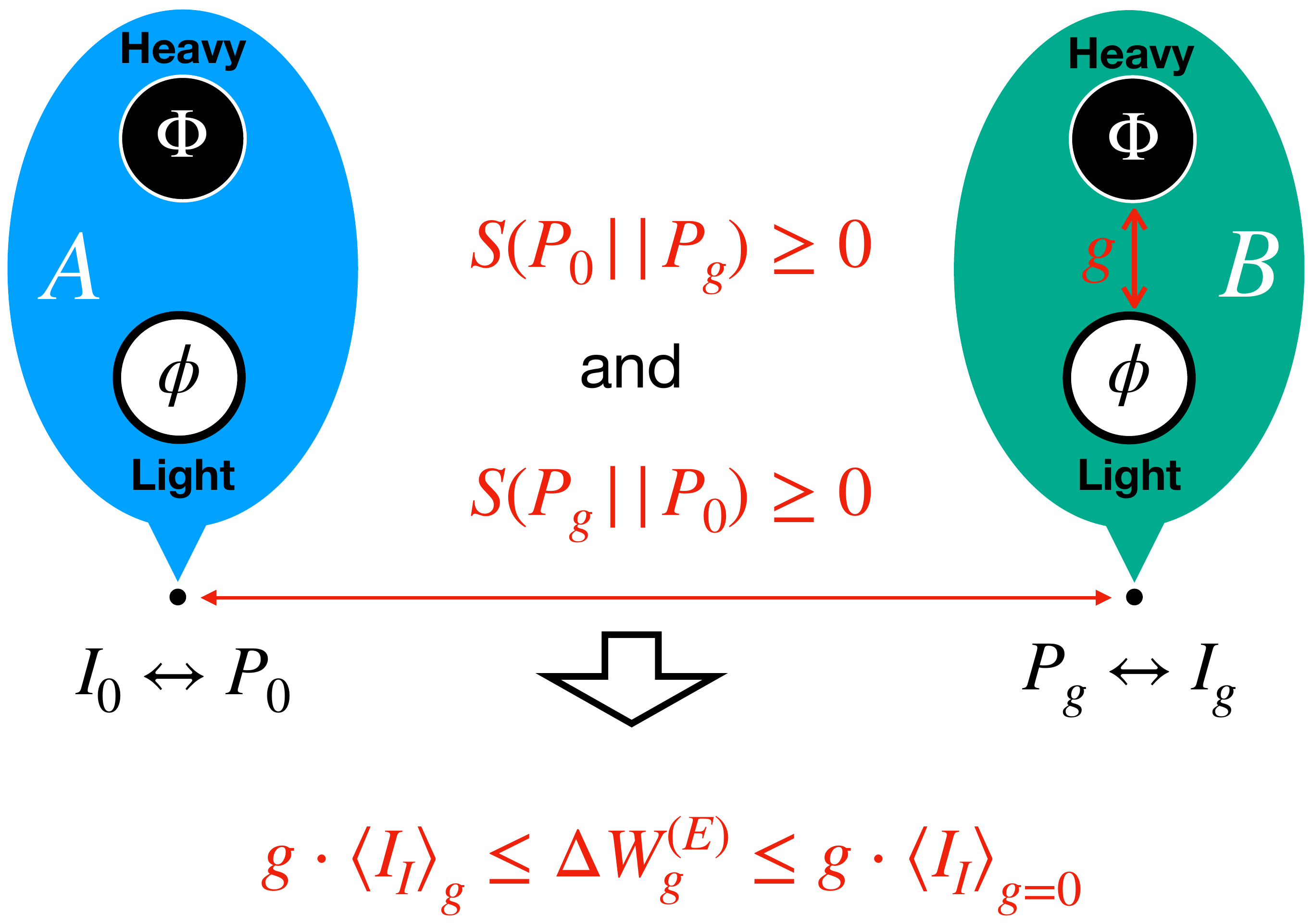}
\end{center}
\vspace{-0.4cm}
\caption{Schematic illustration of the distance between theory A and theory B, defined by the action $I_0$ and $I_g$, respectively. The distance, i.e. the relative entropy between $P_0$ and $P_g$, yields lower and upper bounds on perturbative correction from the interaction between heavy and light degrees of freedom to the Euclidean effective action.}.
\label{fig:dis}
\end{figure}
Consider a field theory that contains a set of light fields $\phi$'s and that of heavy fields $\Phi$'s; see Fig.~\ref{fig:dis}.
We introduce a thermodynamic system $A$ described by the Euclidean action $I_{0}[\phi,\Phi]$, which does not involve interactions between $\phi$'s and $\Phi$'s.
See Appendix~\ref{app:(I)} for the detailed definition of $I_0$\footnote{The theory A is a reference theory to obtain constraints on the low energy theory generated by $I_g$.
%
%
Note that we discuss the constraints on the theory described by $I_g$, not $I_0$.
}.
We define a probability distribution function for the system $A$ as
$
P_0 \equiv e^{-I_{0}}/Z_0[{\beta},\phi],
$
where $\beta$ is an inverse temperature of the system and $\phi$ denotes a background field corresponding to the light field, which is held fixed while the path integral over $\Phi$'s is performed.
Even if the path integral over $\phi$'s is performed, following explanations do not change much; see Appendix~\ref{app:der(5)}.
Note that heavy background fields are expressed by the light ones using the equation of motions.
The partition function is given as the Euclidean path integral 
$Z_0[{\beta},\phi]\equiv\int_{\beta}d[\Phi] e^{-I_{0}}$, which is determined by the Wick-rotated Lagrangian and boundary conditions. The effective action of the system is $W_0[{\beta},\phi]\equiv -\ln Z_0[{\beta},\phi]$.

The system $B$ is defined by $I_{g}[\phi,\Phi]\equiv{I}_{0}[\phi,\Phi]+g \cdot I_I[\phi,\Phi]$, where $I_I$ denotes interactions between $\phi$'s and $\Phi$'s, of which the probability distribution function is 
$
{P}_g \equiv e^{-{I}_{g}}/{Z}_g[{\beta},\phi],
$
where
$
{Z}_g[{\beta},\phi]\equiv \int_{\beta} d[\Phi] e^{-I_g}.
$
The effective action of the system is given by
$W_g[{\beta},\phi]\equiv-\ln Z_g[{\beta},\phi]$.
The coupling $g$ is an auxiliary parameter introduced to characterize the interaction.

The relative entropy between $P_0$ and ${P}_g$ is defined as
\begin{align}
S({P}_0\parallel P_g)&\equiv \int_{\beta} d[\Phi]\left({P}_0\ln {P}_0 -{P}_0\ln {P}_g \right)\geq 0.\label{eq:rel}
\end{align}
It is greater than or equal to zero, with the equality holding if and only if 
$P_g={P}_0$.
Thanks to this non-negativity, the relative entropy is often used as a distance between $P_0$ and $P_g$ even though it is not a symmetric function of the two sets of probabilities $S(P_0\parallel P_g)\neq S(P_g\parallel P_0)$. 
A simple algebra yields
\begin{align}
&S({P}_0\parallel P_g)=\int_{\beta}d[\Phi] \left({P}_0\ln {P}_0 -{P}_0\ln {P}_g \right)\notag
\\
&=-\ln {Z}_0[{\beta},\phi] +\ln Z_g[{\beta},\phi]+g\cdot  \int_{\beta} d[\Phi] {P}_0\cdot I_I \notag
\\
&={W}_0[{\beta},\phi] -W_g[{\beta},\phi] +g\cdot {\langle I_I\rangle}_{g=0},\label{eq:rels0}
\end{align}
where ${\langle I_I\rangle}_{g=0}\equiv \int_{\beta} d[\Phi] {P}_0\cdot I_I$ is an expectation value of the interaction, which satisfies $ ({d W_g}/{d g})_{g=0}=\int_{\beta} d[\Phi] {P}_0\cdot I_I$.
%
%
%
%
Note here that the derivation of Eq.~\eqref{eq:rels0} does not rely on the expansion in $g$.
In Eq.~\eqref{eq:rels0}, the path integral is performed only over the heavy degrees of freedom, and the self-interacting term of the light degrees of freedom cancels in ${\langle I_I\rangle}_{g=0}$.
The path integral over the light degrees of freedom does not change Eq.~\eqref{eq:rels0}; see Appendix~\ref{app:der(5)}.
It follows from the non-negativity of the relative entropy that
\begin{align}
\Delta   W_g^{(E)} \equiv W_g[{\beta},\phi]-W_0[{\beta},\phi] \leq g\cdot {\langle I_I\rangle}_{g=0},\label{eq:end}
\end{align}
where $\Delta W_g^{(E)}$ denotes the difference between the effective actions of the two systems in the Euclidean space.
Another choice of relative entropy,
\begin{equation}
S(P_g\parallel P_0)= W_g[{\beta},\phi]-{W}_0[{\beta},\phi]-g\cdot \int d[\Phi]P_g\cdot I_I, \label{eq:anrel}
\end{equation}
is related to the renormalization group~\cite{Gaite:1995yg}. It provides a lower bound
$
\Delta W_g^{(E)}\geq g \cdot{\langle I_I\rangle}_g
$
with ${\langle I_I\rangle}_{g}\equiv \int_{\beta} d[\Phi] {P}_g\cdot I_I$. We end up with the inequalities
\begin{align}
    g\cdot{\langle I_I\rangle}_g\leq \Delta W_g^{(E)}\leq g\cdot {\langle I_I\rangle}_{g=0},\label{eq:uplow}
\end{align}
which implies that the sign of the interaction controls the sign of perturbative corrections to the Euclidean effective action.
For example, the Euclidean effective action is increased in the theory with $g\cdot {\langle I_I\rangle}_{g}\geq 0$ but decreased in the theory with $g\cdot {\langle  I_I\rangle}_{g=0}\leq 0$. 
We emphasize that the inequalities~\eqref{eq:uplow} are applicable to the UV theories in which one-loop contributions from one-light-particle-irreducible diagrams with heavy-and-light-field mixing appear; then, we perform the path integral over $\phi$'s, i.e., focusing on the case of $d[\Phi]\to d[\phi]d[\Phi]$.
Even for such a case, Eq.~\eqref{eq:uplow} holds; see Appendix~\ref{app:der(5)}.

Here, we provide another explanation of the meaning of the upper bound of Eq.~\eqref{eq:uplow}.
By expanding $W_g[\beta,\phi]$ with respect to $g$, the upper bound of Eq.~\eqref{eq:uplow} yields
\begin{align}
    \frac{g^2}{2}\cdot \left(\frac{d^2 W_g}{dg^2}\right)_{g=0}+\mathcal{O}(g^3)\leq 0,\label{eq:g2}
\end{align}
where $W_g[\beta,\phi]= W_0[\beta,\phi]+g\cdot (dW_g/dg)_{g=0}+g^2\cdot(d^2 W_g/dg^2)_{g=0}/2+\mathcal{O}(g^3)$.
Note here that $g\cdot {\langle I_I\rangle}_{g=0}$ cancels in Eq.~\eqref{eq:g2}.
Therefore, the upper bound of Eq.~\eqref{eq:uplow} means that the Euclidean effective action decrease by the perturbative corrections of the second or higher order corrections for $g$.
As the inequalities~\eqref{eq:uplow} do not rely on either Lorentz symmetry or gauge symmetry, it works for a wide class of quantum theories that consist of both light and heavy degrees of freedom.
Consider thermodynamic systems described by quantum mechanics, which generally do not respect the Lorentz symmetry.
Define the Hamiltonian of the system as $H\equiv H_0+H_I$, where $H_I$ denotes the interaction between light and heavy degrees of freedom, and $H_0$ does not involve the interactions.
Define the theory $A$ as the Hamiltonian $H_0$.
By introducing the auxiliary parameter $g$, also define the theory $B$ as a Hamiltonian $H_g\equiv H_0 +g\cdot H_I$.
The density operators of the theory $A$ and $B$ are respectively defined as follows:
\begin{align}
    \rho_0\equiv \frac{e^{-\beta H_0}}{Z_0(\beta)},~~~\rho_g\equiv\frac{e^{-\beta H_g}}{Z_g(\beta)}
\end{align}
with the partition functions
\begin{align}
    Z_0(\beta)\equiv{\rm Tr}[e^{-\beta H_0}],~~~ Z_g(\beta)\equiv{\rm Tr}[e^{-\beta H_g}].
\end{align}
The non-negativity of relative entropy between $\rho_0$ and $\rho_g$ yields
\begin{align}
    S(\rho_0 ||\rho_g)&={\rm Tr}\left[\rho_0\ln \rho_0-\rho_0\ln \rho_g \right]\notag
    \\
    &=-\ln Z_0(\beta) +\ln Z_g (\beta) +g\cdot \beta {\rm Tr}[\rho_0 H_I]\notag
    \\
    &=W_0(\beta)-W_g (\beta) +g \cdot \beta {\langle H_I\rangle}_{g=0}\geq 0,\label{eq:Hamrel}
\end{align}
where the effective actions, and the expectation value of the interaction are defined as 
\begin{align}
    &W_0(\beta)\equiv-\ln Z_0(\beta),
    \\
    &W_g (\beta)\equiv-\ln Z_0(\beta),
    \\
    &{\langle H_I\rangle}_{g=0}\equiv{\rm Tr}[\rho_0 H_I].
\end{align}
Therefore, Eq.~\eqref{eq:Hamrel} yields
\begin{align}
    \Delta W_g^{(E)} =W_g (\beta)- W_0(\beta)\leq g\cdot \beta {\langle H_I \rangle}_{g=0}.\label{eq:hamup}
\end{align}
This inequality means that the Euclidean action decreases for the non-positive interacting theory defined by $g\cdot {\langle H_I \rangle}_{g=0}\leq 0$.
Also, consider another choice of the relative entropy $S(\rho_g||\rho_0)$ as follows:
\begin{align}
    S(\rho_g||\rho_0)&={\rm Tr}\left[\rho_g \ln \rho_g -\rho_g\ln \rho_0\right]\notag
    \\
    &=-\ln Z_g (\beta) +\ln Z_0 (\beta) -g \cdot \beta {\rm Tr}[\rho_g H_I]\notag
    \\
    &=W_g (\beta) -W_0 (\beta) -g\cdot \beta {\langle H_I\rangle}_{g}\geq 0,\label{eq:anrel2}
\end{align}
where ${\langle H_I\rangle}_{g}\equiv {\rm Tr}[\rho_g H_I]$.
Thus, Eq.~\eqref{eq:anrel2} yields
\begin{align}
    g\cdot \beta {\langle H_I\rangle}_{g}\leq W_g (\beta) -W_0 (\beta)=\Delta W_g^{(E)}.\label{eq:Hamlow}
\end{align}
Combining Eq.~\eqref{eq:hamup} and \eqref{eq:Hamlow}, we obtain
\begin{align}
    g\cdot \beta{\langle H_I\rangle}_g \leq \Delta W_g^{(E)} \leq g\cdot \beta {\langle H_I\rangle}_{g=0}.
\end{align}
This inequality corresponds to Eq.~\eqref{eq:uplow}, and it is clear that the UV properties such as symmetry is not necessary to obtain the entropy constraints of Eq.~\eqref{eq:uplow}.
%
%
%
\\[-3mm]
%

\section{Examples}
\label{sec:3}
%
Equipped with the distance between the two theories, we are ready to discuss the entropy constraints on various EFTs. 
In this section, under the setup of the previous section, i.e., the Euclidean path integral method is valid, and perturbative corrections are generated from the interacting term, we take two different approaches, i.e., the top-down approach and the bottom-up approach.
In the top-down approach, the relative entropy is evaluated in the UV theories with the light and heavy degrees of freedom to check the validity of the inequalities of \eqref{eq:uplow}. 
%
%
In the bottom-up approach, it is supposed that the UV theory is not specified for a given EFT, and the higher-dimensional operators of the EFT are generated by integrating out the heavy fields.
%
%
%
%
%
%
We focus on the EFTs, where the perturbative corrections to the leading terms, such as renormalizable terms
, can be eliminated by the field redefinition and study the constraints on the EFTs.
%

%
%

\subsection{Top-down approach}
%
    %
    We adopt the top-down approach and check the consistency of the entropy constraints by evaluating the effective action of the UV theories.
    The temperature of the system is assumed to be zero in the first four examples.
    
    (a) {A tree level UV completion of the single massless scalar field theory}:
%
%
%
Consider a theory in Minkowski space:
\begin{align}
    I^{(M)}&=\int d^4 x\bigg(\frac{1}{2}(\partial_{\mu}\phi\partial^{\mu}\phi)+\frac{1}{2}(\partial_{\mu}\Phi\partial^{\mu}\Phi)\notag
    \\
    &\quad\quad\quad-\frac{1}{2}m^2\Phi^2+\frac{\alpha}{\Lambda}\cdot \Phi (\partial_{\mu}\phi\partial^{\mu}\phi)+\frac{\beta}{\Lambda^2}\cdot \Phi^2 (\partial_{\mu}\phi\partial^{\mu}\phi)\bigg),
\end{align}
where $\phi$ denotes a massless scalar field, $\Phi$ is a heavy scalar field with mass $m$, $\alpha$ and $\beta$ are dimensionless parameters, and $\Lambda$ is some mass scale.
Define the actions $I_0$ and $I_I$ in Minkowski space as follows:
\begin{align}
    I_0^{(M)}&=\int d^4 x\bigg(\frac{1}{2}(\partial_{\mu}\phi\partial^{\mu}\phi)+\frac{1}{2}(\partial_{\mu}\Phi\partial^{\mu}\Phi)\notag
    \\
    &\quad\quad\quad-\frac{1}{2}m^2\Phi^2\bigg),
    \\
    I_I^{(M)}&=\frac{1}{\Lambda}\cdot\int d^4 x \Phi \left(\partial_{\mu}\phi\partial^{\mu}\phi\right)\left(\alpha+\frac{\beta}{\Lambda}\cdot \Phi\right).
\end{align}
The theory $B$ is defined as $I_g\equiv I_0+g\cdot I_I$ with the parameter $g$.
In this example, higher dimensional operators are generated at tree level, and the interaction $I_I$ does not contribute to the Euclidean effective action at the first order for $g$.
At tree level, the Euclidean effective actions are calculated as follows:
\begin{align}
    &W_g[\phi]=\int (d^4 x)_E \bigg(-\frac{1}{2}(\partial_{\mu}\phi\partial^{\mu}\phi)\notag
    \\
    &\quad\quad\quad-\frac{g^2 \alpha^2}{2\Lambda^2 m^2}(\partial_{\mu}\phi\partial^{\mu}\phi)^2 \bigg),\label{eq:wgtree1}
    \\
    &W_0[\phi]=-\int (d^4 x)_E \frac{1}{2}(\partial_{\mu}\phi\partial^{\mu}\phi).\label{eq:w0tree1}
\end{align}
From Eq.~\eqref{eq:wgtree1} and \eqref{eq:w0tree1}, the shift of the Euclidean effective action is obtained as
\begin{align}
    \Delta W^{(E)}_g=-\frac{g^2 \alpha^2}{2\Lambda^2 m^2}\int (d^4x)_E(\partial_{\mu}\phi\partial^{\mu}\phi)^2.\label{eq:shiftree1}
\end{align}
Then, the expectation value of the interaction is calculated as
\begin{align}
    g\cdot {\langle I_I\rangle}_{g=0}=g\cdot  \left(\frac{dW_g}{dg}\right)_{g=0}=0.\label{eq:IItree1}
\end{align}
Combining the upper bound of Eq.~\eqref{eq:uplow}, \eqref{eq:shiftree1}, and \eqref{eq:IItree1}, we obtain the positivity bound as follows:
\begin{align}
    \frac{g^2 \alpha^2}{2\Lambda^2 m^2}\int (d^4 x)_E(\partial_{\mu}\phi\partial^{\mu}\phi)^2 \geq 0\Rightarrow \frac{g^2 \alpha^2}{2\Lambda^2 m^2}\geq 0.
\end{align}
Therefore, the coefficient of the dimension-eight operator of Eq.~\eqref{eq:wgtree1} is positive because of the non-negativity of relative entropy.

(b) {A tree level UV completion of the single mass less scalar field theory with a linear term}:
We discuss the effects of the linear term of $\Phi$ in $I_0$, which generally generates a non-zero expectation value of the interaction ${\langle I_I\rangle}_{g=0}$.
As shown later, the constraints on EFTs can arise even if ${\langle I_I\rangle}_{g=0}$ takes a non-zero value.
%
%
Consider an action in Minkowski space defined as
\begin{align}
    I^{(M)}&=\int d^4 x \bigg(\frac{1}{2}(\partial_{\mu}\phi \partial^{\mu}\phi)+\frac{1}{2}(\partial_{\mu}\Phi \partial^{\mu}\Phi)\notag
    \\
    &-\frac{1}{2}m^2 \Phi^2+m^2 v \Phi-\frac{1}{2}m^2 v^2\notag
    \\
    &+\frac{\alpha}{\Lambda}\cdot  \Phi(\partial_{\mu}\phi \partial^{\mu}\phi)+\frac{\beta}{\Lambda^2}\cdot \Phi^2 (\partial_{\mu}\phi \partial^{\mu}\phi)\bigg),\label{eq:scaltree}
\end{align}
where $v$ is a dimensionful parameter.
Define $I_0$ and $I_I$ as follows:
\begin{align}
    I_0^{(M)}&=\int d^4 x \bigg(\frac{1}{2}(\partial_{\mu}\phi \partial^{\mu}\phi)+\frac{1}{2}(\partial_{\mu}\Phi \partial^{\mu}\Phi)\notag
    \\
    &-\frac{1}{2}m^2 \Phi^2+m^2 v \Phi-\frac{1}{2}m^2 v^2 \bigg),
    \\
    I_I^{(M)}&=\frac{1}{\Lambda}\cdot \int d^4 x\Phi(\partial_{\mu}\phi \partial^{\mu}\phi)\left(\alpha+\frac{\beta}{\Lambda}\cdot \Phi\right). 
    %
\end{align}
By introducing the parameter $g$, define the theory $B$ as $I_g\equiv I_0 +g \cdot I_I$.
The Euclidean effective actions are calculated as follows:
\begin{align}
    &W_g[\phi]=\int (d^4x)_E\bigg(-\frac{1}{2}(1+2 g\cdot \frac{\alpha v}{\Lambda}+2g\cdot \frac{\beta v^2}{\Lambda^2} )(\partial_{\mu}\phi\partial^{\mu}\phi)\notag
    \\
    &\quad\quad\quad-\frac{g^2 }{2 \Lambda^2 m^2}\left(\alpha+\frac{2v\beta}{\Lambda}\right)^2(\partial_{\mu}\phi\partial^{\mu}\phi)^2
    \bigg),\label{eq:wgli}
    \\
    &W_0[\phi]=-\int (d^4x)_E\frac{1}{2}(\partial_{\mu}\phi\partial^{\mu}\phi).\label{eq:w0li}
\end{align}
From Eq.~\eqref{eq:wgli} and \eqref{eq:w0li}, the shift of the Euclidean effective action is given by
\begin{align}
    \Delta W^{(E)}_g &=\du{-}g\cdot \frac{v}{\Lambda}\left(\alpha+\frac{\beta}{\Lambda}v\right)\int (d^4 x)_E (\partial_{\mu}\phi\partial^{\mu}\phi)\notag
    \\
    &-\frac{g^2 }{2\Lambda^2 m^2}\left(\alpha+\frac{2v\beta}{\Lambda}\right)^2\int (d^4 x)_E (\partial_{\mu}\phi\partial^{\mu}\phi)^2.\label{eq:Deltreelin1}
\end{align}
At tree level, the expectation value of $I_I$ in Euclidean space is calculated as
\begin{align}
    g\cdot {\langle I_I\rangle}_{g=0}&=g\cdot \left(\frac{dW_g}{dg}\right)_{g=0}\notag
    \\
    &=-g\cdot \int d[\Phi]P_0\cdot \int (d^4x)_E \frac{1}{\Lambda}\left(\alpha \Phi+\frac{\beta}{\Lambda}\Phi^2\right)  (\partial_{\mu}\phi \partial^{\mu}\phi)\notag
    \\
    &=-g\cdot  \frac{v}{\Lambda}\left(\alpha +\frac{\beta}{\Lambda}v\right)\int (d^4x)_E   (\partial_{\mu}\phi \partial^{\mu}\phi),\label{eq:INLIN}
\end{align}
where we used $P_0=e^{-I_0}/Z_0$ with $Z_0=\int d[\Phi]e^{-I_0}$.
It is clear that the expectation value $g\cdot {\langle I_I\rangle}_{g=0}$ generally takes a non-zero value even in the tree-level UV completion.
For $v=0$, both linear term of $\Phi$ in Eq.~\eqref{eq:scaltree} and expectation value $g\cdot {\langle I_I\rangle}_{g=0}$ vanish.
This fact holds in general UV theory involving the linear term of $\Phi$.
Here, it should be noted that the constraint on the higher-dimensional operators can be derived even if $g\cdot {\langle I_I\rangle}_{g=0}$ takes a non-zero value.
Combining Eq.~\eqref{eq:Deltreelin1}, \eqref{eq:INLIN}, and the upper bound of Eq.~\eqref{eq:uplow}, the expectation value $g\cdot {\langle I_I\rangle}_{g=0}$ cancels, and we obtain
\begin{align}
    \Delta W^{(E)}_{g=1}\leq {\langle I_I\rangle}_{g=0}&\Rightarrow
    \frac{ g^2}{2\Lambda^2 m^2}\left(\alpha+\frac{2v\beta}{\Lambda}\right)^2\int (d^4 x)_E (\partial_{\mu}\phi\partial^{\mu}\phi)^2\geq 0\notag
    \\
    &\Rightarrow \frac{g^2}{2\Lambda^2 m^2}\left(\alpha+\frac{2v\beta}{\Lambda}\right)^2\geq 0.\label{eq:res1}
\end{align}
Consequently, the relative entropy yields the constraint on the coefficient of the dimension-eight operator of Eq.~\eqref{eq:wgli}.
The reason why the expectation value $ {\langle I_I\rangle}_{g=0}$ cancels in Eq.~\eqref{eq:res1} is the same as that it cancels in Eq.~\eqref{eq:g2}. 

Here, we show that the expectation value $g\cdot {\langle I_I\rangle}_{g=0}$ can be removed by a redefinition of $\Phi$.
%
%
By defining $\Phi\equiv\eta+v$, the action of Eq.~\eqref{eq:scaltree} is expressed as follows:
\begin{align}
    I^{(M)}&= \int d^4x \bigg(
    \frac{1}{2}(\partial_{\mu}\phi\partial^{\mu}\phi)+\frac{1}{2}(\partial_{\mu}\eta\partial^{\mu}\eta)-\frac{1}{2}m^2 \eta^2 \notag
    \\
    &
    +\frac{\alpha}{\Lambda}\cdot v (\partial_{\mu}\phi\partial^{\mu}\phi)+\frac{\beta}{\Lambda^2}\cdot v^2 (\partial_{\mu}\phi\partial^{\mu}\phi)\notag
    \\
    &+\frac{\alpha}{\Lambda}\cdot \eta (\partial_{\mu}\phi\partial^{\mu}\phi)+\frac{\beta}{\Lambda^2}\cdot 2v \eta(\partial_{\mu}\phi\partial^{\mu}\phi)\notag
    \\
    &+\frac{\beta}{\Lambda^2}\cdot \eta^2 (\partial_{\mu}\phi\partial^{\mu}\phi)
    \bigg).\label{eq:Ire}
\end{align}
Note here that the liner term of $\eta$ does not arise in Eq.~\eqref{eq:Ire}.
Then, we define,
\begin{align}
    {I'}^{(M)}_0&\equiv \int d^4x \bigg(
    \frac{1}{2}(\partial_{\mu}\phi\partial^{\mu}\phi)+\frac{1}{2}(\partial_{\mu}\eta\partial^{\mu}\eta)\notag
    \\
    &-\frac{1}{2}m^2 \eta^2 
    +\frac{\alpha}{\Lambda}\cdot v (\partial_{\mu}\phi\partial^{\mu}\phi)+\frac{\beta}{\Lambda^2}\cdot v^2 (\partial_{\mu}\phi\partial^{\mu}\phi)
    \bigg),
    \\
    {I'}^{(M)}_I&\equiv \frac{1}{\Lambda}\cdot\int d^4x \eta  (\partial_{\mu}\phi\partial^{\mu}\phi)\left(\alpha+\frac{\beta}{\Lambda}\cdot 2v +\frac{\beta}{\Lambda}\cdot \eta\right),
\end{align}
where $I'_I$ denotes the interaction, and $I'_0$ does not involve it.
By introducing the parameter $g$, define the theory $B$ as ${I'}_g\equiv {I'}_0 +g\cdot {I'}_I$.
Then, the Euclidean effective actions are calculated as follows:
\begin{align}
    {W'}_g[\phi]&=\int (d^4x)_E \bigg(-\frac{1}{2}\left(1+2\frac{\alpha v}{\Lambda}+2\frac{\beta v^2}{\Lambda^2}\right)(\partial_{\mu}\phi\partial^{\mu}\phi)\notag
    \\
    &-\frac{g^2 }{2\Lambda^2 m^2}\left(\alpha+\frac {2v\beta}{\Lambda}\right)^2(\partial_{\mu}\phi\partial^{\mu}\phi)^2\bigg),
    \\
    {W'}_0[\phi]&=-\int (d^4x)_E \frac{1}{2}\left(1+2\frac{\alpha v}{\Lambda}+2\frac{\beta v^2}{\Lambda^2}\right)(\partial_{\mu}\phi\partial^{\mu}\phi).
\end{align}
The shift of the Euclidean effective action is calculated as
\begin{align}
    \Delta {W'}^{(E)}_{g}&=W'_g[\phi]-W'_0[\phi]\notag
    \\
    &=-\int (d^4x)_E \frac{g^2 }{2\Lambda^2 m^2}\left(\alpha+\frac {2v\beta}{\Lambda}\right)^2(\partial_{\mu}\phi\partial^{\mu}\phi)^2\label{eq:delprW}
\end{align}
Since the linear term of $\eta$ does not arises in Eq.~\eqref{eq:Ire}, the expectation value of the interaction ${I}_I$ in Euclidean space takes a zero value as follows:
\begin{align}
    g\cdot {\langle {I'}_I\rangle}_{g=0} &=g\cdot \left(\frac{d{W'}_g}{dg}\right)_{g=0}=0.\label{eq:IIpri}
\end{align}
Then, from Eq.~\eqref{eq:delprW}, \eqref{eq:IIpri} and the upper bound of Eq.~\eqref{eq:uplow}, we obtain
\begin{align}
    {W'}_{g}[\phi]\leq {\langle I'_I\rangle}_{g=0}&\Rightarrow\frac{g^2 }{2\Lambda^2 m^2}\left(\alpha+\frac {2v\beta}{\Lambda}\right)^2\notag
    \\
    &\times\int (d^4x)_E(\partial_{\mu}\phi\partial^{\mu}\phi)^2\geq 0\notag
    \\
    &\Rightarrow \frac{g^2 }{2\Lambda^2 m^2}\left(\alpha+\frac {2v\beta}{\Lambda}\right)^2\geq 0.\label{eq:res2}
\end{align}
This result is the same as Eq.~\eqref{eq:res1} because the expectation value of the interaction ${I}_I$ cancels in the relative entropy, i.e., the relative entropy is invariant under the redefinition to eliminate the linear term of $\Phi$.  
Therefore, we found that the constraint on the EFT does not depend on the condition of vanishing the linear term.
%
%
%
%

The above explanations are based on the theory of Eq.~\eqref{eq:scaltree}, but the invariance of the inequality of Eq.~\eqref{eq:end} under the field redefinition to eliminate the linear term of $\Phi$ hold in general UV theories.
%
%
Similar to the above explanations, take two different approaches.

First, consider the UV theory with the linear term as follows:
\begin{align}
    I[\phi,\Phi]=I_0^{\rm lin}[\phi,\Phi] +I_I[\phi,\Phi],\label{eq:genI}
\end{align}
where $I_0^{\rm lin}$ involves the linear term of $\Phi$, and $I_I$ is the interacting term.
Consider the classical solution $v$ of $I_0^{\rm lin}$, where indices of the classical solution, such as Lorentz indices, are omitted.
Also, the classical solution of $I$ for $\Phi$ is assumed to be $v+f(\phi)$, where $f$ depends on the light field $\phi$ because of the interacting term $I_I$. 
Note here that $f(\phi)$ vanishes in the limit of $I_I\to 0$.
By introducing the parameter $g$, we define $I_g\equiv I_0^{\rm lin}+g\cdot I_I$.
At tree level, the Euclidean effective actions of $I_{g=1}$ and $I_{g=0}$ are respectively calculated as follows:
\begin{align}
    W_{g=1}[\phi]&=I^{\rm lin}_0[\phi,v+f(\phi)]+ I_I[\phi,v+f(\phi)],
    \\
    W_0[\phi]&=I^{\rm lin}_0[\phi,v].
\end{align}
The shift of the Euclidean effective action is calculated as
\begin{align}
    \Delta W^{(E)}_{g=1}&\equiv W_{g=1}[\phi]-W_0[\phi]\notag
    \\
    &=I^{\rm lin}_0[\phi,v+f(\phi)]+ I_I[\phi,v+f(\phi)]-I^{\rm lin}_0[\phi,v].\label{eq:delWGgen}
\end{align}
The expectation value of the interaction $I_I$ in the Euclidean space is also calculated as
\begin{align}
    {\langle I_I\rangle}_{g=0}=\int d[\Phi]P_0[\Phi] I_I[\phi,\Phi]= I_I[\phi,v],\label{eq:IIgn}
\end{align}
where $P_0[\Phi]\equiv e^{-I_0^{\rm lin}}/Z_0[\phi]$ with $Z_0[\phi]\equiv \int d[\Phi]e^{-I_0^{\rm lin}}$.
Combining Eq.~\eqref{eq:delWGgen}, \eqref{eq:IIgn}, and the inequality of Eq.~\eqref{eq:end}, we obtain
\begin{align}
    &W_{g=1}[\phi]\leq {\langle I_I\rangle}_{g=0}\notag
    \\
    &\Rightarrow I^{\rm lin}_0[\phi,v+f(\phi)]+ I_I[\phi,v+f(\phi)]\notag
    \\
    &-I_I[\phi,v]-I^{\rm lin}_0[\phi,v]\leq 0.\label{eq:re1gn}
\end{align}
This inequality corresponds to Eq.~\eqref{eq:res1}.

Next, consider the field redefinition $\Phi\equiv \eta+v$.
Then, Eq.~\eqref{eq:genI} is expressed as
\begin{align}
    I[\phi,\Phi]&= I_0^{\rm lin}[\phi,\eta+v]+I_I[\phi,\eta+v]\notag
    \\
    &=\left(I_0^{\rm lin}[\phi,\eta+v]+I_{I}[\phi,v] \right)\notag
    \\
    &+\left(I_I[\phi,\eta+v]-I_{I}[\phi,v]\right).\label{eq:linrem1}
\end{align}
For convenience, define
\begin{align}
    &I'_0[\phi,\eta]\equiv I_0^{\rm lin}[\phi,\eta+v]+I_{I}[\phi,v],\label{eq:linrem2}
    \\
    &I'_I[\phi,\eta]\equiv I_I[\phi,\eta+v]-I_{I}[\phi,v],\label{eq:linrem3}
\end{align}
where $I'_0$ does not include the linear term of $\eta$.
By introducing the parameter $g$, we also define $I'_g\equiv I'_0+g\cdot I'_I$.
At tree level, the Euclidean effective actions of $I'_g$ and $I'_0$ are respectively calculated as follows:
\begin{align}
    W'_{g=1}[\phi]&=I'_0[\phi,f(\phi)]+ I'_I[\phi,f(\phi)]\notag
    \\
    &=I_0^{\rm lin}[\phi,v+f(\phi)]+I_{I}[\phi,v]+ I'_I[\phi,f(\phi)],
    \\
    W'_0[\phi]&=I_0^{\rm lin}[\phi,v]+I_{I}[\phi,v].
\end{align}
Note here that the classical solution of $I$ for $\eta$ is $f(\phi)$, and that of $I$ for $\Phi=v+\eta$ is $v+f(\phi)$.  
Similarly, the classical solution of $I'_0$ for $\eta$ is zero, and that of $I'_0$ for $\Phi=v+\eta$ is $v$. 
Then, the shift of the Euclidean effective action is calculated as
\begin{align}
    \Delta W'_{g=1}[\phi]&\equiv W'_{g=1}[\phi]-W'_0[\phi]\notag
    \\
    &=I_0^{\rm lin}[\phi,v+f(\phi)]+I_{I}[\phi,v]+ I'_I[\phi,f(\phi)]\notag
    \\
    &-I_0^{\rm lin}[\phi,v]-I_{I}[\phi,v]\notag
    \\
    &=I_0^{\rm lin}[\phi,v+f(\phi)]+ I'_I[\phi,f(\phi)]-I_0^{\rm lin}[\phi,v].\label{eq:DelWprige}
\end{align}
The expectation value of the interaction $I_I$ in the Euclidean space is calculated as
\begin{align}
     {\langle I'_I\rangle}_{g=0}=\int d[\eta] P'_0[\eta] I'_I[\phi,\eta]=0,\label{eq:Iprigen}
\end{align}
where $P'_0[\eta]\equiv e^{-I'_0[\phi,\eta]}/Z'_0[\phi]$ with $Z'_0[\phi]\equiv \int d[\eta]e^{-I'_0[\phi,\eta]}$.
Combining Eq.~\eqref{eq:DelWprige}, \eqref{eq:Iprigen}, and the inequality of Eq.~\eqref{eq:end}, we obtain
\begin{align}
    &W'_{g=1}[\phi]\leq {\langle I'_I\rangle}_{g=0} \notag
    \\
    &\Rightarrow
    I_0^{\rm lin}[\phi,v+f(\phi)]+ I'_I[\phi,f(\phi)]-I_0^{\rm lin}[\phi,v]\notag
    \\
    &=I_0^{\rm lin}[\phi,v+f(\phi)]+ I_I[\phi,v+f(\phi)]-I_{I}[\phi,v]\notag
    \\
    &-I_0^{\rm lin}[\phi,v]
    \leq 0.
\end{align}
This result is the same as Eq.~\eqref{eq:re1gn}.
Consequently, it is found that the inequality of Eq.~\eqref{eq:end} is invariant under the field redefinition to remove the linear term of $\Phi$.
We often define the heavy fields such that the linear term vanishes for ease of calculation of the relative entropy.
We mention it in the following calculations when such a definition is used.

(c) { Euler-Heisenberg theory}:
The action of quantum electrodynamics of electron field ($\psi$) in Minkowski space is
\begin{align}
I^{( M)}=\int d^4 x \left(-\frac{1}{4}F_{\mu\nu}F^{\mu\nu} +\bar{\psi} (i\slashed{D}-m)\psi\right),
\end{align}
where $D_{\mu}=\partial_{\mu}+ie A_{\mu}$ is the covariant derivative, $m$ is the mass of $\psi$, and $F_{\mu\nu}=\partial_{\mu}A_{\nu}-\partial_{\nu}A_{\mu}$ is the field strength of photon.
Define $I_0$ and $I_I$ as follows:
\begin{align}
    &I_0^{(M)}=\int d^4x \left(-\frac{1}{4}F_{\mu\nu}F^{\mu\nu} +\bar{\psi}(i\slashed{\partial}-m)\psi\right),
    \\
    &I_I^{(M)}=-e\int d^4x \bar{\psi}\gamma_{\mu}\psi A^{\mu}.
\end{align}
By introducing the parameter $g$, the theory $B$ is defined as $I_g\equiv I_0 +g\cdot I_I$.
The Euclidean effective actions of theories $A$ and $B$ are respectively calculated as follows:
\begin{align}
    W_0[\overline{A}]&=\int (d^4x)_E \bigg(\frac{1}{4}\overline{F}_{\mu\nu}\overline{F}^{\mu\nu} \bigg),\label{eq:W0EH}
    \\
    W_g[\overline{A}]&=\int (d^4x)_E \bigg(\frac{1}{4}\overline{F}_{\mu\nu}\overline{F}^{\mu\nu}\notag
    \\
    &-\frac{g^4\cdot{e}^4}{6! \pi^2 m^4}\left(\gamma_{1}(\bar{F}_{\mu\nu}\bar{F}^{\mu\nu})^2 +\gamma_{2} (\bar{F}_{\mu\nu}\tilde{\bar{F}}^{\mu\nu})^2 \right) \bigg),\label{eq:WgEH}
\end{align}
where $\int (d^4 x)_E$ is the volume of Euclidean space-time, $\widetilde{\overline{F}}^{\mu\nu}=\epsilon^{\mu\nu\rho\sigma}\overline{F}_{\rho\sigma}/2$, the Wilson coefficents $\gamma_{1}=1/2$ and $\gamma_{2}=7/8$~\cite{Quevillon:2018mfl}, $\overline{A}_{\mu}$ is the background field satisfing $\partial_{\mu} \overline{F}^{\mu\nu}=0$ with constant $\overline{F}^{\mu\nu}$, and the vacuum energy is omitted because it cancels in relative entropy.
The details of the wave function renormalizations are explained in Appendix~\ref{app:(IV-3)}.
From Eq.~\eqref{eq:W0EH} and \eqref{eq:WgEH}, the difference of the Euclidean effective action at the one loop level is
\begin{align}
\Delta W_g^{(E)}&= -\frac{g^4\cdot{e}^4}{6! \pi^2 m^4} \int (d^4 x)_E \bigg(\gamma_{1}(\bar{F}_{\mu\nu}\bar{F}^{\mu\nu})^2 +\gamma_{2} (\bar{F}_{\mu\nu}\tilde{\bar{F}}^{\mu\nu})^2 \bigg).
\label{eq:lh1}
\end{align}
From Eq.~\eqref{eq:WgEH}, the expectation value of the interaction $I_I$ in the Euclidean space is also calculated as follows:
\begin{align}
    g\cdot{\langle I_I\rangle}_{g=0}&= g\cdot \left(\frac{dW_g}{dg}\right)_{g=0}=0.\label{eq:IIEH}
\end{align}
Combining the inequality~\eqref{eq:uplow}, Eq.~\eqref{eq:IIEH} and \eqref{eq:lh1}, the shift of the Euclidean effective action is given by 
\begin{align}
    &\Delta W_g^{(E)}\leq  g\cdot {\langle I_I\rangle}_{g=0}\notag
    \\
    &\Rightarrow 
    -\frac{g^4\cdot{e}^4}{6! \pi^2 m^4} \int (d^4 x)_E \bigg(\gamma_{1}(\bar{F}_{\mu\nu}\bar{F}^{\mu\nu})^2 +\gamma_{2} (\bar{F}_{\mu\nu}\tilde{\bar{F}}^{\mu\nu})^2 \bigg)\leq 0.\label{eq:DelEHloop}
\end{align}
The left-hand side of Eq.~\eqref{eq:DelEHloop} denotes the linear combination of dimension-eight operators of Eq.~\eqref{eq:WgEH}, and it is found that the constraints on the EFTs arise from the relative entropy. 
%
Consequently, the Euler-Heisenberg theory satisfies the non-negativity of relative entropy because $\gamma_1$ and $\gamma_2$ are positive values.

(d) Massive, gravitationally coupled scalar field at tree level~\cite{Cheung:2018cwt}:
To explain how to define the interaction $I_I$ in gravitational theories, consider a simple theory in Minkowski space:
\begin{align}
    &I^{(M)}[g_{\mu\nu};R_{\mu\nu\rho\sigma},A,\Phi]=\int d^4 x\sqrt{-g}\bigg(\frac{M_{\rm Pl}^2}{2} R-\frac{1}{4}F_{\mu\nu}F^{\mu\nu}\notag
    \\
    &-\left(a_{\Phi} R+b_{\Phi} F_{\mu\nu}F^{\mu\nu}\right)\Phi +\frac{1}{2}g^{\mu\nu}\partial_{\mu}\Phi\partial_{\nu}\Phi-\frac{1}{2}m_{\Phi}^2 \Phi^2 \bigg),
\end{align}
where $R_{\mu\nu\rho\sigma}$ is the Riemann tensor, $R$ is the scalar curvature, and $a_{\Phi}, b_{\Phi}$ are dimensionful coupling constants.
%
%
%
%
Define the non-interacting and interacting terms as follows:
\begin{align}
    I_0^{(M)}[g_{\mu\nu};R_{\mu\nu\rho\sigma},A,\Phi]&=I^{(M)}[g_{\mu\nu};R_{\mu\nu\rho\sigma},A,0]\notag
    \\
    &+I^{(M)}[g_{\mu\nu};0,0,\Phi],\label{eq:EMtopI0}
    \\
    I_I^{(M)}[g_{\mu\nu};R_{\mu\nu\rho\sigma},A,\Phi]&=I^{(M)}[g_{\mu\nu};R_{\mu\nu\rho\sigma},A,\Phi]\notag
    \\
    &-I^{(M)}_0[g_{\mu\nu};R_{\mu\nu\rho\sigma},A,\Phi].
\end{align}
%
%
It should be noted that the theory $A$ does not include the interaction between $\Phi$ and $A_{\mu}$, $R_{\mu\nu\rho\sigma}$, but the interaction between $g_{\mu\nu}$ and $\Phi$. 
The higher-derivative operators generally arise from the interaction between $g_{\mu\nu}$ and $\Phi$, but such effects are discussed later in (h).
%
%
%
The theory $B$ is defined as $I_g=I_0+g\cdot I_I$ with the parameter $g$.
%
%
%
In this example, $I_0$ and $I_I$ are obtained as
\begin{align}
    I_0^{(M)}[g_{\mu\nu};R_{\mu\nu\rho\sigma},A,\Phi]&=\int d^4 x \sqrt{-g}\bigg(\frac{M_{\rm Pl}^2}{2} R-\frac{1}{4}F_{\mu\nu}F^{\mu\nu}\notag
    \\
    &+\frac{1}{2}g^{\mu\nu}\partial_{\mu}\Phi\partial_{\nu}\Phi-\frac{1}{2}m_{\Phi}^2 \Phi^2\bigg),\label{eq:GI0}
    \\
    I_I^{(M)}[g_{\mu\nu};R_{\mu\nu\rho\sigma},A,\Phi]&=-\int d^4x \sqrt{-g}(a_{\Phi} R +b_{\Phi} F_{\mu\nu}F^{\mu\nu})\Phi.\label{eq:GII}
\end{align}
The Euclidean effective actions of theories $A$ and $B$ are respectively calculated as follows:
\begin{align}
    W_0[\overline{g}_{\mu\nu},\overline{A}]&=\int (d^4x)_E \sqrt{\overline{g}}\bigg(-\frac{M_{\rm Pl}^2}{2} \overline{R} +\frac{1}{4}\overline{F}_{\mu\nu}\overline{F}^{\mu\nu}\bigg),\label{eq:Wg0}
    \\
    W_g[\overline{g}_{\mu\nu},\overline{A}]&=\int (d^4x)_E \sqrt{\overline{g}}\bigg(-\frac{M_{\rm Pl}^2}{2} \overline{R} +\frac{1}{4}\overline{F}_{\mu\nu}\overline{F}^{\mu\nu}\notag
    \\
    &-\frac{g^2}{2m_{\Phi}^2}\left(a_{\Phi} \overline{R} +b_{\Phi} \overline{F}_{\mu\nu}\overline{F}^{\mu\nu}\right)^2 \bigg),\label{eq:Wg1}
\end{align}
where $\overline{g}_{\mu\nu}$ and $\overline{A}_{\mu}$ denote the background fields.
From Eq.~\eqref{eq:Wg0} and \eqref{eq:Wg1}, the shift of the Euclidean effective action is calculated as
\begin{align}
    \Delta W^{(E)}_g&=W_g[\overline{g}_{\mu\nu},\overline{A}]-W_0[\overline{g}_{\mu\nu},\overline{A}]\notag
    \\
    &=-\frac{g^2}{2m_{\Phi}^2} \int (d^4x)_E\sqrt{\overline{g}} \left(a_{\Phi} \overline{R} +b_{\Phi} \overline{F}_{\mu\nu}\overline{F}^{\mu\nu}\right)^2.\label{eq:DelEGRtr}
\end{align}
From Eq.~\eqref{eq:Wg1}, the expectation value of the interaction $I_I$ at the tree level is calculated as
\begin{align}
    g\cdot{\langle I_I\rangle}_{g=0}=g\cdot\left(\frac{dW_g}{dg}\right)_{g=0}=0.\label{eq:IIGRTR}
\end{align}
%
%
%
%
Eqs.~\eqref{eq:uplow}, \eqref{eq:DelEGRtr}, and \eqref{eq:IIGRTR} yield
\begin{align}
    &\Delta W^{(E)}_g \leq  g\cdot {\langle I_I\rangle}_{g=0}\notag
    \\
    &\Rightarrow 
    -\frac{g^2}{2m_{\Phi}^2} \int (d^4x)_E\sqrt{\overline{g}} \left(a_{\Phi} \overline{R} +b_{\Phi} \overline{F}_{\mu\nu}\overline{F}^{\mu\nu}\right)^2\leq 0.\label{eq:DelWEMtree}
\end{align}
%
%
%
%
The left-hand side of Eq.~\eqref{eq:DelWEMtree} denotes the linear combination of higher-dimensional operators of Eq.~\eqref{eq:Wg1}, and the constraints on the EFT arise from the relative entropy.
As explained later, the entropy constraints by the relative entropy is a generalization of Ref.~\cite{Cheung:2018cwt}, which includes the result of Ref.~\cite{Cheung:2018cwt}.

(e) {A spin system in one dimension}:
Consider a spin system in one dimension defined by a Hamiltonian
\begin{align}
    H_{\mu}=-J \sum_{i=1}^{N/2} \sigma_{2i-1}\sigma_{2i}-\mu M\sum_{i=1}^N \sigma_i,
\end{align}
where $\sigma_i=\pm 1$ is a spin on site $i$, $J$ is a coupling constant characterizing exchange interactions, $N$ is the number of sites, $\mu$ is a magnetic moment, and $M$ is an external magnetic field.
Then, define $H_0$ and $H_I$ as follows:
\begin{align}
    H_0\equiv -J \sum_{i=1}^{N/2} \sigma_{2i-1}\sigma_{2i},~~~H_I\equiv -\mu M\sum_{i=1}^N \sigma_i.
\end{align}
By introducing the parameter $g$, the theory $B$ is defined as $H_g\equiv H_0+ g\cdot H_I$.
Then, density operators are given by
\begin{align}
    \rho_0=\frac{e^{-\beta H_0}}{Z_0(\beta)},~~~\rho_g=\frac{e^{-\beta H_g}}{Z_g(\beta)},
\end{align}
with the partition functions,
\begin{align}
Z_0(\beta)&={\rm Tr}[e^{-\beta H_0}]\notag
\\
&=\left(2\left\{e^{\beta J}+e^{-\beta J} \right\} \right)^{N/2},
\\
Z_g(\beta)&={\rm Tr}[e^{-\beta H_g}]\notag
\\
&=\left(2\left\{e^{\beta J} \cosh (2\beta g \mu M)+e^{-\beta J} \right\} \right)^{N/2}.
\end{align}
For each of the theories, the effective actions are defined as $W_0(\beta)=-\ln Z_0(\beta)$, and $W_g(\beta)=-\ln Z_g(\beta)$.
The expectation value of the interaction is calculated as ${\rm Tr}[\rho_0 H_I]=0$, and the shift of the Euclidean effective action is given by
\begin{align}
    \Delta W_g^{(E)}&= W_g[\beta]-W_0[\beta]\notag
    \\
    &=\frac{N}{2}\ln \left[\frac{e^{\beta J}+e^{-\beta J}}{e^{\beta J}\cosh(2\beta g\mu M)+e^{-\beta J}}\right]\leq 0.\label{eq:spindel}
\end{align}
This result is consistent with Eq.~\eqref{eq:hamup} because $\cosh(2\beta g\mu M)\geq 1$.
The entropy constraints explain why the free energy of the spin system decreases by the external magnetic field.
    
    \subsection{Bottom-up approach}
    \label{sec:Bott}
    %
    We adopt the bottom-up approach and derive the constraints on a class of EFTs, where corrections to the leading terms, such as the kinetic term and the Einstein-Hilbert term, can be eliminated by the redefinition of the light field.
    For such a class of EFTs, consider the higher-dimensional operators generated by integrating out $\Phi$.
    The interaction of the UV theory is generally expressed as $I_I[\phi,\Phi]=\int (d^4x)_E \mathcal{O}[\Phi] \otimes J[\phi]$.
    Throughout the bottom-up approach, we suppose this general form of interaction for a given EFT.
    Here, assume $J[\phi]$ does not include the higher-dimensional operators.
    %
    In other words, we assume corrections from the interactions involving higher-derivative operators of the light fields are not dominant in the EFTs.
    The assumption is quantitatively reasonable because the higher-dimensional operator $J[\phi]$ is suppressed by a heavier mass than $\Phi$.
    %
    %
    The expectation value of the interaction is calculated as follows:
\begin{align}
    {\langle I_I\rangle}_{g=0}&=\left(\frac{dW_g}{dg}\right)_{g=0}\notag
    \\
    &=\int d[\Phi] P_0[\Phi] I_I[\phi,\Phi]\notag
    \\
    &= \int (d^4x)_E \int d[\Phi] P_0[\Phi]\mathcal{O}[\Phi]\otimes J[\phi] \notag
    \\
    &=\int (d^4x)_E \left(\frac{\delta W_g}{\delta J}\right)_{J=0}\otimes J[\phi]
    .\label{eq:II}
\end{align}
When $J[\phi]$ preserves the symmetries of the EFT, $J[\phi]$ can be proportional to the leading term, such as the kinetic term and the Einstein-Hilbert term, and generally takes a non-zero value.
If $J[\phi]$ is the higher-dimensional operator, the EFT includes terms proportional to $J[\phi]$ generated from degrees of freedom other than $\Phi$.
%
%
Therefore, it would be quantitatively and qualitatively reasonable to impose the above assumption.
As explained later, ${\langle I_I\rangle}_{g=0}$ can take zero value by a suitable field redefinition when $J[\phi]$ does not preserve the symmetries of the EFT, such as the gauge symmetry.
%
%
%

    We focus on two cases: tree-level UV completion and loop-level UV completion.
    In the tree-level UV completion, we assume the tree-level effects dominate the perturbative corrections from the heavy degrees of freedom to the Euclidean effective action.
    On the other hand, in the loop-level UV completion, we assume the loop-level effects dominate the perturbative corrections to the Euclidean effective action.
    %
    %
    %
    %
    %
    %
    %
    %
    %
    %
    %
    %
    For each EFT, we evaluate the relative entropy as follows:

    (f) {Single massless scalar field with dimension-eight term}:
Consider an effective action in Minkowski space defined by
\begin{align}
I_c^{(M)}=\int d^4x \left(\frac{1}{2}(\partial_{\mu} \phi \partial^{\mu} \phi)+\frac{c}{\Lambda^4}(\partial_{\mu} \phi \partial^{\mu} \phi)^2 \right),\label{eq:sinphi}
\end{align}
where we used a metric signature convention, $g_{\mu\nu}={\rm diag}(+1,-1,-1,-1)$, and the second term is induced by integrating out heavy fields.
Because of the shift symmetry: $\phi\to\phi+{\rm const.}$, Eq.~\eqref{eq:sinphi} involves only the kinetic term as the renormalizable term, and corrections to the kinetic term can be removed by the field redefinition of $\phi$.
We suppose that the dimension-six operators are eliminated by demanding $\partial_{\mu}\partial^{\mu}\widetilde{\phi}=0$ with constant $\partial^{\mu}\widetilde{\phi}$.
Because of the assumption, i.e., $J[\phi]$ does not include the higher-derivative operators, $J[\phi]$ can be $\partial_{\mu}\phi$ or $\partial_{\mu}\phi\partial^{\mu}\phi$, which preserve the shift symmetry, but $\partial_{\mu}\phi$ effects on ${\langle I_I\rangle}_{g=0}$ vanish because ${\langle I_I\rangle}_{g=0}$ preserves the Lorentz symmetry.
When we suppose that the EFT arises from integrating out heavy degrees of freedom, the first order corrections for $g$ to the Euclidean effective action are expressed as
\begin{align}
    {\langle I_I\rangle}_{g=0}&=\left(\frac{dW_g}{dg}\right)_{g=0}\notag
    \\
    &=\int (d^4x)_E\left(\frac{\delta W_g}{\delta J}\right)_{J=0}J[\phi]\notag
    \\
    &\propto \int (d^4x)_E (\partial_{\mu}\phi\partial^{\mu}\phi).\label{eq:phiLI}
\end{align}
%
%
%
For each tree and loop-level UV completions, we evaluate the constraint from the relative entropy as follows:

\begin{itemize}
    \item Tree-level UV completion ---
    First, consider the EFT generated by the tree-level UV completion.
    Not depending on details of the UV theory, up to the dimension-eight operator, the Euclidean effective action of the theory $B$ is calculated as follows:
    \begin{align}
        W_g[\phi]&=\int (d^4 x)_E \bigg(\frac{1}{2}(1+\alpha_2^{\rm tree})(\partial_{\mu}\phi\partial^{\mu}{\phi})\notag
        \\
        &-\beta_2^{\rm tree}(\partial_{\mu}{\phi}\partial^{\mu}{\phi})^2 \bigg),\label{eq:}
    \end{align}
    where $\alpha_2^{\rm tree}$ and $\beta_2^{\rm tree}$ denote the second or higher order corrections for $g$.
    Note here that $\beta_2^{\rm tree}$ does not include the first order correction for $g$ because of the assumption, i.e., $J[\phi]$ does not include the higher-dimensional operators.
    It is assumed that $\alpha_2^{\rm tree}$ and $\beta_2^{\rm tree}$ are generated at the tree level. 
    Also, according to the procedure in Eq.~\eqref{eq:linrem1}, \eqref{eq:linrem2}, and \eqref{eq:linrem3}, the first order correction for $g$ is removed in $\alpha_2^{\rm tree}$. 
    We choose the background fields as  $\partial_{\mu}\widetilde{\phi}={\rm const.}$ to remove the dimension-six operators.
    The Euclidean effective actions of the theory $B$ and $A$ are respectively obtained as
    \begin{align}
        W_g[\widetilde{\phi}]&=\int (d^4 x)_E \bigg(\frac{1}{2}(\partial_{\mu}\widetilde{\phi}\partial^{\mu}\widetilde{\phi})\notag
        \\
        &-\beta_2^{\rm tree}\cdot\left(1+\alpha^{\rm tree}_2 \right)^{-2}(\partial_{\mu}\widetilde{\phi}\partial^{\mu}\widetilde{\phi})^2 
        \bigg),\label{eq:sigWgM}
        \\
        W_0[\widetilde{\phi}]&=\int (d^4 x)_E \bigg(\frac{1}{2}(\partial_{\mu}\widetilde{\phi}\partial^{\mu}\widetilde{\phi}) 
        \bigg),
    \end{align}
    where the wave function renormalization is performed in Eq.~\eqref{eq:sigWgM}; see Appendix~\ref{app:(IV-3)}.
    Note here that $\widetilde{\phi}$ is also a classical solution of $W_0[\phi]$.
    Then, the shift of the Euclidean effective action is calculated as
    \begin{align}
        \Delta W_g^{(E)}&=W_g[\widetilde{\phi}]-W_0[\widetilde{\phi}]\notag
        \\
        &=-\beta_2^{\rm tree}\cdot\left(1+\alpha^{\rm tree}_2 \right)^{-2} \int (d^4x)_E (\partial_{\mu}\widetilde{\phi}\partial^{\mu}\widetilde{\phi})^2.\label{eq:DelWgsigM}
    \end{align}
    Also, from Eq.~\eqref{eq:sigWgM}, we obtain
    \begin{align}
        \left(\frac{dW_g}{dg}\right)_{g=0}&=0.\label{eq:sigdWgdgM}
    \end{align}
    The detail derivation of Eq.~\eqref{eq:sigdWgdgM} is provided in Appendix~\ref{app:(IV-3)}.
    From Eq.~\eqref{eq:uplow} or \eqref{eq:dyligh}, combining Eq.~\eqref{eq:DelWgsigM} and \eqref{eq:sigdWgdgM} yields
    \begin{align}
        &\Delta W_g^{(E)}\leq g\cdot{\langle I_I\rangle}_{g=0}\notag
        \\
        &\Rightarrow -\beta_2^{\rm tree}\cdot\left(1+\alpha^{\rm tree}_2 \right)^{-2} \int (d^4x)_E (\partial_{\mu}\widetilde{\phi}\partial^{\mu}\widetilde{\phi})^2\leq 0 \notag
        \\
        &\Rightarrow \beta_2^{\rm tree}\cdot\left(1+\alpha^{\rm tree}_2 \right)^{-2}\geq 0.\label{eq:treesin1M}
    \end{align}
    Equation~\eqref{eq:treesin1M} denotes the constraint on the coefficient of dimension-eight operator of Eq.~\eqref{eq:sigWgM}.
    
    \item Loop-level UV completion ---
    Next, consider the EFT generated by the loop-level UV completion.
    The Euclidean effective action of the theory $B$ is calculated as follows:
    %
    %
    %
    \begin{align}
        W_g[\phi]&=\int (d^4 x)_E \bigg(\frac{1}{2}(1+ \alpha_1^{\rm loop}+\alpha_2^{\rm loop})(\partial_{\mu}\phi\partial^{\mu}\phi)\notag
        \\
        &-\beta_2^{\rm loop}(\partial_{\mu}\phi\partial^{\mu}\phi)^2+E_{\rm vac} \bigg)
        ,\label{eq:Wgloop1M}
    \end{align}
    where $\alpha_1^{\rm loop}$ is the first order correction for $g$, $\alpha_2^{\rm loop}$ and $\beta_2^{\rm loop}$ are the second or higher order correction for $g$, 
    and $E_{\rm vac}$ is the vacuum energy coming from $\Phi$ and $\phi$.
    %
    %
    %
    It is assumed that $\alpha_1^{\rm loop}$, $\alpha_2^{\rm loop}$, and $\beta_2^{\rm loop}$ are generated from the loop corrections of $\Phi$.
    %
    %
    %
    %
    %
    %
    We choose $\partial_{\mu}\widetilde{\phi}={\rm const.}$ to remove the dimension-six operators. 
    Since the background field $\widetilde{\phi}$ is also a classical solution of $W_0[\phi]$, the Euclidean effective action for the theory $B$ and $A$ are respectively obtained as Eq.~\eqref{eq:wgloop2},
    \begin{align}
        W_g[\widetilde{\phi}]
        &=\int (d^4 x)_E \bigg(\frac{1}{2}(1+\alpha^{\rm loop}_1)(\partial_{\mu}\widetilde{\phi}\partial^{\mu}\widetilde{\phi})\notag
        \\
        &-\beta_2^{\rm loop}(\partial_{\mu}\widetilde{\phi}\partial^{\mu}\widetilde{\phi})^2+E_{\rm vac} 
        \bigg),\label{eq:wgloop2M}
        \\
        W_0[\widetilde{\phi}]
        &=\int (d^4 x)_E \bigg(\frac{1}{2}(\partial_{\mu}\widetilde{\phi}\partial^{\mu}\widetilde{\phi})+E_{\rm vac} 
        \bigg).
    \end{align}
    %
    %
    Then, the shift of the Euclidean effective action is obtained as
    \begin{align}
    \Delta W_g^{(E)}&=W_g[\widetilde{\phi}]-W_0[\widetilde{\phi}]\notag
    \\
    &=\frac{1}{2}\alpha^{\rm loop}_1\int (d^4 x)_E (\partial_{\mu}\widetilde{\phi}\partial^{\mu}\widetilde{\phi})-\beta_2^{\rm loop}\int (d^4 x)_E (\partial_{\mu}\widetilde{\phi}\partial^{\mu}\widetilde{\phi})^2.\label{eq:DelWgloop3M}
    \end{align}
    Also, from Eq.~\eqref{eq:phiLI} and \eqref{eq:wgloop2M}, we obtain
    \begin{align}
        \left(\frac{dW_g}{dg}\right)_{g=0}&=\frac{1}{2}\frac{d\alpha^{\rm loop}_1}{dg}\cdot\int (d^4x)_E (\partial_{\mu}\widetilde{\phi}\partial^{\mu}\widetilde{\phi})
        ,\label{eq:dWgdgloop2M}
    \end{align}
    where $\alpha_1^{\rm loop}$ denotes the first order correction for $g$ and satisfies a relation of the form $g\cdot({d\alpha^{\rm loop}_1}/{dg})=\alpha^{\rm loop}_1$.
    From Eq.~\eqref{eq:uplow} or \eqref{eq:dyligh}, combining Eq.~\eqref{eq:DelWgloop3M} and \eqref{eq:dWgdgloop2M} yields
    \begin{align}
        &\Delta W_g^{(E)}\leq g \cdot{\langle I_I \rangle}_{g=0} \notag
        \\
        &\Rightarrow -\beta_2^{\rm loop}\int (d^4 x)_E (\partial_{\mu}\widetilde{\phi}\partial^{\mu}\widetilde{\phi})^2\leq 0\notag
        \\
        &\Rightarrow \beta_2^{\rm loop}\geq 0
        .\label{eq:loopboubd1M}
    \end{align}
    Equation~\eqref{eq:loopboubd1M} yields the constraint on the dimension-eight operator generated at the loop level.
\end{itemize}

For both tree and loop-level UV completion, demanding $\partial_{\mu}\partial^{\mu}\widetilde{\phi}=0$ with constant $\partial^{\mu}\widetilde{\phi}$, after Wick-rotation the inequality~\eqref{eq:uplow} gives rise to 
\begin{align}
\frac{c}{\Lambda^4}\int d^4x_E (\partial_{\mu} \widetilde{\phi} \partial^{\mu} \widetilde{\phi})^2\geq  0\Rightarrow \frac{c}{\Lambda^4}\geq 0.\label{eq:sclpos}
\end{align}
Consequently, the coefficient $c$ must be positive to respect the entropy constraints, when it arises from integrating out the heavy fields.
This result is the same as the positivity bound from the unitarity and causality.

%
%
%
%

(g) {Standard Model EFT (SMEFT) dimension-eight $SU(N)$ gauge bosonic operators}:
Consider an effective action in Minkowski space defined by
\begin{align}
I_{\rm SMEFT}^{({\rm M})}=\int d^4 x \left(-\frac{1}{4}F^a_{\mu\nu}F^{a,\mu\nu}+\frac{1}{\Lambda^4}\sum_i c_i\mathcal{O}_i\right),
\end{align}
where the dimensional-eight operators $\mathcal{O}_i$'s are~\cite{Remmen:2019cyz}
\begin{align}
&\mathcal{O}_1^{F^4} =(F^a_{\mu\nu}F^{a,\mu\nu})(F^b_{\rho\sigma}F^{b,\rho\sigma}),
\\
&\mathcal{O}_2^{F^4} =(F^a_{\mu\nu}\tilde{F}^{a,\mu\nu})(F^b_{\rho\sigma}\tilde{F}^{b,\rho\sigma}),
\\
&\mathcal{O}_3^{F^4} =(F^a_{\mu\nu}{F}^{b,\mu\nu})(F^a_{\rho\sigma}F^{b,\rho\sigma}),
\\
&{\mathcal{O}}_4^{F^4} =(F^a_{\mu\nu}\tilde{F}^{b,\mu\nu})(F^a_{\rho\sigma}\tilde{F}^{b,\rho\sigma}),
\\
&{\mathcal{O}}_5^{F^4} =d^{abe}d^{cde}(F^a_{\mu\nu}F^{b,\mu\nu})(F^c_{\rho\sigma}F^{d,\rho\sigma}),
\\
&{\mathcal{O}}_6^{F^4} =d^{abe}d^{cde}(F^a_{\mu\nu}\tilde{F}^{b,\mu\nu})(F^c_{\rho\sigma}\tilde{F}^{d,\rho\sigma}),
\\
&{\mathcal{O}}_7^{F^4} =d^{ace}d^{bde}(F^a_{\mu\nu}F^{b,\mu\nu})(F^c_{\rho\sigma}{F}^{d,\rho\sigma}),
\\
&{\mathcal{O}}_8^{F^4} =d^{ace}d^{bde}(F^a_{\mu\nu}\tilde{F}^{b,\mu\nu})(F^c_{\rho\sigma}\tilde{F}^{d,\rho\sigma}),
\\
&\tilde{\mathcal{O}}_1^{F^4} =(F^a_{\mu\nu}F^{a,\mu\nu})(F^b_{\rho\sigma}\tilde{F}^{b,\rho\sigma}),
\\
&\tilde{\mathcal{O}}_2^{F^4} =(F^a_{\mu\nu}F^{b,\mu\nu})(F^a_{\rho\sigma}\tilde{F}^{b,\rho\sigma}),
\\
&\tilde{\mathcal{O}}_3^{F^4} =d^{abe}d^{cde}(F^a_{\mu\nu}F^{b,\mu\nu})(F^c_{\rho\sigma}\tilde{F}^{d,\rho\sigma}),
\\
&\tilde{\mathcal{O}}_4^{F^4} =d^{ace}d^{bde}(F^a_{\mu\nu}F^{b,\mu\nu})(F^c_{\rho\sigma}\tilde{F}^{d,\rho\sigma}),
\label{eq:SMEFT1}
\end{align} 
where $F^a_{\mu\nu}\equiv \partial_{\mu}A^a_{\nu}-\partial_{\nu}A^a_{\mu}+g f^{abc}A^b_{\mu}A^c_{\nu}$ is the field strength of the gauge field $A^a_{\mu}$ and $g$ denotes the gauge coupling of $SU(N)$. 
The Greek letters stand for Lorentz indices, the Italic letters represent $SU(N)$ color indices, and totally antisymmetric and symmetric structure constants are defined by $[T^a,T^b]=if^{abc}T^c$ and $\{T^a,T^b\}=\delta^{ab}\hat{1}/N+d^{abc} T^c$ with $T^a$ the generator of $SU(N)$ Lie algebra.
To avoid the effect from the dimension-six operators $f^{abc}F^{a\nu}_{\mu}F^{b\rho}_{\nu}F^{c\mu}_{\rho}$ and $f^{abc}F^{a\nu}_{\mu}F^{b\rho}_{\nu}\tilde{F}^{c\mu}_{\rho}$, we follow~\cite{Remmen:2019cyz} to choose a background field satisfying the leading-order equation of motion, $\partial^{\mu}F^a_{\mu\nu}+g f^{abc}A^{\mu b}F^c_{\mu\nu}=0$, in Minkowski space as
$
\overline{A}^a_{\mu} =u^a_1 \epsilon_{1\mu}w_1 +u^a_2 \epsilon_{2\mu}w_2
$
with $f^{abc}u^a_1 u^b_2=0$, where $u_{1,2}$ is a constant real vector in $SU(N)$ color space, $\epsilon_{1,2}$ is a constant four-vector, and $w_{1,2}$ is an arbitrary Cartesian coordinate in spacetime satisfying $\partial_{\mu} w_{1}=l_{\mu}$ and $\partial_{\mu} w_{2}=k_{\mu}$ with $l_{\mu}$ and $k_{\mu}$ being constant four-vectors. 

When $J[A^a_{\mu}]$ does not include the higher-dimensional operators, there are two cases: (i) $J[A^a_{\mu}]$ preserves the gauge symmetry or (ii) not.
For case (i), $J[A^a_{\mu}]\propto F^a_{\mu\nu}F^{a,\mu\nu}$ holds.
The $CP$ violating term generally arises, but we assume such a term is removed by axion-like degrees of freedom in the UV theory.
Then, from Eq.~\eqref{eq:II}, the first order corrections for $g$ to the Euclidean effective action are expressed as
\begin{align}
    {\langle I_I\rangle}_{g=0}&=\left(\frac{dW_g}{dg}\right)_{g=0}\notag
    \\
    &=\int (d^4x)_E \left(\frac{\delta W_g}{\delta J}\right)_{J=0} J[A^a_{\mu}]\notag
    \\
    &\propto \int (d^4x)_E F^a_{\mu\nu}F^{a,\mu\nu}.\label{eq:IISMEFTM}
\end{align}
For case (ii), $J[A^a_{\mu}]$ can be proportional to $A^a_{\mu}$, or $A^a_{\mu}A^{a,\mu}$ because of the covariant derivative of the kinetic term.
Since corrections from the interacting terms of the higher-dimensional operators would not be dominant effects, we focus on corrections from the kinetic terms.   
However, $J[A^a_{\mu}]\propto A^a_{\mu}$ vanishes because ${\langle I_I\rangle}_{g=0}$ keeps the Lorentz symmetry.
Although $J[A^a_{\mu}]\propto A^a_{\mu}A^{a,\mu}$ generally remains, it can be eliminated by the gauge fixing condition, which is called a nonlinear gauge; see Ref.~\cite{Nambu:1968qk,Cao:2022ajt}.
Therefore, we focus on the case of Eq.~\eqref{eq:IISMEFTM} below.
For each tree and loop-level UV completions, the constraints on the SMEFT from the relative entropy are evaluated as follows:

\begin{itemize}
    \item Tree-level UV completion ---
    Consider the EFT generated by the tree-level UV completion.
    The Euclidean effective action of the theory $B$ is generally calculated as follows:
    %
    %
    %
    %
    \begin{align}
        W_g[A]=\int (d^4x)_E \left(\frac{1}{2}(1+\alpha^{\rm tree}_2)F^a_{\mu\nu}F^{a,\mu\nu}-\sum_i\beta_{i,2}^{\rm tree}\mathcal{O}_i[A] \right),
    \end{align}
     where $\alpha_2^{\rm tree}$ and $\beta_{i,2}^{\rm tree}$ denote the second or higher order corrections for $g$, and $\beta_{i,2}^{\rm tree}$ does not include the first order correction for $g$ because of Eq.~\eqref{eq:IISMEFTM}.
    The corrections $\alpha^{\rm tree}_{2}$ and $\beta^{\rm tree}_{i,2}$ are assumed to be generated at the tree-level. 
    According to the procedure in Eq.~\eqref{eq:linrem1}, \eqref{eq:linrem2}, and \eqref{eq:linrem3}, the first order correction for $g$ is eliminated in $\alpha_2^{\rm tree}$. 
    The background fields are chosen to hold $ \overline{F}_{\mu\nu}={\rm const.}$.
    %
    %
    %
    Since $\overline{A}^a_{\mu}$ is also a classical solution of $W_0[A]$, the Euclidean effective actions of the theory $B$ and $A$ are respectively obtained as follows:
    \begin{align}
        &W_g[\overline{A}]=\int (d^4x)_E \bigg(\frac{1}{2}\overline{F}^a_{\mu\nu}\overline{F}^{a,\mu\nu}\notag
        \\
        &-\sum_i\beta_{i,2}^{\rm tree}\cdot (1+a^{\rm tree}_2)^{-2}\mathcal{O}_i[\overline{A}] \bigg),\label{eq:WgsmefttrreeMMa}
        \\
        &W_0[\overline{A}]=\int (d^4x)_E \left(\frac{1}{2}\overline{F}^a_{\mu\nu}\overline{F}^{a,\mu\nu} \right),
    \end{align}
    where the wave function renormalization is performed in Eq.~\eqref{eq:WgsmefttrreeMMa}; see Eq.~\eqref{eq:Wgsmefttrree}.
    Then, the shift of the Euclidean effective action is calculated as follows:
    \begin{align}
        \Delta W_g^{(E)}&=W_g[\overline{A}]-W_0[\overline{A}]\notag
        \\
        &=-\sum_i\beta_{i,2}^{\rm tree}\cdot (1+a^{\rm tree}_2)^{-2}\int (d^4x)_E \mathcal{O}_i[\overline{A}]. \label{eq:DeltreeSMEFTM}
    \end{align}
    From Eq.~\eqref{eq:WgsmefttrreeMMa}, the first order correction for $g$ is calculated as,
    \begin{align}
        \left(\frac{dW_g}{dg}\right)_{g=0}&=0.\label{eq:sigdWgdgSMEFTM}
    \end{align}
     From Eq.~\eqref{eq:uplow} or \eqref{eq:dyligh}, combining Eq.~\eqref{eq:DeltreeSMEFTM} and \eqref{eq:sigdWgdgSMEFTM} yields
    \begin{align}
        &\Delta W_g^{(E)}\leq g\cdot {\langle I_I\rangle}_{g=0}\notag
        \\
        &\Rightarrow \sum_i\beta_{i,2}^{\rm tree}\cdot (1+a^{\rm tree}_2)^{-2}\int (d^4x)_E \mathcal{O}_i[\overline{A}]\geq 0.\label{eq:boundSMEFT1treeMa}
    \end{align}
    The left-hand side of Eq.~\eqref{eq:boundSMEFT1treeMa} denotes the coefficients of the dimension-eight operators of Eq.~\eqref{eq:WgsmefttrreeMMa}.
    Therefore, the relative entropy yields the constraints on the linear combination of the the dimension-eight operators.
    
    \item Loop-level UV completion ---
    Consider the SMEFT generated by the loop-level UV completion.
    The Euclidean effective action of the theory $B$ is generally calculated as follows:
    %
    %
    %
    %
    \begin{align}
        W_g[{A}]&=\int (d^4x)_E \bigg(
        \frac{1}{2} \left(1+\alpha^{\rm loop}_1+\alpha^{\rm loop}_2\right)F^a_{\mu\nu}F^{a,\mu\nu}\notag
        \\
        &-\sum_i\beta^{\rm loop}_{2,i}\mathcal{O}_i[{A}]+E_{\rm vac}\bigg)
        \\
        ,\label{eq:SMEFTWgloop0}
    \end{align}
    where $\alpha_1^{\rm loop}$ is the first order correction for $g$, $\alpha_2^{\rm loop}$ and $\beta_{2,i}^{\rm loop}$ are the second or higher order correction for $g$, 
    and $E_{\rm vac}$ is the vacuum energy coming from $\Phi$ and $A^a_{\mu}$.
    %
    %
    %
    It is assumed that $\alpha_1^{\rm loop}$, $\alpha_2^{\rm loop}$, and $\beta_{2,i}^{\rm loop}$ arise from the loop corrections of $\Phi$.
    %
    %
    We choose the background field satisfying $\overline{F}_{\mu\nu}^a={\rm const.}$ to remove the dimension-six operators.
    $\overline{A}^a_{\mu}$ is also a classical solution of $W_0[A]$, and the Euclidean effective actions of the theory $B$ and $A$ are respectively obtained as Eq.~\eqref{eq:appSMEFTWGloop}, 
    \begin{align}
        W_g[\overline{A}]&=\int (d^4x)_E \bigg(
        \frac{1}{2} \left(1+\alpha^{\rm loop}_1\right)\overline{F}^a_{\mu\nu}\overline{F}^{a,\mu\nu}\notag
        \\
        &-\sum_i\beta^{\rm loop}_{2,i}\mathcal{O}_i[\overline{A}]+E_{\rm vac}\bigg),\label{eq:WgSM}
        \\
        W_0[\overline{A}]&=\int (d^4x)_E \bigg(
        \frac{1}{2} \overline{F}^a_{\mu\nu}\overline{F}^{a,\mu\nu}+E_{\rm vac}\bigg),
    \end{align}
    where the wave function renormalization is performed in Eq.~\eqref{eq:WgSM}; see Eq.~\eqref{eq:appZgloopSMEFT}.
    %
    %
    Then, the shift of the Euclidean effective action is calculated as follows:
    \begin{align}
        \Delta W^{(E)}_g &= W_g[\overline{A}]-W_0[\overline{A}]\notag
        \\
        &=\int (d^4x)_E \bigg(
        \frac{1}{2} \alpha^{\rm loop}_1\overline{F}^a_{\mu\nu}\overline{F}^{a,\mu\nu}-\sum_i\beta^{\rm loop}_{2,i}\mathcal{O}_i[\overline{A}]\bigg).\label{eq:loopDelWgSMM}
    \end{align}
    Also, from Eq.~\eqref{eq:WgSM}, the first order corrections for $g$ is calculated as
    \begin{align}
        \left(\frac{dW_g}{dg}\right)_{g=0}&=\frac{1}{2} \frac{d\alpha^{\rm loop}_1}{dg}\int (d^4x)_E 
        \overline{F}^a_{\mu\nu}\overline{F}^{a,\mu\nu},\label{eq:loopdelWdgSMM}
    \end{align}
    where $g\cdot (d\alpha_1^{\rm loop}/dg)=\alpha_1^{\rm loop}$.
    From Eq.~\eqref{eq:uplow} or \eqref{eq:dyligh}, combining Eq.~\eqref{eq:loopDelWgSMM} and \eqref{eq:loopdelWdgSMM} yields
    \begin{align}
        &\Delta W_g^{(E)}\leq g \cdot{\langle I_I\rangle}_{g=0}\notag
        \\
        &\Rightarrow \sum_i \beta^{\rm loop}_{2,i}\int (d^4x)_E 
        \mathcal{O}_i[\overline{A}] \geq 0.
        \label{eq:SMEFTloopconsM}
    \end{align}
    %
    %
    Equation~\eqref{eq:SMEFTloopconsM} yields the constraint on the dimension-eight operator generated at the loop-level.

\end{itemize}

It is found that, for both tree and loop-level UV completion, the inequality~\eqref{eq:uplow} gives rise to
\begin{align}
    \frac{1}{\Lambda^4} \sum_i c_i \mathcal{O}_i[\overline{A}]\geq 0.\label{eq:SMEFTcom}
\end{align}
After Wick-rotation, Eq.~\eqref{eq:SMEFTcom} yields,
\begin{align}
A\cdot\cos^2\psi  +B\cdot \sin^2{\psi} +C\cdot \sin{\psi}\cos{\psi} \geq 0,
\end{align}
where 
\begin{align}
A&=N\left[(2c_1+c_3)(u_1\cdot u_2)^2+c_3u_1^2u_2^2 +2(c_5+c_7)U^2\right]\notag
\\
&+2 c_7 \left[(u_1\cdot u_2)^2-u_1^2u_2^2\right]
, 
\nonumber\\
B&=N\left[(2c_2+c_4)(u_1\cdot u_2)^2+c_4 u_1^2u_2^2 +2(c_6+c_8)U^2\right]\notag
\\
&+2 c_8 \left[(u_1\cdot u_2)^2-u_1^2u_2^2\right],
\nonumber\\
C&=N\left[(2\tilde{c}_1+\tilde{c}_2)(u_1\cdot u_2)^2+\tilde{c}_2 u_1^2 u_2^2 +2(\tilde{c}_3+\tilde{c}_4)U^2\right]\notag
\\
&+2 \tilde{c}_4 \left[(u_1\cdot u_2)^2-u_1^2 u_2^2\right]
\end{align}
with $U^a=d^{abc}u_1^b u_2^c$, $l^{\mu}=(1,0,-\sin\psi,\cos\psi)/\sqrt{2}$, $k^{\mu}=(1,1,0,0)/\sqrt{2}$, $\epsilon^{\mu}_1\propto(0,1,0,0)$, and $\epsilon^{\mu}_2\propto(0,0,0,1)$.
We end up with positivity bounds as follows:
\begin{align}
A\geq 0,~~~B\geq 0,~~~C^2\leq 4 AB,
\end{align}
which are completely consistent with the positivity bounds from unitarity and causality~\cite{Remmen:2019cyz,Cao:2022ajt}. 
More comprehensive constraints are studied in Ref.~\cite{Cao:2022ajt} by considering more general solutions, which yield additional constraints on the Wilson coefficients of $SU(3)$ gauge bosonic operators.

(h) {Einstein-Maxwell theory with higher-derivative operators}:
Consider a gravitational effective action in Minkowski space defined by
\begin{align}
    I^{\rm (M)}_{\rm EM}&=\int d^4 x\sqrt{-g} \bigg(\frac{M_{\rm Pl}^2}{2} R -\frac{1}{4}F_{\mu\nu}F^{\mu\nu}\notag
    \\
    &+\frac{\alpha_1}{4 M_{\rm Pl}^2} (F_{\mu\nu}F^{\mu\nu})^2+\frac{\alpha_2}{4 M_{\rm Pl}^4} (F_{\mu\nu} \tilde{F}^{\mu\nu})^2\notag
    \\
    &+\frac{\alpha_3}{2 M_{\rm Pl}^2}F_{\mu\nu}F_{\rho\sigma} R^{\mu\nu\rho\sigma} \bigg),
\end{align}
where other operators up to four-derivative are eliminated by the field redefinition of $g_{\mu\nu}$. 
Also, the Gauss-Bonnet combination, i.e., $R_{\mu\nu\rho\sigma}R^{\mu\nu\rho\sigma}-4 R_{\mu\nu}R^{\mu\nu}+R^2$, is a total derivative and vanishes in four dimensions.
Consider the higher-derivative operators generated from the UV theory defined by $I[g_{\mu\nu};R_{\mu\nu\rho\sigma},A,\Phi]$, where $g_{\mu\nu}$ is the metric of space-time, $R_{\mu\nu\rho\sigma}$ is the Riemann tensor, $A_{\mu}$ is the $U(1)$ gauge boson, and $\Phi$ is the heavy degrees of freedom.
Then, the non-interacting and interacting terms are defined as Eq.~\eqref{eq:GI0} and \eqref{eq:GII}. 
It should be noted that the theory of $I_0$ does not include the interaction between $A_{\mu}, R_{\mu\nu\rho\sigma}$ and $\Phi$, but the interaction between $g_{\mu\nu}$ and $\Phi$.
The gravitational operators up to four-derivative such as $R_{\mu\nu}^2$ are generated from $I_0$ and can contribute to $\alpha_1$ and $\alpha_2$  by the field redefinition of $g_{\mu\nu}$.
Our entropy consideration does not constrain such effects because the relative entropy constrains only the higher-derivative operators generated from the interaction $I_I$.
In the following explanations, especially for loop-level UV theory, we suppose that the $R_{\mu\nu}^2$ operator effects are not dominant by assuming a large charge-to-mass ratio of the particle integrated out.

Similar to the SMEFT, when $J[g_{\mu\nu};R_{\mu\nu\rho\sigma},A_{\mu}]$ does not include the higher-derivative operators, there are two cases: (i) $J[g_{\mu\nu};R_{\mu\nu\rho\sigma},A_{\mu}]\propto F_{\mu\nu}F^{\mu\nu}$ or $R$, and (ii) $J[g_{\mu\nu};R_{\mu\nu\rho\sigma},A_{\mu}]\propto A_{\mu}$, or $A_{\mu}A^{\mu}$.
Because of the same reason as the SMEFT, we focus on the following case,
\begin{align}
{\langle I_I\rangle}_{g=0}&=\left(\frac{dW_g}{dg}\right)_{g=0}\notag
\\
&=\int (d^4x)_E\left(\frac{\delta W_g}{\delta J}\right)_{J=0}J[g_{\mu\nu};R_{\mu\nu\rho\sigma},A_{\mu}]\notag
\\
&\propto \int (d^4 x)_E \sqrt{g} F_{\mu\nu}F^{\mu\nu}~{\rm or}~\int (d^4 x)_E \sqrt{g} R.\label{eq:IIEMM}
\end{align}
For each tree and loop-level UV completions, the constraints on the EFT from the relative entropy are evaluated as follows:

\begin{itemize}
    \item Tree-level UV completion ---
    Consider the EFT generated by the tree-level UV completion.
    Then, by integrating out the heavy fields, the Euclidean effective action is generally calculated as follows:
    \begin{align}
        W_g[g_{\mu\nu},A]&=\int (d^4x)_E \sqrt{g}\bigg(-\frac{M^2_{\rm Pl}}{2}(1+\alpha_{2,R}^{\rm tree})R \notag
        \\
        &+\frac{1}{4}(1+\alpha_{2,F}^{\rm tree})F_{\mu\nu}F^{\mu\nu}-\beta_{2,1}^{\rm tree}(F_{\mu\nu}F^{\mu\nu})^2\notag
        \\
        &-\beta_{2,2}^{\rm tree}(F_{\mu\nu}\widetilde{F}^{\mu\nu})^2-\beta_{2,3}^{\rm tree} F_{\mu\nu}F_{\rho\sigma}R^{\mu\nu\rho\sigma}\bigg),
    \end{align}
    where $\alpha_{2,R}^{\rm tree}$, $\alpha_{2,F}^{\rm tree}$, $\beta_{2,1}^{\rm tree}$, $\beta_{2,2}^{\rm tree}$ and $\beta_{2,3}^{\rm tree}$ denote the second or higher order corrections for $g$.
    Note here that $\beta_{2,1}^{\rm tree}$, $\beta_{2,2}^{\rm tree}$ and $\beta_{2,3}^{\rm tree}$ do not include the first order correction for $g$ because of Eq.~\eqref{eq:IIEMM}.
    %
    %
    According to the procedure in Eq.~\eqref{eq:linrem1}, \eqref{eq:linrem2}, and \eqref{eq:linrem3}, the first order correction for $g$ is eliminated in $\alpha_{2,R}^{\rm tree}$ and $\alpha_{2,F}^{\rm tree}$. 
    Since the gravitational higher-derivative operators involving the Riemann tensor can be removed by field redefinition, and the Riemann-squared operator vanishes in four dimensions, we omit such terms.
    %
    %
    %
    The effective actions of the theory $B$ and $A$ are respectively obtained as Eq.~\eqref{eq:WgEMth},
    \begin{align}
        W_g[\overline{g}_{\mu\nu},\overline{A}]&=\int (d^4x)_E \sqrt{\overline{g}}\bigg(-\frac{M^2_{\rm Pl}}{2}\overline{R} +\frac{1}{4}\overline{F}_{\mu\nu}\overline{F}^{\mu\nu}\notag
        \\
        &-\beta_{2,1}^{\rm tree}\left(1+\frac{2}{3}\alpha^{\rm tree}_{2,R}-2 \alpha^{\rm tree}_{2,F}\right)(\overline{F}_{\mu\nu}\overline{F}^{\mu\nu})^2\notag
        \\
        &-\beta_{2,2}^{\rm tree}\left(1+2\alpha^{\rm tree}_{2,R}-2\alpha^{\rm tree}_{2,F}\right)(\overline{F}_{\mu\nu}\widetilde{\overline{F}}^{\mu\nu})^2\notag
        \\
        &-\beta_{2,3}^{\rm tree}\left(1+\frac{1}{3}\alpha^{\rm tree}_{2,R}-\alpha^{\rm tree}_{2,F}\right) \overline{F}_{\mu\nu}\overline{F}_{\rho\sigma}\overline{R}^{\mu\nu\rho\sigma}\bigg),\label{eq:EMMainWg}
        \\
        W_0[\overline{g}_{\mu\nu},\overline{A}]
        &=\int (d^4x)_E \sqrt{\overline{g}}\bigg(-\frac{M^2_{\rm Pl}}{2}\overline{R} +\frac{1}{4}\overline{F}_{\mu\nu}\overline{F}^{\mu\nu}\bigg),
    \end{align}
    where $\overline{A}_{\mu}$ and $\overline{g}_{\mu\nu}$ include the effects of the higher-derivative terms.  
    It should be noted that the first order correction for the higher-derivative terms vanishes in $W_0$ by using the equation of motion.
    %
    %
    Then, the shift of the Euclidean effective action $\Delta W^{(E)}_g$ denotes corrections from the higher-derivative terms.
    Also, from Eq.~\eqref{eq:EMMainWg}, the first order correction for $g$ is calculated as
    \begin{align}
        \left(\frac{dW_g}{dg}\right)_{g=0}&=0.\label{eq:EMdWdgtreeM}
    \end{align}
    %
    %
    From Eq.~\eqref{eq:uplow} or \eqref{eq:dyligh}, combining Eq.~\eqref{eq:EMMainWg} and \eqref{eq:EMdWdgtreeM} yields
    \begin{align}
        &\Delta W^{(E)}_g \leq  0.
    \end{align}
    This inequality means that the relative entropy yields the negative shift of the Euclidean effective action by the higher-derivative operators generated at tree level.

    \item Loop-level UV completion ---
    Next, consider the EFT generated by the loop-level UV completion.
    %
    %
    %
    The Euclidean effective actions of the theory $B$ and $A$ are respectively obtained as Eq.~\eqref{eq:WgEM} and \eqref{eq:W0EM},
    \begin{align}
        W_g[\overline{g}_{\mu\nu},\overline{A}]&=\int (d^4x)_E \sqrt{\overline{g}}\bigg(\Lambda_{0,\Phi}^{\rm loop}-\frac{M^2_{\rm Pl}}{2}(1+\alpha^{\rm loop}_{1,R})\overline{R}\notag
        \\
        &+\frac{1}{4}(1+\alpha^{\rm loop}_{1,F})\overline{F}_{\mu\nu}\overline{F}^{\mu\nu}-\beta_{2,1}^{\rm loop}(\overline{F}_{\mu\nu}\overline{F}^{\mu\nu})^2\notag
        \\
        &-\beta_{2,2}^{\rm loop}(\overline{F}_{\mu\nu}\widetilde{\overline{F}}^{\mu\nu})^2-\beta_{2,3}^{\rm loop} \overline{F}_{\mu\nu}\overline{F}_{\rho\sigma}\overline{R}^{\mu\nu\rho\sigma}\notag
        \\
        &+({\rm correction~from}~R~{\rm and}~F_{\mu\nu}F^{\mu\nu})\bigg),\label{eq:WgEMMain}
        \\
        W_0[\overline{g}_{\mu\nu},\overline{A}]&=\int (d^4x)_E \sqrt{\overline{g}}\bigg(\Lambda_{0,\Phi}^{\rm loop}-\frac{M^2_{\rm Pl}}{2}\overline{R}+\frac{1}{4}\overline{F}_{\mu\nu}\overline{F}^{\mu\nu}\notag
        \\
        &+({\rm correction~from}~R~{\rm and}~F_{\mu\nu}F^{\mu\nu})\bigg).\label{eq:W0EMMain}
    \end{align}
    where $\beta_{2,1}^{\rm loop}$, $\beta_{2,2}^{\rm loop}$ and $\beta_{2,3}^{\rm loop}$ are the second or higher order corrections for $g$, $\alpha_{1,R}^{\rm loop}$ and $\alpha_{1,F}^{\rm loop}$ are the first order corrections for $g$, and
    $\Lambda_{0,\Phi}^{\rm loop}$ is the vacuum energy coming from $\Phi$.
    The last terms of Eq.~\eqref{eq:WgEMMain} and \eqref{eq:W0EMMain} arise from the one-loop correction of light fields in $M^2_{\rm Pl}R/2$ and $F_{\mu\nu}F^{\mu\nu}/4$. 
    %
    %
    Since such a correction does not depend on $g$, they cancel in relative entropy.
    %
    %
    %
    %
    From Eq.~\eqref{eq:WgEMMain},  the first order correction for $g$ is calculated as
    \begin{align}
        \left(\frac{dW_g}{dg}\right)_{g=0}
        &=\int (d^4x)_E \sqrt{\overline{g}}\bigg(-\frac{M^2_{\rm Pl}}{2} \frac{d \alpha^{\rm loop}_{1,R}}{dg}\overline{R}\notag
        \\
        &+\frac{1}{4} \frac{d \alpha^{\rm loop}_{1,F}}{dg}\overline{F}_{\mu\nu}\overline{F}^{\mu\nu}
        \bigg),\label{eq:EMFIRgMain}
    \end{align}
    where $g\cdot({d \alpha^{\rm loop}_{1,R}}/{dg})=\alpha^{\rm loop}_{1,R}$ and $g\cdot({d \alpha^{\rm loop}_{1,F}}/{dg})=\alpha^{\rm loop}_{1,F}$.
    %
    %
    %
    From Eq.~\eqref{eq:uplow} or \eqref{eq:dyligh}, Eq.~\eqref{eq:WgEMMain}, \eqref{eq:W0EMMain} and \eqref{eq:EMFIRgMain} yields
     \begin{align}
        &\Delta W_g^{(E)}\leq g\cdot{\langle I_I\rangle}_{g=0}\notag
        \\
        &\Rightarrow 
        W_g^{\rm non\text{-}lin}[\overline{g}_{\mu\nu},\overline{A}]-W_0[\overline{g}_{\mu\nu},\overline{A}]\leq 0.\label{eq:EMboundma}
    \end{align}
    Here, we defined the effective action without the first order corrections for $g$ as follows:
        \begin{align}
        W_g^{\rm non\text{-}lin}[\overline{g}_{\mu\nu},\overline{A}]&=\int (d^4x)_E \sqrt{\overline{g}}\bigg(\Lambda_{0,\Phi}^{\rm loop}-\frac{M^2_{\rm Pl}}{2}\overline{R} +\frac{1}{4}\overline{F}_{\mu\nu}\overline{F}^{\mu\nu}\notag
        \\
        &-\beta_{2,1}^{\rm loop}(\overline{F}_{\mu\nu}\overline{F}^{\mu\nu})^2-\beta_{2,2}^{\rm loop}(\overline{F}_{\mu\nu}\widetilde{\overline{F}}^{\mu\nu})^2\notag
        \\
        &-\beta_{2,3}^{\rm loop} \overline{F}_{\mu\nu}\overline{F}_{\rho\sigma}\overline{R}^{\mu\nu\rho\sigma}\notag
        \\
        &+({\rm correction~from}~R~{\rm and}~F_{\mu\nu}F^{\mu\nu})\bigg).\label{eq:wgnonli}
    \end{align}
%
        %
        It should be noted that the one-loop corrections from $R$ and $F_{\mu\nu}F^{\mu\nu}$ cancel in Eq.~\eqref{eq:EMboundma}.
        Therefore, $W_g^{\rm non\text{-}lin}[\overline{g}_{\mu\nu},\overline{A}]-W_0[\overline{g}_{\mu\nu},\overline{A}]$ denotes the shift of the Euclidean effective action by the higher-derivative operators.
        Consequently, even for the loop-level UV completion, the relative entropy yields the negative shift of the Euclidean effective action by the higher-derivative operators.
    
\end{itemize}

For both tree and loop-level UV completion, it is found that the non-negativity of relative entropy yields the negative shift of the Euclidean effective action by the higher-derivative operators.
As explained in the next section, this result is closely related to the WGC-like behavior.

Here, we consider the relative entropy when additional higher derivative operators are added to theory A.
In Eq.~\eqref{eq:EMboundma}, the loop effects from light fields cancel in the relative entropy, and the relative entropy does not depend on whether the higher derivative operators are added to the theory A or not.
Consider the action of theory A with the additional higher derivative operators as follows:
\begin{align}
    I_0\to I'_0=I_0+I_c, 
\end{align}
where $I_c$ denotes the additional higher derivative operators consisting of light fields.
Then, the Euclidean effective action of theory A of Eq.~\eqref{eq:W0EMMain} is modified as follows:
\begin{align}
 W_0[\overline{g}_{\mu\nu},\overline{A}]&\to W'_0[\overline{g}_{\mu\nu},\overline{A}]=  W_0[\overline{g}_{\mu\nu},\overline{A}]+I_c[\overline{g}_{\mu\nu},\overline{A}],\label{eq:w0shift}
\end{align}
where $I_c$ eliminates the divergences of loop effects from the light fields and would make the probability distribution function well-defined.
Note here that $I_c$ does not depend on the parameter $g$ because the theory A is defined from the action $I_g$ by taking the limit of $g=0$.
In other words, the action of theory B is also modified as follows:
\begin{align}
    I_{g}\to I'_{g}=I_{g}+I_c.
\end{align}
Then, the Euclidean effective action of the theory B of Eq.~\eqref{eq:WgEMMain} is also rewritten as follows:
\begin{align}
&W_g[\overline{g}_{\mu\nu},\overline{A}]\to  W'_g[\overline{g}_{\mu\nu},\overline{A}]=W_g[\overline{g}_{\mu\nu},\overline{A}]+I_c[\overline{g}_{\mu\nu},\overline{A}],\label{eq:wgshift}
\end{align}
where $I_c$ also eliminates the divergences coming from the loop effects from the light fields in the effective action of theory B.
Then, the relative entropy of Eq.~\eqref{eq:rels0} is modified as follows:
\begin{align}
    S(P_0||P_g)&= W_0 -W_g +g\cdot \left(\frac{dW_g}{dg}\right)_{g=0}\notag
    \\
    &\to W'_0 -W'_g +g\cdot \left(\frac{dW'_g}{dg}\right)_{g=0},\label{eq:relshift}
\end{align}
where we used ${\langle I_I\rangle}_{g=0}=\int_{\beta} d[\Phi] P_0\cdot I_I =(dW_g/dg)_{g=0}$ in the first line.
Substituting Eqs.~\eqref{eq:w0shift} and \eqref{eq:wgshift} into Eq.~\eqref{eq:relshift}, we obtain
\begin{align}
    W'_0-W'_g+g\cdot \left(\frac{dW'_g}{dg}\right)_{g=0}&=W_0-W_g+g\cdot \left(\frac{dW_g}{dg}\right)_{g=0}\notag
    \\
    &=S(P_0||P_g),\label{eq:relshiftcanc}
\end{align}
where $I_c$ cancels in $W'_0-W_g$, and $(dW'_g/dg)_{g=0}=(dW_g/dg)_{g=0}$ holds because $I_c$ does not depend on the parameter $g$.
Therefore, the relative entropy of Eq.~\eqref{eq:rels0} does not depend on whether the higher derivative operators consisting of the light fields are added to the theory A.
%
%
%
%

(i) { Weak Gravity Conjecture}:
Last but not the least, we discuss the close connection between the entropy inequality~\eqref{eq:uplow} and the WGC. 
The WGC states that quantum gravity theories have to contain {\it a charged particle} with the charge-to-mass ratio larger than unity, which is motivated by a gedanken experiment of the decay of an extremal black hole. 
The extremality bound, $M\geq M_{\rm ext}=Q$ where $M$ and $Q$ denote the mass and charge of the black hole described by the Einstein-Maxwell theory and $M_{\rm ext}$ represents the minimum mass, would indicate existence of a particle with the charge-to-mass ratio larger than unity.
The extremality bound is modified by a perturbative correction in the Einstein-Maxwell theory; however, the conclusion of the above gedanken experiment remains for an extremal BH of arbitrary large size if the perturbative correction does reduce $M_{\rm ext}$ at fixed charge.
Based on thermodynamic, Ref.~\cite{Goon:2019faz} generalizes a relation between the perturbative corrections to the black hole entropy and the extremality bound~\cite{Cheung:2018cwt} to a wide class of thermodynamic system as
\begin{align}
\left(\frac{\partial M_{\rm ext}}{\partial \epsilon}\right)_{\vec{Q}}=\lim_{M\to M_{\rm ext}(\vec{Q},\epsilon)} -\frac{1}{\beta}\left(\frac{\partial S}{\partial \epsilon}\right)_{M,\vec{Q}},\label{eq:Mext}    
\end{align}
where $\epsilon$ is the parameter introduced to characterize the perturbative corrections in the system, and $\vec{Q}$ is the charge. 
Note here that the extremal limit is taken in Eq.~\eqref{eq:Mext}.
From Eq.~\eqref{eq:Mext}, if $\epsilon\cdot(\partial S/\partial \epsilon)_{M,\vec{Q}}>0$, then a perturbed extremal system is less massive than its unperturbed counterpart at fixed charge.

Consider the effective action including the perturbative correction as
\begin{align}
    W_{\epsilon}[\beta,\phi]= W_{0}[\beta,\phi]+\epsilon \cdot (\partial W_{\epsilon}/\partial \epsilon)_{\epsilon=0},\label{eq:Wep1}
\end{align}
where $ \epsilon \cdot (\partial W_{\epsilon}/\partial \epsilon)_{\epsilon=0}\equiv\Delta W_g^{(E)}\leq g\cdot {\langle I_I\rangle}_{g=0}$ in accord to the inequality~\eqref{eq:end}.
%
%
%
%
%
Note that $\epsilon \cdot (\partial W_{\epsilon}/\partial \epsilon)_{\epsilon=0}$ contains the higher order correction of $\mathcal{O}(g^2)$.
Again, the parameter $\epsilon$ characterizes the perturbative corrections and we consider the leading term of $\epsilon$ hereafter.
The free energy of the thermodynamic system, $ G\equiv  M-S/\beta - \vec{Q}\cdot\vec{\mu}$, is expressed as 
\begin{align}
{G}[\beta,\vec{\mu},\epsilon]=\beta^{-1}\cdot W_{\epsilon}[\beta,\widetilde{\phi}_{\epsilon}],
\end{align}
where $\widetilde{\phi}_{\epsilon}$ is a local minimum of $W_{\epsilon}$, $\beta$ is the inverse temperature, $S$ is the thermodynamic entropy, and $\vec{\mu}$ is the chemical potential.
Therefore, the difference in the free energy between the two theories is 
\begin{align}
\hspace*{-2mm}\Delta G\equiv G[\beta,\vec{\mu},\epsilon]-{G}[\beta,\vec{\mu},0]=  \frac{1}{\beta}\epsilon \cdot \left(\frac{\partial W_{\epsilon}}{\partial \epsilon}\right)_{\epsilon=0}=\frac{\Delta W_g^{(E)}}{\beta},
\end{align}
where $W_{\epsilon}[\beta,\widetilde{\phi}_{\epsilon}]=W_{\epsilon}[\beta,\widetilde{\phi}_{0}] +\mathcal{O}(\epsilon^2)$ is used because $\widetilde{\phi}_{\epsilon}$ is a local minimum of $W_{\epsilon}$.
In gravitational EFTs, this point has been mentioned in Ref.~\cite{Goon:2019faz} with special attention to contributions from boundary terms.
From the relation $(\partial S/\partial\epsilon)_{M,\vec{Q}}=-\beta(\partial G/\partial \epsilon)_{\beta,\vec{\mu}}$ in Refs.~\cite{Loges:2019jzs,Goon:2019faz}, we obtain 
\begin{align}
\epsilon\cdot \frac{1}{\beta}\left(\frac{\partial S}{\partial \epsilon}\right)_{M,\vec{Q}}&=-\epsilon\cdot\left(\frac{\partial G}{\partial \epsilon}\right)_{\beta,\vec{\mu}}=- \frac{\Delta W_g^{(E)}}{\beta}.\label{eq:SWg}
\end{align}
Combining Eqs.~\eqref{eq:uplow} and \eqref{eq:SWg}, lower and upper bounds on the perturbative correction to entropy are given by
\begin{align}
-\frac{1}{\beta}g\cdot{\langle I_I\rangle}_{g}\geq \epsilon\cdot \frac{1}{\beta}\left(\frac{\partial S}{\partial \epsilon}\right)_{M,\vec{Q}}\geq -\frac{1}{\beta}g\cdot{\langle I_I\rangle}_{g=0}.\label{eq:delS}
\end{align}
For the EFTs discussed in \ref{sec:Bott}, under the assumption that $J$ does not include the higher-derivative operators, the shift of the Euclidean effective action by the higher-derivative operators becomes non-positive at zero temperature.
When we substitute such non-positive perturbative corrections from the higher-derivative operators into $\epsilon\cdot (\partial W_{\epsilon}/\partial\epsilon)_{\epsilon=0}$ in Eq.~\eqref{eq:Wep1}, the right-hand side of Eq.~\eqref{eq:SWg} takes a non-negative value up to the first order of the higher-derivative operators, and the WGC-like behavior arises in the EFTs discussed in \ref{sec:Bott}.
In particular, to derive the above arguments for the Einstein-Maxwell theory with higher-derivative operators, it is also supposed that the $R_{\mu\nu}^2$ operator effects are not dominant because of a large charge-to-mass ratio of the particle integrated out.
%
Note here that the exception is possible because the entropy constraints rely on the Euclidean path integral method.
Some conditions to apply the entropy constraints are explained in Appendix~\ref{app:(IV)}.
Although the entropy constraint is a generalization of Ref.~\cite{Cheung:2018cwt}, investigations of the adaption range of the entropy constraint on the WGC is one of our future directions.

We comment on a connection between this work and Ref.~\cite{Cheung:2018cwt}.
In Ref.~\cite{Cheung:2018cwt}, it is demonstrated that the Euclidean effective action decreases by higher-derivative operators generated at tree level.
For convenience, we briefly review it. 
At finite temperature $\beta$, consider the actions $I_0$ and $I_g$.
The saddle point approximation yields
\begin{align}
    I_0[\widetilde{\phi}_{0},0]=I_g[\widetilde{\phi}_{0},0]\geq I_g[\widetilde{\phi}_{g},\widetilde{\Phi}_{g}],\label{eq:prewo1}
\end{align}
where $\widetilde{\phi}_{0}$ is the classical solution of $I_0$, $\widetilde{\phi}_{g}$ and $\widetilde{\Phi}_{g}$ are those of $I_g$, and $I_g[\widetilde{\phi}_{0},0]=I_0[\widetilde{\phi}_{0},0]$ holds because the interacting term of $I_g$ vanishes for $\Phi=0$.
It should be noted that the relation
\begin{align}
\lim_{\widetilde{\phi}_{g}\to\widetilde{\phi}_{0},\widetilde{\Phi}_{g}\to 0}I_g[\widetilde{\phi}_{g},\widetilde{\Phi}_{g}]=I_0[\widetilde{\phi}_{0},0]=\lim_{g\to 0}I_g[\widetilde{\phi}_{g},\widetilde{\Phi}_{g}]
\end{align}
is derived by taking the limit of $g=0$ in this work.
Thus, $I_0[\widetilde{\phi}_{0},0]$ and $I_g[\widetilde{\phi}_{g},\widetilde{\Phi}_{g}]$ are the Euclidean effective action of the theory $A$ and $B$, respectively.
Since $\widetilde{\Phi}_g$ denotes the local minimum of $I_g$ and would take a small value because of heavy field mass suppressions, the inequality of \eqref{eq:prewo1} arises by the saddle point approximation.
The action $I_0$ does not generate the higher-dimensional operators, but the action $I_g$ yields them through the interacting term between $\phi$ and $\Phi$. 
Therefore, the inequality~\eqref{eq:prewo1} means that the Euclidean effective action decreases by higher-dimensional operators generated at tree level.
In other words, at fixed temperature $\beta$, the free energy decreases by higher-dimensional operators generated at tree level.
Note here that the inequality of \eqref{eq:prewo1} does not need the extremal limit to be valid.
Although the origin of the inequality is slightly different, Ref.~\cite{Cheung:2018cwt} is essentially the same as this work at the tree-level.
\\[-3mm]

\section{Implication of entropy constraint}
\label{sec:Result}
The entropy constraint is intimately connected to the unitarity of time evolution.
In the study, the canonical distributions are adopted as the density operator, which is a positive semidefinite (Hermitian) operator with trace one.
In other words, the Hamiltonians of the two theories are Hermitian to ensure the non-negativity of relative entropy.
Therefore, the entropy constraint on the EFTs is consistent with the positivity bound obtained from unitarity considerations.

So far we have studied the constraints on theories from the non-negativity of relative entropy, however, the second law of thermodynamics is also intimately connected with the non-negativity of relative entropy~\cite{2012}. 
For example, consider a thermodynamic system consisting of a system, and an external heat bath system described by the Hamiltonian $H_B$.
We assume that the initial state of the entire system is  ${\rho}_S\otimes e^{-\beta H_B}/Z_B$, where $\rho_S$ is a quantum state of the system, $\beta$ is an inverse temperature of the external heat bath system, and $Z_B\equiv{\rm Tr}_B[e^{-\beta H_B}]$ is obtained by tracing out the heat bath system degrees of freedom.
After the time evolution by the unitary operator $U$, the final state of the entire system becomes $U\rho_g\otimes e^{-\beta H_B}/Z_B U^{\dagger}$.
Then, the final state of the system is obtained as $\tilde{\rho}_S\equiv{\rm Tr}_B[U\rho_g\otimes e^{-\beta H_B}/Z_B U^{\dagger}]$ by tracing out the heat bath system.
The definition of relative entropy Eq.~\eqref{eq:rel} yields \cite{2012} 
\begin{align}
S(U{\rho}_S\otimes e^{-\beta H_B}/Z_B U^{\dagger}\parallel\tilde{\rho}_S&\otimes e^{-\beta H_B}/Z_B)\notag
\\
&=\Delta s-\beta \cdot \Delta q \geq 0,
\end{align}
where $\Delta s\equiv - {\rm Tr}_S[\tilde{\rho}_S\ln \tilde{\rho}_S]+{\rm Tr}_S[{\rho}_S\ln {\rho}_S]$ denotes the difference in the thermodynamic entropy of the system, $\Delta q\equiv {\rm Tr}[\rho_S\otimes e^{-\beta H_B}/Z_B H_B]-{\rm Tr}[\rho_S\otimes e^{-\beta H_B}/Z_B U^{\dagger}H_B U]$ is a heat exchange between the system and the external heat bath system, and the second term $-\beta \cdot \Delta q$ represents the difference in the thermodynamic entropy of external heat bath systems at inverse temperature $\beta$.
Therefore, the non-negativity of relative entropy yields the second law of thermodynamics, and any theory violating the non-negativity of relative entropy does not respect the second law of thermodynamics.
It is remarkable that the non-negativity of relative entropy yields a unified understanding of various phenomena, e.g., the positivity bounds on EFTs, the WGC-like behavior in thermodynamics, and the second law of thermodynamics.
\\[-3mm]

\section{Concluding remarks}
\label{sec:conclusion}
%
In this Letter, we have studied the positivity bounds on EFTs, and the WGC-like behavior in thermodynamics in terms of the non-negativity of relative entropy.
Form the relative entropy, we obtained the lower and upper bounds on perturbative corrections from the interaction between heavy and light degrees of freedom to the Euclidean effective action.
We argued that the bounds are applicable in both field theoretical systems and quantum mechanical systems.
Focusing on the class of EFTs, e.g., the single massless scalar field with dimension-eight term, SMEFT $SU(N)$ gauge bosonic operators, and Einstein-Maxwell theory with higher-derivative operators, generated by the interactions, we found that the upper bound yields the positivity bounds as the same as those derived by unitarity and causality in the conventional EFT study~\cite{Remmen:2019cyz}. 
This argument holds when the corrections from the interactions involving higher-derivative operators of the light fields are not dominant in the EFTs.
By combining the entropy constraints and pure thermodynamics, it is also shown that the WGC-like behavior arises in some EFTs, e.g., the single massless scalar field with dimension-eight term, SMEFT $SU(N)$ gauge bosonic operators, and Einstein-Maxwell theory with higher-derivative operators, up to the first order of the higher-derivative operators.
Finally, we remark that the positivity bounds on EFTs, the WGC-like behavior in thermodynamics, and the second law of thermodynamics are intimately connected by the non-negativity of relative entropy.

\subsection*{Acknowledgement}
D.U. thanks the KEK theory group and the University of Tokyo for their hospitality, where most of this study was done. We would like to thank 
Yuta Hamada for worthwhile discussions of the WGC. 
We greatly appreciate many valuable conversations with our colleagues, Yasuhito Sakaki, Kiyoharu Kawana, Teppei Kitahara, Yoshinori Tomiyoshi, Katsuya Hashino, Hikaru Ohta, Naoto Kan, Ryota Kojima, Sumito Yokoo, and Takumi Kuwahara. The work is supported in part by the National Science Foundation of China under Grant Nos. 11725520, 11675002, and 11635001.

\appendix
\begin{widetext}
%
\section{Definition of $I_0$ and $I_g$}
\label{app:(I)}
We study the relative entropy between a theory with and without interactions between heavy and light degrees of freedom.
In this section we provide a definition of the two theories by using some examples.
Let us consider a theory defined by an action $I_0[\phi,\Phi]+I_I[\phi,\Phi]$.
Throughout this Letter, it is supposed that $I_0$ does not involve interactions between $\phi$'s and $\Phi$'s, and $I_I$ denotes the interactions.
To characterize the interaction, we introduce an auxiliary parameter $g$ and define an action $I_g[\phi,\Phi]\equiv I_0[\phi,\Phi]+g\cdot I_I[\phi,\Phi]$.
The theory $A$ and $B$ are defined as $I_0[\phi,\Phi]$ and $I_{g}[\phi,\Phi]$, respectively.
Note here that, for $g=1$, the theory $B$ reproduces the original action defined by $I_0[\phi,\Phi]+I_I[\phi,\Phi]$.
For ease of understanding, we provide the definitions of $I_0$ and $I_g$ by using some examples.
\\

(A) A loop level UV completion of the single massless scalar field theory with dimension-eight term:
Let us consider a theory in Minkowski space:
\begin{align}
    I^{(M)}=\int d^4 x\left(\frac{1}{2}(\partial_{\mu}\phi \partial^{\mu}\phi)+\bar{\psi}\left(i\slashed{\partial}-\frac{1}{v}\slashed{\partial}\phi\gamma^5-m\right)\psi \right),
\end{align}
where $\phi$ denotes a massless scalar field, and $\psi$ is a heavy fermion feild with mass $m$.
We define an action $I_g\equiv I_0+g \cdot I_I$ with
\begin{align}
    I_0^{(M)}=\int d^4 x\left(\frac{1}{2}(\partial_{\mu}\phi \partial^{\mu}\phi)+\bar{\psi}\left(i\slashed{\partial}-m\right)\psi \right),~~~I_I^{(M)}=-\frac{1}{v}\int d^4 x\bar{\psi}\slashed{\partial}\phi\gamma^5\psi.
\end{align}
Then, the upper bound of Eq.~\eqref{eq:uplow} becomes zero, because a tadpole diagram proportional to $g$ vanishes.
Consequently, the positivity bound of Eq.~\eqref{eq:sclpos} arises.

(B) Massive scalar field theory in linearized gravity:
Let us consider following theory in Minkowski space:
\begin{align}
    I^{(M)}=\int d^4 x\sqrt{-g}\left(\frac{M_{\rm Pl}^2}{2} R -\frac{1}{4}F_{\mu\nu}F^{\mu\nu}+g^{\mu\nu}D_{\mu}\Phi D_{\nu}\Phi^{\ast}-m^2|\Phi|^2 -V(\Phi)+\frac{1}{2}\xi R|\Phi|^2 \right),\label{eq:lin1}
\end{align}
where $\Phi$ is a massive charged scalar field, $D_{\mu}\Phi=(\partial_{\mu}+ie A_{\mu})\Phi$, and $\xi$ is a dimensionless coupling constant. 
We can define as
\begin{align}
    I_0^{(M)}&=\int d^4 x\sqrt{-g}\left(\frac{M^2_{\rm Pl}}{2} R- \frac{1}{4}F^{\mu\nu}F_{\mu\nu}\right)
    +\int d^4 x\sqrt{-\eta}\left(\eta^{\mu\nu}\partial_{\mu}\Phi\partial_{\nu}\Phi-m^2|\Phi|^2-V(\Phi)\right),
    \\
    I_I^{(M)}&=I^{(M)}-I^{(M)}_0.
\end{align}
Then, the theory $B$ is defined as $I_g=I_0+g\cdot I_I$ by introducing the coupling $g$.
Let us consider a classical fluctuation of the metric $h_{\mu\nu}$ around the flat metric $\eta_{\mu\nu}$:
\begin{align}
    g_{\mu\nu}=\eta_{\mu\nu}+h_{\mu\nu}.
\end{align}
At the linearized level, we obtain
\begin{align}
    &I_0^{(M)}=\int d^4 x\sqrt{-g}\left(\frac{M^2_{\rm Pl}}{2} R- \frac{1}{4}F^{\mu\nu}F_{\mu\nu}\right)+\int d^4x \left(\eta^{\mu\nu}\partial_{\mu}\Phi \partial_{\nu}\Phi^{\ast}-m^2|\Phi|^2-V(\Phi)\right),
    \\
    &I_I^{(M)}=\int d^4x\bigg(
    \eta^{\mu\nu}\left(-ie A_{\nu} (\partial_{\mu}\Phi)\Phi^{\ast}+ie A_{\mu} \Phi \partial_{\nu}\Phi^{\ast} \right)+e^2 A_{\mu}A_{\nu}|\Phi|^2\notag
    \\
    &+\frac{1}{2}\eta^{\mu\nu}h_{\mu\nu}\left(\eta^{\mu\nu}D_{\mu}\Phi D_{\nu}\Phi^{\ast}-m^2|\Phi|^2-V(\Phi) \right)+h^{\mu\nu}D_{\mu}\Phi D_{\nu}\Phi^{\ast}+\frac{1}{2}\xi R|\Phi|^2 \bigg).
\end{align}
It is clear that the $I_I$ denotes the interaction between graviton, photon and massive scalar field, and $I_0$ does not involve it. 
%
%
%
\\

\section{Derivation of Eq.~\eqref{eq:uplow}}
\label{app:der(5)}
For convenience, we would like to provide details of the derivation of Eq.~\eqref{eq:uplow}.
The relative entropy is calculated as follows,
\begin{align}
    S(P_0||P_g)&\equiv \int d[\Phi]\left(P_0 \ln P_0- P_0\ln P_g\right),\notag
    \\
    &=\int d[\Phi]\left(
    P_0\left(-I_0[\phi,\Phi]-\ln Z_0[\beta,\phi]\right)-P_0 \left(-I_g[\phi,\Phi]-\ln Z_g[\beta,\phi]\right)
    \right),\notag
    \\
    &=-\ln Z_0[\beta,\phi]+\ln Z_g[\beta,\phi]+\int d[\Phi]P_0 \left(I_g[\phi,\Phi]-I_0[\phi,\Phi]\right),\notag
    \\
    &=W_0[\beta,\phi]-W_g[\beta,\phi]+g\cdot {\langle I_I\rangle}_{g=0},\notag
    \\
    &=-\Delta W^{(E)}_g+g\cdot {\langle I_I\rangle}_{g=0}\geq 0,\label{eq:appS1}
\end{align}
where the probability distributions are defined as
\begin{align}
    &P_0\equiv \frac{e^{-I_0[\phi,\Phi]}}{Z_0[\beta,\phi]},~~~P_g\equiv \frac{e^{-I_g[\phi,\Phi]}}{Z_g[\beta,\phi]}.
\end{align}
The first line denotes the definition of the relative entropy.
In the second line, we used following relations,
\begin{align}
    &\ln P_0=-I_0[\phi,\Phi]-\ln Z_0[\beta,\phi],
    \\
    &\ln P_g=-I_g[\phi,\Phi]-\ln Z_g[\beta,\phi].
\end{align}
In the fourth line, we used the following definitions,
\begin{align}
    &W_0[\beta,\phi]\equiv -\ln Z_0[\beta,\phi],
    \\
    &W_g[\beta,\phi]\equiv -\ln Z_g[\beta,\phi],
    \\
    &g\cdot {\langle I_I\rangle}_{g=0}\equiv\int d[\Phi] P_0 \left(I_g[\phi,\Phi]-I_0[\phi,\Phi]\right)=\int d[\Phi] P_0 g\cdot I_I[\phi,\Phi].
\end{align}
In the last line, we used the non-negativity of relative entropy and $\Delta W^{(E)}_g\equiv W_g[\beta,\phi]-W_0[\beta,\phi]$.
From Eq.~\eqref{eq:appS1}, the upper bound of the shift of the Euclidean effective action is expressed as
\begin{align}
    \Delta W_g^{(E)}\leq g\cdot {\langle I_I\rangle}_{g=0}.\label{eq:upa}
\end{align}
Similarly, another choice of the relative entropy is calculated as follows,
\begin{align}
    S(P_g||P_0)&\equiv \int d[\Phi]\left(P_g \ln P_g -P_g\ln P_0\right),\notag
    \\
    &=\int d[\Phi]\left(P_g \left(-I_g[\phi,\Phi]-\ln Z_g[\beta,\phi]\right)-P_g \left(-I_0[\phi,\Phi]-\ln Z_0[\beta,\phi]\right) \right),\notag
    \\
    &=-\ln Z_g[\beta,\phi]+\ln Z_0[\beta,\phi]-\int d[\Phi] P_g \left(I_g[\phi,\Phi]-I_0[\phi,\Phi]\right),\notag
    \\
    &=W_g[\beta,\phi]-W_0[\beta,\phi]-g\cdot {\langle I_I\rangle}_g,\notag
    \\
    &=\Delta W^{(E)}_g -g\cdot  {\langle I_I\rangle}_g\geq 0.
\end{align}
In the fourth line, we used
\begin{align}
    g \cdot {\langle I_I\rangle}_g=\int d[\Phi] P_g \left(I_g[\phi,\Phi]-I_0[\phi,\Phi]\right)=\int d[\Phi] P_g g\cdot I_I[\phi,\Phi].
\end{align}
The last line yields the lower bound of the shift of the Euclidean effective action as follows,
\begin{align}
    g \cdot{\langle I_I\rangle}_g\leq \Delta W^{(E)}_g.\label{eq:lowa}
\end{align}
Combining Eq.~\eqref{eq:upa} and \eqref{eq:lowa}, we get Eq.~\eqref{eq:uplow}.
Note here that the derivation of Eq.~\eqref{eq:uplow} does not depend on the detail form of $I_g$. 
Since, however, the relative entropy is calculated based on the Euclidean path integral method, Eq.~\eqref{eq:uplow} may be broken when the Euclidean path integral method does not work, see Appendix~\ref{app:(IV)}.

For the dynamical light fields, similar to Eq.~\eqref{eq:rels0}, the relative entropy is calculated as follows, 
\begin{align}
S({P}_0\parallel P_g)&=\int_{\beta}d[\phi]d[\Phi] \left({P}_0\ln {P}_0 -{P}_0\ln {P}_g \right)\notag
\\
&=-\ln {Z}_0[{\beta},\widetilde{\phi}_0] +\ln Z_g[{\beta},\widetilde{\phi}_g]+g  \int_{\beta} d[\phi]d[\Phi] {P}_0\cdot I_I \notag
\\
&={W}_0[{\beta},\widetilde{\phi}_0] -W_g[{\beta},\widetilde{\phi}_g] +g \cdot {\langle I_I\rangle}_{g=0}\notag
\\
&=-\Delta W_g^{(E)}+g \cdot{\langle I_I\rangle}_{g=0}\geq 0,\label{eq:dyligh}
\end{align}
where ${W}_0[{\beta},\widetilde{\phi}_0]\equiv -\ln {Z}_0[{\beta},\widetilde{\phi}_0]$, $W_g[{\beta},\widetilde{\phi}_g]\equiv -\ln Z_g[{\beta},\widetilde{\phi}_g]$, $\Delta W_g^{(E)}\equiv W_g[{\beta},\widetilde{\phi}_g]-{W}_0[{\beta},\widetilde{\phi}_0]$, and the partition functions are defined as
\begin{align}
    &Z_0[\beta,\widetilde{\phi}_0]\equiv\int_{\beta}d[\phi]d[\Phi] e^{-I_0[\phi,\Phi]},
    \\
    &Z_g[\beta,\widetilde{\phi}_g]\equiv\int_{\beta}d[\phi]d[\Phi] e^{-I_g[\phi,\Phi]}.
\end{align}
The expectation value of the interaction is expressed as
\begin{align}
    {\langle I_I\rangle}_{g=0}=\left(\frac{\partial W_g[\beta,\widetilde{\phi}_g]}{\partial g}\right)_{g=0}, \label{eq:par0}
\end{align}
where the partial derivative is performed with the fixed classical solution $\widetilde{\phi}_g$.

Also, the another choice of relative entropy of Eq.~\eqref{eq:anrel} is calculated as follows,
\begin{align}
    S(P_g\parallel P_0)&=\int_{\beta}d[\phi]d[\Phi] \left(P_g \ln P_g- P_g\ln P_0\right)\notag
    \\
    &=-\ln Z_g[\beta,\widetilde{\phi}_g]+\ln Z_0[\beta,\widetilde{\phi}_0]-g \int_{\beta} d[\phi]d[\Phi]P_g\cdot I_I\notag
    \\
    &=W_g[{\beta},\widetilde{\phi}_g]-{W}_0[{\beta},\widetilde{\phi}_0]-g\cdot  {\langle I_I\rangle}_g\notag
    \\
    &=\Delta W_g^{(E)}-g \cdot {\langle I_I\rangle}_g\geq 0,
\end{align}
where the expectation value of the interaction is expressed as
\begin{align}
    {\langle I_I\rangle}_g =\left(\frac{\partial W_g[\beta,\widetilde{\phi}_g]}{\partial g}\right)_g.
\end{align}
Here, similar to Eq.~\eqref{eq:par0}, the partial derivative is performed with the fixed classical solution.

\section{Relative entropy under field redefinition}
\label{app:(II)}
To demonstrate how to use the entropy constraints, let us consider theories described by the following functions:
\begin{align}
    I_0[x_l,x_h]=m_h^2 x_h^2 +m_l^2 x_l^2,~~~I_I[x_l,x_h]=c\cdot x_l x_h,\label{eq:Gau_int}
\end{align}
where $x_l$ and $x_h$ denote the light and heavy degrees of freedom, respectively, and $m_h, m_l$, and $c$ are coupling constants.
We define $I_g\equiv I_0+g\cdot I_I$ with the parameter $g$.
Then, probability distribution functions are defined as follows:
\begin{align}
    P_0[x_l,x_h]\equiv \frac{e^{-I_0[x_l,x_h]}}{Z_0[x_l]},~~~P_g[x_l,x_h]\equiv\frac{e^{-I_g[x_l,x_h]}}{Z_g[x_l]},
\end{align}
with the partition functions
\begin{align}
    &Z_0[x_l]=\int_{-\infty}^{\infty}dx_h e^{-I_0[x_l,x_h]}=e^{-m_l^2 x_l^2}\sqrt{\frac{\pi}{m_h^2}},\label{eq:G_Z0}
    \\
    &Z_g[x_l]=\int_{-\infty}^{\infty}dx_h e^{-I_g[x_l,x_h]}=Z_0[x_l]\cdot e^{g^2 c^2 x_l^2/4m_h^2}.\label{eq:G_Zg}
\end{align}
The expectation value of the interaction is calculated as
\begin{align}
    g\cdot {\langle I_I\rangle}_{g=0}=\int_{-\infty}^{\infty} dx_h P_0[x_l,x_h] I_I[x_l,x_h]=0.\label{eq:G_II}
\end{align}
Combining Eq.~\eqref{eq:G_Z0}, \eqref{eq:G_Zg}, and \eqref{eq:G_II}, the relative entropy between $P_0$ and $P_g$ is calculated as
\begin{align}
    S(P_0||P_g)&=\int_{-\infty}^{\infty} dx_h \left(P_0\ln P_0 -P_0\ln P_g\right),\notag
    \\
    &=-\ln Z_0[x_l]+\ln Z_g[x_l]+g\cdot  {\langle I_I\rangle}_{g=0},\notag
    \\
    &=-\ln Z_0[x_l]+\ln Z_g[x_l],\notag
    \\
    &=\frac{g^2 c^2 x_l^2}{4m_h^2}\geq 0.
\end{align}
It is clear that entropy constraint is satisfied in systems described by the Gaussian distributions.
Note here that the relative entropy is invariant under the field redefinition of $x_h$.
Although the definition of the interaction of Eq.~\eqref{eq:Gau_int} is not invariant under the redefinition of $x_h$, the definition of the relative entropy of Eq.~\eqref{eq:rel} and the integral of the Gaussian distributions do not change under the field redefinition.
%
%

To see the invariant formulation under the field redefinition, let us consider a tree level UV completion described by the following action in Euclidean space:
\begin{align}
    I^{(E)}=\int (d^4 x)_E\left(\frac{1}{4}F_{\mu\nu}F^{\mu\nu}+m_A^2\phi_A^2 -\frac{1}{M} \phi_A F_{\rho\sigma}F^{\rho\sigma} \right),
\end{align}
where $\phi_A$ is an auxiliary field.
We define the theory $B$ as $I_g= I_0+g\cdot I_I$ with the parameter $g$, and
\begin{align}
    &I_0^{(E)}=\int (d^4 x)_E \left(\frac{1}{4}F_{\mu\nu}F^{\mu\nu}+m_A^2\phi_A^2 \right),
    \\
    &I_I^{(E)}=-\int (d^4 x)_E\left(\frac{1}{M} \phi_A F_{\rho\sigma}F^{\rho\sigma} \right).
\end{align}
At tree level, the expectation value of the interaction $I_I$ is calculated as
\begin{align}
    {\langle I_I\rangle}_{g=0}=\frac{dW_g}{dg}\bigg|_{g=0}=\int d[\phi_A] P_0 I_I=0.
\end{align}
Therefore, the definition of the relative entropy yields
\begin{align}
    S(P_0||P_g) &=W_0[\beta,\phi]-W_g[\beta,\phi]+g\cdot  {\langle I_I\rangle}_{g=0}\notag
    \\
    &=W_0[\beta,\phi]-W_g[\beta,\phi]\notag
    \\
    &=g^2\cdot\int (d^4 x)_E \left( \frac{1}{4 m_A^2 M^2}(F_{\rho\sigma}F^{\rho\sigma})^2 \right)\geq 0,\label{eq:relex1}
\end{align}
where we used
\begin{align}
    &W_g[\beta,\phi]=\int (d^4 x)_E \left(\frac{1}{4}F_{\mu\nu} F^{\mu\nu}-g^2\cdot \frac{1}{4 m_A^2 M^2}(F_{\rho\sigma}F^{\rho\sigma})^2 \right),
    \\
    &W_0[\beta,\phi]=\int (d^4 x)_E \left(\frac{1}{4}F_{\mu\nu} F^{\mu\nu}\right).
\end{align}

Here, consider a field redefinition:
\begin{align}
    \phi_A\to \phi_A+g\cdot\frac{1}{2 m_A^2 M}F_{\rho\sigma}F^{\rho\sigma}.
\end{align}
Under this field redefinition, the actions are transformed as
\begin{align}
    I_0^{(E)}&\to {I'}^{(E)}_0=\int (d^4x)_E \left(\frac{1}{4}F_{\mu\nu}F^{\mu\nu}+m_A^2 \phi_A^2+g\cdot  \frac{1}{M}\phi_A F_{\rho\sigma}F^{\rho\sigma} +g^2 \cdot \frac{1}{4 m_A^2 M^2}( F_{\rho\sigma}F^{\rho\sigma})^2 \right), \\
    g\cdot I_I^{(E)} &\to g\cdot {I'}^{(E)}_I= g\cdot I_I^{(E)} -g^2\cdot \int (d^4 x)_E \frac{1}{2m_A^2 M^2}(F_{\rho\sigma}F^{\rho\sigma})^2,\label{eq:fireIN}
    \\
    I_g^{(E)} &\to {I'}^{(E)}_g=\int (d^4 x)_E \left(\frac{1}{4}F_{\mu\nu}F^{\mu\nu}+m_A^2 \phi_A^2 -g^2 \cdot \frac{1}{4 m_A^2 M^2}( F_{\rho\sigma}F^{\rho\sigma})^2 \right).
\end{align}
Similarly, the relative entropy is transformed as
\begin{align}
    S(P_0||P_g)\to S(P'_0||P'_g) 
\end{align}
where $P'_0=e^{-I'_0}/Z'_0[\beta,\phi]$ and $P'_g=e^{-I'_g}/Z'_g[\beta,\phi]$ with $Z'_0[\beta,\phi]=\int d[\phi_A] e^{-I'_0}$ and $Z'_g[\beta,\phi]=\int d[\phi_A] e^{-I'_g}$.
Then, the relative entropy $ S(P'_0||P'_g)$ is calculated as
\begin{align}
    S(P'_0||P'_g)&=\int d[\phi_A] \left(P'_0\ln P'_0-P'_0\ln P'_g \right)
    \\
    &=-\ln Z'_0[\beta,\phi]+\ln Z'_g[\beta,\phi]+ \int d[\phi_A] P'_0 \left(I'_g -I'_0 \right)\notag
    \\
    &=W'_0[\beta,\phi]- W'_g[\beta,\phi]-g^2 \cdot \frac{1}{2 m_A^2 M^2}(F_{\rho\sigma}F^{\rho\sigma})^2-g\cdot \int d[\phi_A] P'_0 \int (d^4 x)_E \frac{1}{M} \phi_A (F_{\rho\sigma}F^{\rho\sigma})\notag
    \\
    &=W'_0[\beta,\phi]- W'_g[\beta,\phi]\notag
    \\
    &=g^2\cdot\int (d^4 x)_E \left( \frac{1}{4 m_A^2 M^2}(F_{\rho\sigma}F^{\rho\sigma})^2 \right)\geq 0,\label{eq:relex2}
\end{align}
where we used following relations:
\begin{align}
    &\int d[\phi_A] P'_0 \int (d^4 x)_E \frac{1}{M} \phi_A (F_{\rho\sigma}F^{\rho\sigma})=-g \cdot \frac{1}{2 m_A^2 M^2}(F_{\rho\sigma}F^{\rho\sigma})^2,
    \\
    &W'_0[\beta,\phi]=W_0[\beta,\phi],
    \\
    &W'_g[\beta,\phi]=W_g[\beta,\phi].
\end{align}
Comparing Eq.~\eqref{eq:relex1} and \eqref{eq:relex2}, we found that the relative entropy is invariant under the field redefinition.
%
%
Although, in this Letter, we focus on the case that $I_I$ represents the interactions between the heavy and light degrees of freedom, the formulation using the relative entropy does not depend on whether $I_I$ represents the interactions.
In fact, as shown in Eq.~\eqref{eq:fireIN}, non-interacting terms arise after the field redefinition.
%
%
The key point is that the formulation using the relative entropy is invariant under the field redefinition once $I_0$ and $I_I$ are defined.
\\

\section{Wave function renormalization in relative entropy}
\label{app:(IV-3)}
We evaluate the entropy constraints on EFTs paying particular attention to the wave function renormalization.
    To clarify the wave function renormalization, we suppose that the light fields are dynamical and evaluate the relative entropy by the procedure of Appendix~\ref{app:der(5)}.
    We focus on two cases: tree-level UV completion and loop-level UV completion.
    In the tree-level UV completion, we assume the tree-level effects dominate the perturbative corrections from the heavy degrees of freedom to the effective actions.
    On the other hand, in the loop-level UV completion, we assume the loop-level effects dominate the perturbative corrections to effective actions.
    %
    %
    %
    %
    %
    %
    %
    %
    %
    %
    %
    %
    For the two cases, we calculate the relative entropy of each EFTs as follows,
\subsection{Single massless scalar field with dimension-eight term}
According to the assumptions, i.e., $J[\phi]$ does not iclude the higher-dimensional operators, $J[\phi]\propto \partial_{\mu}\phi\partial^{\mu}\phi$ may hold.
Then, from Eq.~\eqref{eq:II}, the first order corrections for $g$ to the Euclidean effective action are expressed as
\begin{align}
    {\langle I_I\rangle}_{g=0}=\left(\frac{dW_g}{dg}\right)_{g=0}\propto \int (d^4x)_E (\partial_{\mu}\phi\partial^{\mu}\phi).\label{eq:IIsing}
\end{align}
Note here that $J[\phi]$ can be proportional to $\phi$, $\partial_{\mu}\phi$, and so on, but the EFT does not respect the symmetry of the EFTs, such as the Lorentz symmetry and the global shift symmetry.
%
%
%
\begin{itemize}
    \item Tree-level UV completion ---
    First, consider the EFT generated by the tree-level UV completion.
    Then, not depending on the details of the UV theory, the partition function is calculated as follows,
    \begin{align}
        Z_g[\widetilde{\phi}]&=\int d[\phi]d[\Phi] e^{-I_g[\phi,\Phi]}\notag
        \\
        &=\int d[\phi]{\rm exp}\bigg[
        -\int (d^4 x)_E \bigg(\frac{1}{2}(1+\alpha_2^{\rm tree})(\partial_{\mu}\phi\partial^{\mu}{\phi})-\beta_2^{\rm tree}(\partial_{\mu}{\phi}\partial^{\mu}{\phi})^2 \bigg)
        \bigg]\notag
        \\
        &={\rm exp}\bigg[
        -\int (d^4 x)_E \bigg(\frac{1}{2}(1+\alpha_2^{\rm tree})(\partial_{\mu}\widetilde{\phi'}\partial^{\mu}\widetilde{\phi'})-\beta_2^{\rm tree}(\partial_{\mu}\widetilde{\phi'}\partial^{\mu}\widetilde{\phi'})^2 
        \bigg)
        \bigg]\notag
        \\
        &={\rm exp}\bigg[
        -\int (d^4 x)_E \bigg(\frac{1}{2}(\partial_{\mu}\widetilde{\phi}\partial^{\mu}\widetilde{\phi})-\beta_2^{\rm tree}\cdot\left(1+\alpha^{\rm tree}_2 \right)^{-2}(\partial_{\mu}\widetilde{\phi}\partial^{\mu}\widetilde{\phi})^2 
        \bigg)
        \bigg],
    \end{align}
    where $\alpha_2^{\rm tree}$ and $\beta_2^{\rm tree}$ denote the second or higher order corrections for $g$.
    Note here that $\beta_2^{\rm tree}$ does not include the first order correction for $g$ because of Eq.~\eqref{eq:IIsing}.
    It is assumed that $\alpha_2^{\rm tree}$, and $\beta_2^{\rm tree}$ are generated at the tree-level. 
    Also, in the second line, according to the procedure in Eq.~\eqref{eq:linrem1}, \eqref{eq:linrem2}, and \eqref{eq:linrem3}, the first order correction for $g$ is eliminated in $\alpha_2^{\rm tree}$. 
    The background field $\widetilde{\phi'}$ denotes the classical solution of the effective action of
    \begin{align}
        W_g[\phi]=\int (d^4 x)_E \bigg(\frac{1}{2}(1+\alpha_2^{\rm tree})(\partial_{\mu}\phi\partial^{\mu}{\phi})-\beta_2^{\rm tree}(\partial_{\mu}{\phi}\partial^{\mu}{\phi})^2 \bigg).\label{eq:}
    \end{align}
    The equation of motion of $W_g[\phi]$ is expressed as
    \begin{align}
        (1+\alpha_2^{\rm tree})\partial_{\mu}\partial^{\mu}\phi-\beta_2^{\rm tree} \partial_{\mu}\left(\partial_{\nu}\phi\partial^{\nu}\phi\partial^{\mu}\phi\right)=0.
    \end{align}
    To remove the dimension-six operators, we choose the background fields as follows,
    \begin{align}
        \widetilde{\phi'}=\left(1+\alpha^{\rm tree}_2 \right)^{-1/2}\cdot\widetilde{\phi},
    \end{align}
    where $\partial_{\mu}\widetilde{\phi}={\rm const.}$.
    Note here that the background field $\widetilde{\phi}$ is also a classical solution of $W_0[\phi]$.
    Therefore, the Euclidean effective actions of theories $B$ and $A$ are respectively obtained as
    \begin{align}
        &W_g[\widetilde{\phi}]=-\ln Z_g[\widetilde{\phi}]=\int (d^4 x)_E \bigg(\frac{1}{2}(\partial_{\mu}\widetilde{\phi}\partial^{\mu}\widetilde{\phi})-\beta_2^{\rm tree}\cdot\left(1+\alpha^{\rm tree}_2 \right)^{-2}(\partial_{\mu}\widetilde{\phi}\partial^{\mu}\widetilde{\phi})^2 
        \bigg),\label{eq:sigWg}
        \\
        &W_0[\widetilde{\phi}]=-\ln Z_0[\widetilde{\phi}]=\int (d^4 x)_E \bigg(\frac{1}{2}(\partial_{\mu}\widetilde{\phi}\partial^{\mu}\widetilde{\phi}) 
        \bigg).
    \end{align}
    Then, the shift of the Euclidean effective action is calculated as
    \begin{align}
        \Delta W_g^{(E)}&=W_g[\widetilde{\phi}]-W_0[\widetilde{\phi}]\notag
        \\
        &=-\beta_2^{\rm tree}\cdot\left(1+\alpha^{\rm tree}_2 \right)^{-2} \int (d^4x)_E (\partial_{\mu}\widetilde{\phi}\partial^{\mu}\widetilde{\phi})^2.\label{eq:DelWgsig}
    \end{align}
    Also, from Eq.~\eqref{eq:sigWg}, we obtain the following relation
    \begin{align}
        \left(\frac{dW_g}{dg}\right)_{g=0}&=\left(\frac{\partial W_g}{\partial g}\right)_{g=0}+\int (d^4 x)_E \left(\frac{\delta W_g}{\delta \widetilde{\phi'}}\right)\cdot \left(\frac{d\widetilde{\phi'}}{dg}\right)_{g=0}\notag
        \\
        &=\left(\frac{\partial W_g}{\partial g}\right)_{g=0}=0,\label{eq:sigdWgdg}
    \end{align}
    where $({d\widetilde{\phi'}}/{dg})_{g=0}=0$ because $\alpha_2^{\rm tree}$ denotes the second or higher order corrections for $g$.
    From Eq.~\eqref{eq:uplow} or \eqref{eq:dyligh}, combining Eq.~\eqref{eq:DelWgsig} and \eqref{eq:sigdWgdg} yields
    \begin{align}
        \Delta W_g^{(E)}\leq g\cdot {\langle I_I\rangle}_{g=0}&\Rightarrow -\beta_2^{\rm tree}\cdot\left(1+\alpha^{\rm tree}_2 \right)^{-2} \int (d^4x)_E (\partial_{\mu}\widetilde{\phi}\partial^{\mu}\widetilde{\phi})^2\leq 0 \notag
        \\
        &\Rightarrow \beta_2^{\rm tree}\cdot\left(1+\alpha^{\rm tree}_2 \right)^{-2}\geq 0.\label{eq:treesin1}
    \end{align}
    Equation~\eqref{eq:treesin1} denotes the constraint on the coefficient of dimension-eight operator of Eq.~\eqref{eq:sigWg}.
    
    \item Loop-level UV completion ---
    Next, consider the EFT generated by the loop-level UV completion.
    Then, the partition function is calculated as follows,
        \begin{align}
        Z_g[\widetilde{\phi}]&=\int d[\phi]d[\Phi] e^{-I_g[\phi,\Phi]}\notag
        \\
        &=\int d[\phi] {\rm exp}\bigg[
        -\int (d^4 x)_E \bigg(\frac{1}{2}(1+ \alpha_1^{\rm loop}+\alpha_2^{\rm loop})(\partial_{\mu}\phi\partial^{\mu}\phi)-\beta_2^{\rm loop}(\partial_{\mu}\phi\partial^{\mu}\phi)^2 +E_{\rm vac}^{\Phi}\bigg)
        \bigg]\notag
        \\
        &={\rm exp}\bigg[
        -\int (d^4 x)_E \bigg(\frac{1}{2}(1+\alpha^{\rm loop}_1+\alpha_2^{\rm loop})(\partial_{\mu}\widetilde{\phi'}\partial^{\mu}\widetilde{\phi'})-\beta_2^{\rm loop}(\partial_{\mu}\widetilde{\phi'}\partial^{\mu}\widetilde{\phi'})^2+E_{\rm vac} \bigg)
        \bigg]\notag
        \\
        &={\rm exp}\bigg[
        -\int (d^4 x)_E \bigg(\frac{1}{2}(1+\alpha^{\rm loop}_1)(\partial_{\mu}\widetilde{\phi}\partial^{\mu}\widetilde{\phi})-\beta_2^{\rm loop}(\partial_{\mu}\widetilde{\phi}\partial^{\mu}\widetilde{\phi})^2+E_{\rm vac} 
        \bigg)
        \bigg],\label{eq:appZgSMEFT}
    \end{align}
    %
    %
    where $\alpha_1^{\rm loop}$ is the first order correction for $g$, $\alpha_2^{\rm loop}$ and $\beta_2^{\rm loop}$ are the second or higher order correction for $g$, 
    $E_{\rm vac}^{\Phi}$ is the vacuum energy coming from the loop-level correction of $\Phi$, and $E_{\rm vac}$ is the vacuum energy of $\Phi$ and $\phi.$
    %
    %
    %
    It is assumed that $\alpha_1^{\rm loop}$, $\alpha_2^{\rm loop}$, and $\beta_2^{\rm loop}$ are generated from the loop corrections of $\Phi$.
    The background field $\widetilde{\phi'}$ denotes the classical solution of the effective action of
    \begin{align}
        W_g[\phi]=\int (d^4 x)_E \bigg(\frac{1}{2}(1+ \alpha_1^{\rm loop}+\alpha_2^{\rm loop})(\partial_{\mu}\phi\partial^{\mu}\phi)-\beta_2^{\rm loop}(\partial_{\mu}\phi\partial^{\mu}\phi)^2 \bigg).\label{eq:Wgloop1}
    \end{align}
    %
    The equation of motion of Eq.~\eqref{eq:Wgloop1} is expressed as follows,
    \begin{align}
        (1+ \alpha_1^{\rm loop}+\alpha_2^{\rm loop}) \partial_{\mu}\partial^{\mu}\phi-\beta_2^{\rm loop} \partial_{\mu}(\partial_{\nu}\phi\partial^{\nu}\phi \partial^{\mu}\phi)=0.
    \end{align}
    %
    We choose the background field as follows,
    \begin{align}
        \widetilde{\phi'}=\left(1-\frac{1}{2}\alpha_2^{\rm loop}\right)\cdot\widetilde{\phi},
    \end{align}
    %
    where $\partial_{\mu}\widetilde{\phi}={\rm const.}$ to remove the dimension-six operators.
    Since the background field $\widetilde{\phi}$ is also a classical solution of $W_0[\phi]$, the Euclidean effective actions of theories $B$ and $A$ are respectively obtained as
        \begin{align}
        W_g[\widetilde{\phi}]&=-\ln Z_g[\tilde{\phi}]\notag
        \\
        &=\int (d^4 x)_E \bigg(\frac{1}{2}(1+\alpha^{\rm loop}_1)(\partial_{\mu}\widetilde{\phi}\partial^{\mu}\widetilde{\phi})-\beta_2^{\rm loop}(\partial_{\mu}\widetilde{\phi}\partial^{\mu}\widetilde{\phi})^2+E_{\rm vac} 
        \bigg),\label{eq:wgloop2}
        \\
        W_0[\widetilde{\phi}]&=-\ln Z_0[\tilde{\phi}]\notag
        \\
        &=\int (d^4 x)_E \bigg(\frac{1}{2}(\partial_{\mu}\widetilde{\phi}\partial^{\mu}\widetilde{\phi})+E_{\rm vac} 
        \bigg).
    \end{align}
    %
    %
    Then, the shift of the Euclidean effective action is calculated as
    %
    \begin{align}
    \Delta W_g^{(E)}&=W_g[\widetilde{\phi}]-W_0[\widetilde{\phi}]\notag
    \\
    &=\frac{1}{2}\alpha^{\rm loop}_1\cdot\int (d^4 x)_E (\partial_{\mu}\widetilde{\phi}\partial^{\mu}\widetilde{\phi})-\beta_2^{\rm loop}\int (d^4 x)_E (\partial_{\mu}\widetilde{\phi}\partial^{\mu}\widetilde{\phi})^2.\label{eq:DelWgloop3}
    \end{align}
    Also, from Eq.~\eqref{eq:wgloop2}, we obtain
    \begin{align}
        \left(\frac{dW_g}{dg}\right)_{g=0}&=\left(\frac{\partial W_g}{\partial g}\right)_{g=0} +\int (d^4x)_E \left(\frac{\delta W_g}{\delta \widetilde{\phi'}}\right)\cdot \left(\frac{d \widetilde{\phi'}}{dg}\right)_{g=0}\notag
        \\
        &=\left(\frac{\partial W_g}{\partial g}\right)_{g=0}\notag
        \\
        &=\frac{1}{2}\frac{d\alpha^{\rm loop}_1}{dg}\cdot\int (d^4x)_E (\partial_{\mu}\widetilde{\phi}\partial^{\mu}\widetilde{\phi})
        ,\label{eq:dWgdgloop2}
    \end{align}
    where $({d \widetilde{\phi'}}/{dg})_{g=0}=0$ is used.
    Note here that $\alpha_1^{\rm loop}$ denotes the first order correction for $g$ and satisfies a relation of the form $g\cdot ({d\alpha^{\rm loop}_1}/{dg})=\alpha^{\rm loop}_1$.
    From Eq.~\eqref{eq:uplow} or \eqref{eq:dyligh}, combining Eq.~\eqref{eq:DelWgloop3} and \eqref{eq:dWgdgloop2} yields
    \begin{align}
        \Delta W_g^{(E)}\leq g\cdot  {\langle I_I \rangle}_{g=0} &\Rightarrow -\beta_2^{\rm loop}\int (d^4 x)_E (\partial_{\mu}\widetilde{\phi}\partial^{\mu}\widetilde{\phi})^2\leq 0\notag
        \\
        &\Rightarrow \beta_2^{\rm loop}\geq 0.\label{eq:loopboubd1}
    \end{align}
    %
    In the loop-level UV completions, Eq.~\eqref{eq:loopboubd1} yields the constraint on the dimension-eight operator generated at the loop-level.

\end{itemize}

\subsection{SMEFT dimension-eight $SU(N)$ gauge bosonic operators}
When $J[A^a_{\mu}]$ does not include the higher-dimensional operators, there are two cases: (i) $J[A^a_{\mu}]$ preserves the gauge symmetry or (ii) not.
For case (i), $J[A^a_{\mu}]\propto F^a_{\mu\nu}F^{a,\mu\nu}$ holds.
In general, the $CP$ violating term arises, but we supposed that such a term is removed by axion-like degrees of freedom in the UV theory.
Then, from Eq.~\eqref{eq:II}, the first order corrections for $g$ to the Euclidean effective action are expressed as
\begin{align}
    {\langle I_I\rangle}_{g=0}=\left(\frac{dW_g}{dg}\right)_{g=0}\propto \int (d^4x)_E F^a_{\mu\nu}F^{a,\mu\nu}.\label{eq:IISMEFT}
\end{align}
For case (ii), $J[A^a_{\mu}]$ can be proportional to $A^a_{\mu}$, and $A^a_{\mu}A^{a,\mu}$ because of the covariant derivative of the kinetic term.
Since corrections from the interacting terms of the higher-dimensional operators would not be dominant effects, we focus on corrections from the kinetic terms.   
Then, $J[A^a_{\mu}]\propto A^a_{\mu}$ vanishes because ${\langle I_I\rangle}_{g=0}$ keeps the Lorentz symmetry.
Although $J[A^a_{\mu}]\propto A^a_{\mu}A^{a,\mu}$ generally remains, it can be eliminated by the gauge fixing condition.
Therefore, we focus on the case of Eq.~\eqref{eq:IISMEFT} below. 


\begin{itemize}
    \item Tree-level UV completion ---
    Consider the EFT generated by the tree-level UV completion.
    The partition function is generally calculated as follows, 
    \begin{align}
        Z_g[\overline{A}]&=\int d[A] d[\Phi] e^{-I_g[A,\Phi]}\notag
        \\
        &=\int d[A] {\rm exp}\left[
        -\int (d^4x)_E \left(\frac{1}{2}(1+\alpha^{\rm tree}_2)F^a_{\mu\nu}F^{a,\mu\nu}-\sum_i\beta_{i,2}^{\rm tree}\mathcal{O}_i[A] \right)
        \right]\notag
        \\
        &= {\rm exp}\left[
        -\int (d^4x)_E \left(\frac{1}{2}(1+\alpha^{\rm tree}_2)\overline{F'}^a_{\mu\nu}\overline{F'}^{a,\mu\nu}-\sum_i\beta_{i,2}^{\rm tree}\mathcal{O}_i[\overline{A'}] \right)
        \right]\notag
        \\
        &= {\rm exp}\left[
        -\int (d^4x)_E \left(\frac{1}{2}\overline{F}^a_{\mu\nu}\overline{F}^{a,\mu\nu}-\sum_i\beta_{i,2}^{\rm tree}\cdot (1+\alpha^{\rm tree}_2)^{-2}\mathcal{O}_i[\overline{A}] \right)
        \right],
    \end{align}
    where $\alpha_2^{\rm tree}$ and $\beta_{i,2}^{\rm tree}$ denote the second or higher order corrections for $g$, and $\beta_{i,2}^{\rm tree}$ does not include the first order correction for $g$ because of Eq.~\eqref{eq:IISMEFT}.
    The corrections $\alpha^{\rm tree}_{2}$ and $\beta^{\rm tree}_{i,2}$ are assumed to be generated at the tree-level. 
    According to the procedure in Eq.~\eqref{eq:linrem1}, \eqref{eq:linrem2}, and \eqref{eq:linrem3}, the first order correction for $g$ is eliminated in $\alpha_2^{\rm tree}$. 
    The background field $\overline{A'}_{\mu}^a$ denotes the classical solution of the effective action of
    \begin{align}
        W_g[A]=\int (d^4x)_E \left(\frac{1}{2}(1+\alpha^{\rm tree}_2)F^a_{\mu\nu}F^{a,\mu\nu}-\sum_i\beta_{i,2}^{\rm tree}\mathcal{O}_i[A] \right).
    \end{align}
    The background fields are chosen as follows,
    \begin{align}
        \overline{A'}^a_{\mu}=(1+\alpha_2^{\rm tree})^{-1/2} \cdot \overline{A}^a_{\mu},
    \end{align}
    where $ \overline{F}_{\mu\nu}={\rm const.}$
    Since $\overline{A}^a_{\mu}$ is also a classical solution of $W_0[A]$, the Euclidean effective actions of theories $B$ and $A$ are respectively obtained as follows, 
    \begin{align}
        &W_g[\overline{A}]=-\ln Z_g[\overline{A}]=\int (d^4x)_E \left(\frac{1}{2}\overline{F}^a_{\mu\nu}\overline{F}^{a,\mu\nu}-\sum_i\beta_{i,2}^{\rm tree}\cdot (1+a^{\rm tree}_2)^{-2}\mathcal{O}_i[\overline{A}] \right),\label{eq:Wgsmefttrree}
        \\
        &W_0[\overline{A}]=-\ln Z_0[\overline{A}]=\int (d^4x)_E \left(\frac{1}{2}\overline{F}^a_{\mu\nu}\overline{F}^{a,\mu\nu} \right),
    \end{align}
    Then, the shift of the Euclidean effective action is calculated as follows,
    \begin{align}
        \Delta W_g^{(E)}&=W_g[\overline{A}]-W_0[\overline{A}]\notag
        \\
        &=-\sum_i\beta_{i,2}^{\rm tree}\cdot (1+a^{\rm tree}_2)^{-2}\int (d^4x)_E \mathcal{O}_i[\overline{A}]. \label{eq:DeltreeSMEFT}
    \end{align}
    Also, the first order corrections for $g$ is calculated as
    \begin{align}
        \left(\frac{dW_g}{dg}\right)_{g=0}&=\left(\frac{\partial W_g}{\partial g}\right)_{g=0}+
        \int (d^4x)_E \left(\frac{\delta W_g}{\delta \overline{A'}}\right)\cdot \left(\frac{d \overline{A'}}{dg}\right)_{g=0}\notag
        \\
        &=\left(\frac{\partial W_g}{\partial g}\right)_{g=0}=0,\label{eq:sigdWgdgSMEFT}
    \end{align}
    where $({d \overline{A'}}/{dg})_{g=0}=0$ is used.
     From Eq.~\eqref{eq:uplow} or \eqref{eq:dyligh}, combining Eq.~\eqref{eq:DeltreeSMEFT} and \eqref{eq:sigdWgdgSMEFT} yields
    \begin{align}
        \Delta W_g^{(E)}\leq g\cdot {\langle I_I\rangle}_{g=0}&\Rightarrow -\sum_i\beta_{i,2}^{\rm tree}\cdot (1+a^{\rm tree}_2)^{-2}\int (d^4x)_E \mathcal{O}_i[\overline{A}] \leq 0\notag
        \\
        &\Rightarrow \sum_i\beta_{i,2}^{\rm tree}\cdot (1+a^{\rm tree}_2)^{-2}\int (d^4x)_E \mathcal{O}_i[\overline{A}]\geq 0.\label{eq:boundSMEFT1tree}
    \end{align}
    The left-hand side of Eq.~\eqref{eq:boundSMEFT1tree} denotes a linear combination of coefficients of the dimension-eight operators of Eq.~\eqref{eq:Wgsmefttrree}.
    %
    
    \item Loop-level UV completion ---
    Consider the SMEFT generated by the loop-level UV completion.
    The partition function is generally calculated as follows,
        \begin{align}
        Z_g[\overline{A}]&= \int d[A] d[\Phi] e^{-I_g[A,\Phi]}\notag
        \\
        &=\int d[A] {\rm exp}\left[
        -\int (d^4x)_E \left(
        \frac{1}{2} \left(1+\alpha^{\rm loop}_1+\alpha^{\rm loop}_2\right)F^a_{\mu\nu}F^{a,\mu\nu}
        -\sum_i \beta^{\rm loop}_{2,i}\mathcal{O}_i[{A}]+E^{\Phi}_{\rm vac}\right)
        \right]\notag
        \\
        &={\rm exp}\bigg[
        -\int (d^4x)_E \bigg(
        \frac{1}{2} \left(1+\alpha^{\rm loop}_1+\alpha^{\rm loop}_2\right)\overline{F'}^a_{\mu\nu}\overline{F'}^{a,\mu\nu}-\sum_i\beta^{\rm loop}_{2,i}\mathcal{O}_i[\overline{A}']+E_{\rm vac}\bigg)
        \bigg]\notag
        \\
        &={\rm exp}\bigg[
        -\int (d^4x)_E \bigg(
        \frac{1}{2} \left(1+\alpha^{\rm loop}_1\right)\overline{F}^a_{\mu\nu}\overline{F}^{a,\mu\nu}-\sum_i\beta^{\rm loop}_{2,i} \mathcal{O}_i[\overline{A}]+E_{\rm vac}\bigg)
        \bigg].\label{eq:appZgloopSMEFT}
    \end{align}
    %
    where $\alpha_1^{\rm loop}$ is the first order correction for $g$, $\alpha_2^{\rm loop}$ and $\beta_{2,i}^{\rm loop}$ are the second or higher order correction for $g$, 
    $E_{\rm vac}^{\Phi}$ is the vacuum energy coming from the loop-level correction of $\Phi$, and $E_{\rm vac}$ is the vacuum energy of $\Phi$ and $A^a_{\mu}$.
    %
    %
    %
    It is assumed that $\alpha_1^{\rm loop}$, $\alpha_2^{\rm loop}$, and $\beta_{2,i}^{\rm loop}$ are generated from the loop corrections of $\Phi$.
    The background field $\widetilde{A'}^a_{\mu}$ denotes the classical solution of the effective action of
    \begin{align}
        W_g[{A}]=\int (d^4x)_E \left(
        \frac{1}{2} \left(1+\alpha^{\rm loop}_1+\alpha^{\rm loop}_2\right)F^a_{\mu\nu}F^{a,\mu\nu}
        -\sum_i\beta^{\rm loop}_{2,i}\mathcal{O}_i[{A}]+E^{\Phi}_{\rm vac}\right).
    \end{align}
    We choose the background field as follows,
    \begin{align}
        \overline{A'}^a_{\mu}=\left(1-\frac{1}{2}\alpha_2^{\rm loop} \right)\overline{A}^a_{\mu},
    \end{align}
    where $\overline{F}_{\mu\nu}={\rm const.}$ to remove the dimension-six operators.
    $\overline{A}^a_{\mu}$ is also a classical solution of $W_0[A]$, and the Euclidean effective actions of theories $B$ and $A$ are respectively obtained as follows, 
    \begin{align}
        W_g[\overline{A}]&=\int (d^4x)_E \bigg(
        \frac{1}{2} \left(1+\alpha^{\rm loop}_1\right)\overline{F}^a_{\mu\nu}\overline{F}^{a,\mu\nu}-\sum_i\beta^{\rm loop}_{2,i} \mathcal{O}_i[\overline{A}]+E_{\rm vac}\bigg),\label{eq:appSMEFTWGloop}
        \\
        W_0[\overline{A}]&=\int (d^4x)_E \bigg(
        \frac{1}{2} \overline{F}^a_{\mu\nu}\overline{F}^{a,\mu\nu}+E_{\rm vac}\bigg).
    \end{align}
    Then, the shift of the Euclidean effective action is calculated as follows,
    \begin{align}
        \Delta W^{(E)}_g &= W_g[\overline{A}]-W_0[\overline{A}]\notag
        \\
        &=\int (d^4x)_E \bigg(
        \frac{1}{2} \alpha^{\rm loop}_1\overline{F}^a_{\mu\nu}\overline{F}^{a,\mu\nu}-\sum_i\beta^{\rm loop}_{2,i} \mathcal{O}_i[\overline{A}]\bigg).\label{eq:loopDelWgSM}
    \end{align}
    Also, the first order corrections for $g$ is calculated as
    \begin{align}
        \left(\frac{dW_g}{dg}\right)_{g=0}&=\left(\frac{\partial W_g}{\partial g}\right)_{g=0}+\int (d^4x)_E \left(\frac{\delta W_g}{\delta \overline{{A}'}}\right)\cdot \left(\frac{d \overline{{A}'}}{d g}\right)_{g=0}\notag
        \\
        &=\left(\frac{\partial W_g}{\partial g}\right)_{g=0}\notag
        \\
        &=\frac{1}{2} \frac{d\alpha^{\rm loop}_1}{dg}\int (d^4x)_E 
        \overline{F}^a_{\mu\nu}\overline{F}^{a,\mu\nu},\label{eq:loopdelWdgSM}
    \end{align}
    where $({d \overline{{A}'}}/{d g})_{g=0}=0$ is used.
    From Eq.~\eqref{eq:uplow} or \eqref{eq:dyligh}, combining Eq.~\eqref{eq:loopDelWgSM} and \eqref{eq:loopdelWdgSM} yields
    \begin{align}
        \Delta W_g^{(E)}\leq g {\langle I_I\rangle}_{g=0}&\Rightarrow -\sum_i\beta^{\rm loop}_{2,i}\int (d^4x)_E 
        \mathcal{O}_i[\overline{A}]\leq 0\notag
        \\
        &\Rightarrow \sum_i \beta^{\rm loop}_{2,i}\int (d^4x)_E 
        \mathcal{O}_i[\overline{A}] \geq 0.\label{eq:SMEFTloopcons}
    \end{align}
    where $g\cdot (d\alpha_1^{\rm loop}/dg)=\alpha_1^{\rm loop}$ is used.
    In the loop-level UV completion, 
    Eq.~\eqref{eq:SMEFTloopcons} yields the constraint on the dimension-eight operator generated at the loop-level.

\end{itemize}

\subsection{Einstein-Maxwell theory with higher-dimensional operators}
Consider the Einstein-Maxwell theory with higher-dimensional operators generated from the UV theory defined by $I[g_{\mu\nu};R_{\mu\nu\rho\sigma},A,\Phi]$, where $g_{\mu\nu}$ is the metric of space-time, $R_{\mu\nu\rho\sigma}$ is the Riemann tensor, $A_{\mu}$ is the $U(1)$ gauge boson, and $\Phi$ is the heavy degrees of freedom.
Define the non-interacting and interacting terms as follows, 
\begin{align}
    &I_0[g_{\mu\nu};R_{\mu\nu\rho\sigma},A,\Phi]=I[g_{\mu\nu};R_{\mu\nu\rho\sigma},A,0]+I[g_{\mu\nu};0,0,\Phi],
    \\
    &I_I[g_{\mu\nu};R_{\mu\nu\rho\sigma},A,\Phi]=I[g_{\mu\nu};R_{\mu\nu\rho\sigma},A,\Phi]-I_0[g_{\mu\nu};R_{\mu\nu\rho\sigma},A,\Phi],
\end{align}
where the cosmological constant is omitted because it cancels in the relative entropy.
It should be noted that the theory of $I_0$ does not include the interaction between $\Phi$ and $A_{\mu}, R_{\mu\nu\rho\sigma}$, but the interaction between $g_{\mu\nu}$ and $\Phi$.
Note that gravitational operators such as $R_{\mu\nu}^2$ can be generated from $I_0$.
Also, the Gauss-Bonnet combination, i.e., $R_{\mu\nu\rho\sigma}R^{\mu\nu\rho\sigma}-4 R_{\mu\nu}R^{\mu\nu}+R^2$, is a total derivative and vanishes in four dimensions.
In this work, we focus on the higher-dimensional operators generated from the interaction between $\Phi$ and $A_{\mu}, R_{\mu\nu\rho\sigma}$.

Similar to the SMEFT, when $J[g_{\mu\nu};R_{\mu\nu\rho\sigma},A_{\mu}]$ does not include the higher-derivative operators, there are two cases: (i) $J[g_{\mu\nu};R_{\mu\nu\rho\sigma},A_{\mu}]\propto F_{\mu\nu}F^{\mu\nu}$ or $R$, and (ii) $J[g_{\mu\nu};R_{\mu\nu\rho\sigma},A_{\mu}]\propto A_{\mu}$ or $A_{\mu}A^{\mu}$.
Because of the same reason as the SMEFT, we focus on the following case,
\begin{align}
{\langle I_I\rangle}_{g=0}=\left(\frac{dW_g}{dg}\right)_{g=0}\propto \int (d^4 x)_E \sqrt{g} F_{\mu\nu}F^{\mu\nu}~{\rm or}~\int (d^4 x)_E \sqrt{g} R.\label{eq:IIEM}
\end{align}
For each of the tree and loop level UV completion, the constraints on the EFTs are evaluated as follows,

\begin{itemize}
    \item Tree-level UV completion ---
    Consider the EFT generated at the tree-level UV completion.
    Then, the partition function is generally calculated as follows,
    \begin{align}
        Z_g[\overline{g}_{\mu\nu},\overline{A}]&=\int d [g]d[A]d[\Phi]e^{-I_g[g_{\mu\nu};R_{\mu\nu\rho\sigma},A,\Phi]}\notag
        \\
        &=\int d [g]d[A] {\rm exp}\bigg[
        -\int (d^4x)_E \sqrt{g}\bigg(-\frac{M^2_{\rm Pl}}{2}(1+\alpha_{2,R}^{\rm tree})R +\frac{1}{4}(1+\alpha_{2,F}^{\rm tree})F_{\mu\nu}F^{\mu\nu}-\beta_{2,1}^{\rm tree}(F_{\mu\nu}F^{\mu\nu})^2\notag
        \\
        &-\beta_{2,2}^{\rm tree}(F_{\mu\nu}\widetilde{F}^{\mu\nu})^2-\beta_{2,3}^{\rm tree} F_{\mu\nu}F_{\rho\sigma}R^{\mu\nu\rho\sigma}\bigg)
        \bigg]\notag
        \\
        &={\rm exp}\bigg[
        -\int (d^4x)_E \sqrt{\overline{g'}}\bigg(-\frac{M^2_{\rm Pl}}{2}(1+\alpha_{2,R}^{\rm tree})\overline{R'} +\frac{1}{4}(1+\alpha_{2,F}^{\rm tree})\overline{F'}_{\mu\nu}\overline{F'}^{\mu\nu}-\beta_{2,1}^{\rm tree}(\overline{F'}_{\mu\nu}\overline{F'}^{\mu\nu})^2\notag
        \\
        &-\beta_{2,2}^{\rm tree}(\overline{F'}_{\mu\nu}\widetilde{\overline{F'}}^{\mu\nu})^2-\beta_{2,3}^{\rm tree} \overline{F'}_{\mu\nu}\overline{F'}_{\rho\sigma}\overline{R'}^{\mu\nu\rho\sigma}\bigg)
        \bigg]\notag
        \\
        &={\rm exp}\bigg[
        -\int (d^4x)_E \sqrt{\overline{g}}\bigg(-\frac{M^2_{\rm Pl}}{2}\overline{R} +\frac{1}{4}\overline{F}_{\mu\nu}\overline{F}^{\mu\nu}-\beta_{2,1}^{\rm tree}\left(1+\frac{2}{3} \alpha^{\rm tree}_{2,R}-2 \alpha^{\rm tree}_{2,F}\right)(\overline{F}_{\mu\nu}\overline{F}^{\mu\nu})^2\notag
        \\
        &-\beta_{2,2}^{\rm tree}\left(1+2 \alpha^{\rm tree}_{2,R}-2\alpha^{\rm tree}_{2,F}\right)(\overline{F}_{\mu\nu}\widetilde{\overline{F}}^{\mu\nu})^2-\beta_{2,3}^{\rm tree}\left(1+\frac{1}{3}\alpha^{\rm tree}_{2,R}-\alpha^{\rm tree}_{2,F}\right) \overline{F}_{\mu\nu}\overline{F}_{\rho\sigma}\overline{R}^{\mu\nu\rho\sigma}\bigg)
        \bigg],
    \end{align}
    where $\alpha_{2,R}^{\rm tree}$, $\alpha_{2,F}^{\rm tree}$, $\beta_{2,1}^{\rm tree}$, $\beta_{2,2}^{\rm tree}$ and $\beta_{2,3}^{\rm tree}$ denote the second or higher order corrections for $g$.
    Note here that $\beta_{2,1}^{\rm tree}$, $\beta_{2,2}^{\rm tree}$ and $\beta_{2,3}^{\rm tree}$ do not include the first order correction for $g$ because of Eq.~\eqref{eq:IIEM}.
    %
    %
    According to the procedure in Eq.~\eqref{eq:linrem1}, \eqref{eq:linrem2}, and \eqref{eq:linrem3}, the first order correction for $g$ is eliminated in $\alpha_{2,R}^{\rm tree}$ and $\alpha_{2,F}^{\rm tree}$. 
    Since the gravitational operators only involving the Riemann tensors can be removed by field redefinition, and the Riemann-squared operator can be dropped in four dimensions, we omit such terms.  
    The background fields $\overline{A'}_{\mu}$ and $\overline{g'}_{\mu\nu}$ denote the classical solutions of the effective action of
    \begin{align}
        W_g[g_{\mu\nu},A]&=\int (d^4x)_E \sqrt{g}\bigg(-\frac{M^2_{\rm Pl}}{2}(1+\alpha_{2,R}^{\rm tree})R +\frac{1}{4}(1+\alpha_{2,F}^{\rm tree})F_{\mu\nu}F^{\mu\nu}-\beta_{2,1}^{\rm tree}(F_{\mu\nu}F^{\mu\nu})^2\notag
        \\
        &-\beta_{2,2}^{\rm tree}(F_{\mu\nu}\widetilde{F}^{\mu\nu})^2-\beta_{2,3}^{\rm tree} F_{\mu\nu}F_{\rho\sigma}R^{\mu\nu\rho\sigma}\bigg).
    \end{align}
    We choose the background field as follows,
    \begin{align}
        &\overline{A'}_{\mu}=\left(1+\frac{1}{2}\left(\frac{4}{3}\alpha_{2,R}^{\rm tree}-\alpha_{2,F}^{\rm tree}\right)\right)\overline{A}_{\mu},
        \\
        &\overline{g'}_{\mu\nu}=\left(1-\frac{1}{3}\alpha_{2,R}^{\rm tree}\right)\overline{g}_{\mu\nu},~~~\overline{g'}^{\mu\nu}=\left(1+\frac{1}{3}\alpha_{2,R}^{\rm tree}\right)\overline{g}^{\mu\nu}.
    \end{align}
    The effective actions of theories $B$ and $A$ are respectively obtained as follows,
    \begin{align}
        W_g[\overline{g}_{\mu\nu},\overline{A}]&=\int (d^4x)_E \sqrt{\overline{g}}\bigg(-\frac{M^2_{\rm Pl}}{2}\overline{R} +\frac{1}{4}\overline{F}_{\mu\nu}\overline{F}^{\mu\nu}-\beta_{2,1}^{\rm tree}\left(1+\frac{2}{3}\alpha^{\rm tree}_{2,R}-2 \alpha^{\rm tree}_{2,F}\right)(\overline{F}_{\mu\nu}\overline{F}^{\mu\nu})^2\notag
        \\
        &-\beta_{2,2}^{\rm tree}\left(1+2\alpha^{\rm tree}_{2,R}-2\alpha^{\rm tree}_{2,F}\right)(\overline{F}_{\mu\nu}\widetilde{\overline{F}}^{\mu\nu})^2-\beta_{2,3}^{\rm tree}\left(1+\frac{1}{3}\alpha^{\rm tree}_{2,R}-\alpha^{\rm tree}_{2,F}\right) \overline{F}_{\mu\nu}\overline{F}_{\rho\sigma}\overline{R}^{\mu\nu\rho\sigma}\bigg),\label{eq:WgEMth}
        \\
        W_0[\overline{g}_{\mu\nu},\overline{A}]&=\int (d^4x)_E \sqrt{\overline{g}}\bigg(-\frac{M^2_{\rm Pl}}{2}\overline{R} +\frac{1}{4}\overline{F}_{\mu\nu}\overline{F}^{\mu\nu}\bigg),
    \end{align}
    %
    %
    where $\overline{A}_{\mu}$ and $\overline{g}_{\mu\nu}$ include the effects of the higher-derivative terms.  
    It should be noted that the first order correction for the higher-derivative terms vanishes in $W_0$ by using the equation of motion.
    Then, $\Delta W_g^{(E)}=W_g[\overline{g}_{\mu\nu},\overline{A}]-W_0[\overline{g}_{\mu\nu},\overline{A}]$ denotes the shift of the Euclidean effective action by the higher-derivative terms.
    %
    %
    Also, from Eq.~\eqref{eq:WgEMth}, the first order correction for $g$ is calculated as
    \begin{align}
        \left(\frac{dW_g}{dg}\right)_{g=0}&=\left(\frac{\partial W_g}{\partial g}\right)_{g=0}
        +\int (d^4x)_E\sqrt{-g} \bigg(\left(\frac{\delta W_g}{\delta \overline{A'}}\right)\cdot \left(\frac{d \overline{A'}}{dg}\right)_{g=0}+\left(\frac{\delta W_g}{\delta \overline{g'}_{\mu\nu}}\right)\cdot \left(\frac{d \overline{g'}_{\mu\nu}}{dg}\right)_{g=0}\bigg)\notag
        \\
        &=\left(\frac{\partial W_g}{\partial g}\right)_{g=0}=0,\label{eq:EMdWdgtree}
    \end{align}
    where $(d\overline{A'}/dg)_{g=0}=0$ and $(d\overline{g'}_{\mu\nu}/dg)_{g=0}=0$ are used.
    From Eq.~\eqref{eq:uplow} or \eqref{eq:dyligh}, Eq.~\eqref{eq:EMdWdgtree} yields
    \begin{align}
        \Delta W^{(E)}_g \leq 0.
    \end{align}
    Consequently, it is found that the relative entropy yields the negative shift of the effective action by the higher derivative terms generated at the tree-level.
    

    \item Loop-level UV completion ---
    Next, consider the EFT generated by the loop-level UV completion.
    %
    %
    The partition function is generally calculated as follows,
    \begin{align}
        Z_g[\overline{g}_{\mu\nu},\overline{A}]&=\int d [g]d[A]d[\Phi]e^{-I_g[g_{\mu\nu};R_{\mu\nu\rho\sigma},A,\Phi]}\notag
        \\
        &=\int d [g]d[A] {\rm exp}\bigg[
        -\int (d^4x)_E \sqrt{g}\bigg(\Lambda_{0,\Phi}^{\rm loop}-\frac{M^2_{\rm Pl}}{2}(1+\alpha_{1,R}^{\rm loop}+\alpha_{2,R}^{\rm loop})R +\frac{1}{4}(1+\alpha_{1,F}^{\rm loop}+\alpha_{2,F}^{\rm loop})F_{\mu\nu}F^{\mu\nu}\notag
        \\
        &-\beta_{2,1}^{\rm loop}(F_{\mu\nu}F^{\mu\nu})^2-\beta_{2,2}^{\rm loop}(F_{\mu\nu}\widetilde{F}^{\mu\nu})^2-\beta_{2,3}^{\rm loop} F_{\mu\nu}F_{\rho\sigma}R^{\mu\nu\rho\sigma}\bigg)
        \bigg]\notag
        \\
        &={\rm exp}\bigg[
        -\int (d^4x)_E \sqrt{\overline{g'}}\bigg(\Lambda_{0,\Phi}^{\rm loop}-\frac{M^2_{\rm Pl}}{2}(1+\alpha_{1,R}^{\rm loop}+\alpha_{2,R}^{\rm loop})\overline{R'} +\frac{1}{4}(1+\alpha_{1,F}^{\rm loop}+\alpha_{2,F}^{\rm loop})\overline{F'}_{\mu\nu}\overline{F'}^{\mu\nu}\notag
        \\
        &-\beta_{2,1}^{\rm loop}(\overline{F'}_{\mu\nu}\overline{F'}^{\mu\nu})^2-\beta_{2,2}^{\rm loop}(\overline{F'}_{\mu\nu}\widetilde{\overline{F'}}^{\mu\nu})^2-\beta_{2,3}^{\rm loop} \overline{F'}_{\mu\nu}\overline{F'}_{\rho\sigma}\overline{R'}^{\mu\nu\rho\sigma}
        +({\rm correction~from}~R~{\rm and}~F_{\mu\nu}F^{\mu\nu})\bigg)
        \bigg]\label{eq:senEMFT}
        \\
        &={\rm exp}\bigg[
        -\int (d^4x)_E \sqrt{\overline{g}}\bigg(\Lambda_{0,\Phi}^{\rm loop}-\frac{M^2_{\rm Pl}}{2}\left(1+\alpha^{\rm loop}_{1,R}\right)\overline{R} +\frac{1}{4}\left(1+\alpha^{\rm loop}_{1,F}\right)\overline{F}_{\mu\nu}\overline{F}^{\mu\nu}\notag
        \\
        &-\beta_{2,1}^{\rm loop}(\overline{F}_{\mu\nu}\overline{F}^{\mu\nu})^2-\beta_{2,2}^{\rm loop}(\overline{F}_{\mu\nu}\widetilde{\overline{F}}^{\mu\nu})^2-\beta_{2,3}^{\rm loop} \overline{F}_{\mu\nu}\overline{F}_{\rho\sigma}\overline{R}^{\mu\nu\rho\sigma}+({\rm correction~from}~R~{\rm and}~F_{\mu\nu}F^{\mu\nu})\bigg)
        \bigg],
    \end{align}
    where $\alpha_{2,R}^{\rm loop}$, $\alpha_{2,F}^{\rm loop}$, $\beta_{2,1}^{\rm loop}$, $\beta_{2,2}^{\rm loop}$ and $\beta_{2,3}^{\rm loop}$ are the second or higher order corrections for $g$, $\alpha_{1,R}^{\rm loop}$ and $\alpha_{1,F}^{\rm loop}$ are the first order corrections for $g$,
    and $\Lambda_{0,\Phi}^{\rm loop}$ is the vacuum energy coming from $\Phi$.
    %
    %
    %
    %
    %
    %
    %
    The last term of Eq.~\eqref{eq:senEMFT} arises from loop corrections of light fields in $M^2_{\rm Pl}R/2$ and $F_{\mu\nu}F^{\mu\nu}/4$. 
    %
    %
    Since these corrections do not depend on $g$, they cancel in relative entropy.
    The background fields $\overline{A'}_{\mu}$ and $\overline{g'}_{\mu\nu}$ denote the classical solution of the effective action of
    \begin{align}
        W_{g}[g_{\mu\nu},A]&=\int (d^4x)_E \sqrt{g}\bigg(\Lambda_{0,\Phi}^{\rm loop}-\frac{M^2_{\rm Pl}}{2}(1+\alpha_{1,R}^{\rm loop}+\alpha_{2,R}^{\rm loop})R +\frac{1}{4}(1+\alpha_{1,F}^{\rm loop}+\alpha_{2,F}^{\rm loop})F_{\mu\nu}F^{\mu\nu}\notag
        \\
        &-\beta_{2,1}^{\rm loop}(F_{\mu\nu}F^{\mu\nu})^2-\beta_{2,2}^{\rm loop}(F_{\mu\nu}\widetilde{F}^{\mu\nu})^2-\beta_{2,3}^{\rm loop} F_{\mu\nu}F_{\rho\sigma}R^{\mu\nu\rho\sigma}\bigg).
    \end{align}
    We choose the background field as follows,
    \begin{align}
        &\overline{A'}_{\mu}=\left(1+\frac{1}{2}\left(\frac{4}{3}\alpha_{2,R}^{\rm loop}-\alpha_{2,F}^{\rm loop}\right)\right)\overline{A}_{\mu},
        \\
        &\overline{g'}_{\mu\nu}=\left(1-\frac{1}{3}\alpha_{2,R}^{\rm loop}\right)\overline{g}_{\mu\nu},~~~\overline{g'}^{\mu\nu}=\left(1+\frac{1}{3}\alpha_{2,R}^{\rm loop}\right)\overline{g}^{\mu\nu}.
    \end{align}
    The effective action for the theory $B$ and $A$ are respectively obtained as follows,
    \begin{align}
        W_g[\overline{g}_{\mu\nu},\overline{A}]&=\int (d^4x)_E \sqrt{\overline{g}}\bigg(\Lambda_{0,\Phi}^{\rm loop}-\frac{M^2_{\rm Pl}}{2}(1+\alpha^{\rm loop}_{1,R})\overline{R} +\frac{1}{4}(1+\alpha^{\rm loop}_{1,F})\overline{F}_{\mu\nu}\overline{F}^{\mu\nu}\notag
        \\
        &-\beta_{2,1}^{\rm loop}(\overline{F}_{\mu\nu}\overline{F}^{\mu\nu})^2-\beta_{2,2}^{\rm loop}(\overline{F}_{\mu\nu}\widetilde{\overline{F}}^{\mu\nu})^2-\beta_{2,3}^{\rm loop} \overline{F}_{\mu\nu}\overline{F}_{\rho\sigma}\overline{R}^{\mu\nu\rho\sigma}+({\rm correction~from}~R~{\rm and}~F_{\mu\nu}F^{\mu\nu})\bigg),\label{eq:WgEM}
        \\
        W_0[\overline{g}_{\mu\nu},\overline{A}]&=\int (d^4x)_E \sqrt{\overline{g}}\bigg(\Lambda_{0,\Phi}^{\rm loop}-\frac{M^2_{\rm Pl}}{2}\overline{R} +\frac{1}{4}\overline{F}_{\mu\nu}\overline{F}^{\mu\nu}+({\rm correction~from}~R~{\rm and}~F_{\mu\nu}F^{\mu\nu})\bigg).\label{eq:W0EM}
    \end{align}
    Similar to the case of the tree-level UV completion, the first order correction for the higher-derivative terms vanish in $W_0$ by using the equation of motion.
    %
    %
    %
    Also, from Eq.~\eqref{eq:WgEM},  the first order correction for $g$ is calculated as
    \begin{align}
        \left(\frac{dW_g}{dg}\right)_{g=0}&=\left(\frac{\partial W_g}{\partial g}\right)_{g=0}
        +\int (d^4x)_E\sqrt{-g} \bigg(\left(\frac{\delta W_g}{\delta \overline{A'}}\right)\cdot \left(\frac{d \overline{A'}}{dg}\right)_{g=0}+\left(\frac{\delta W_g}{\delta \overline{g'}_{\mu\nu}}\right)\cdot \left(\frac{d \overline{g'}_{\mu\nu}}{dg}\right)_{g=0}\bigg)\notag
        \\
        &=\left(\frac{\partial W_g}{\partial g}\right)_{g=0}\notag
        \\
        &=\int (d^4x)_E \sqrt{\overline{g}}\bigg(-\frac{M^2_{\rm Pl}}{2} \frac{d \alpha^{\rm loop}_{1,R}}{dg}\overline{R}+\frac{1}{4} \frac{d \alpha^{\rm loop}_{1,F}}{dg}\overline{F}_{\mu\nu}\overline{F}^{\mu\nu}
        \bigg),\label{eq:EMFIRg}
    \end{align}
    where $(d\overline{A'}_{\mu}/dg)_{g=0}=0$ and $(d\overline{g'}_{\mu\nu}/dg)_{g=0}=0$ are used.
    Note here that the last term of Eq.~\eqref{eq:WgEM} does not depend on $g$.
    From Eq.~\eqref{eq:uplow} or \eqref{eq:dyligh}, Eq.~\eqref{eq:WgEM}, \eqref{eq:W0EM} and \eqref{eq:EMFIRg} yields
     \begin{align}
        \Delta W_g^{(E)}\leq g\cdot {\langle I_I\rangle}_{g=0} &\Rightarrow 
        W_g^{\rm non\text{-}lin}[\overline{g}_{\mu\nu},\overline{A}]-W_0[\overline{g}_{\mu\nu},\overline{A}]\leq 0.\label{eq:EMbound}
    \end{align}
    Here, we define the effective action without the first order corrections for $g$ as follows,
        \begin{align}
        W_g^{\rm non\text{-}lin}[\overline{g}_{\mu\nu},\overline{A}]&=\int (d^4x)_E \sqrt{\overline{g}}\bigg(\Lambda_{0,\Phi}^{\rm loop}-\frac{M^2_{\rm Pl}}{2}\overline{R} +\frac{1}{4}\overline{F}_{\mu\nu}\overline{F}^{\mu\nu}\notag
        \\
        &-\beta_{2,1}^{\rm loop}(\overline{F}_{\mu\nu}\overline{F}^{\mu\nu})^2-\beta_{2,2}^{\rm loop}(\overline{F}_{\mu\nu}\widetilde{\overline{F}}^{\mu\nu})^2-\beta_{2,3}^{\rm loop} \overline{F}_{\mu\nu}\overline{F}_{\rho\sigma}\overline{R}^{\mu\nu\rho\sigma}+({\rm correction~from}~R~{\rm and}~F_{\mu\nu}F^{\mu\nu})\bigg).
    \end{align}
         Therefore, $W_g^{\rm non\text{-}lin}[\overline{g}_{\mu\nu},\overline{A}]-W_0[\overline{g}_{0,\mu\nu},\overline{A}_0]$ denotes the corrections from the higher-derivative terms to the Euclidean effective action.
        It should be noted that the one-loop correction from $R$ and $F_{\mu\nu}F^{\mu\nu}$ cancels in Eq.~\eqref{eq:EMbound}.

\end{itemize}

\section{Loophole of entropy constraints}
\label{app:(IV)}
We discuss the loophole of the entropy constraints.
As discussed in Ref.~\cite{Hamada:2018dde} and \cite{Arkani-Hamed:2021ajd}, positive perturvative corrections to the Euclidean action can arise in some examples.
We show that the loophole arises because the entropy constraints are based on the saddle point approximation in the Euclidean path integral method.
First, we consider the entropy constraints on tree-level UV completions, and clarify a relation between this work and Ref.~\cite{Cheung:2018cwt}.
The relative entropy of Eq.~\eqref{eq:rel} is calculated as
\begin{align}
    S(P_0||P_g)&=\int_{\beta} d[\Phi] \left(P_0 \ln P_0-P_0\ln P_g \right)\notag
    \\
    &=-\ln Z_0[\beta,\phi]+\ln Z_g[\beta,\phi]+g \int_{\beta} d[\Phi] P_0\cdot I_I\notag
    \\
    &=-\ln Z_0[\beta,\phi]+\ln Z_g[\beta,\phi]\notag
    \\
    &\simeq I_0[{\phi},\widetilde{\Phi}_{0}]-I_g[{\phi},\widetilde{\Phi}_{g}]\geq 0,\label{eq:sad}
\end{align}
where in the third line we used $\int_{\beta} d[\Phi] P_0\cdot I_I=0$ at tree-level by using a suitable definition of $\Phi$, in the last line the saddle point approximation is used, and $\widetilde{\Phi}_{0}$ and $\widetilde{\Phi}_{g}$ are classical solutions of $I_0$ and $I_g$, respectively.
By the definition of $\Phi$, $
I_0[\phi,0]= I_0[\phi,\widetilde{\Phi}_{0}]
$ is satisfied.
Then, Eq.~\eqref{eq:sad} yields
\begin{align}
    I_g[\phi,0]=I_0[\phi,0]= I_0[\phi,\tilde{\Phi}_{0}]\geq I_g[\phi,\tilde{\Phi}_{g}],\label{eq:tree1}
\end{align}
where we used $I_g[\phi,0]=I_0[\phi,0]$ similar to Ref.~\cite{Cheung:2018cwt}.
This inequality has been provided in Ref.~\cite{Cheung:2018cwt}, and it is clear that the entropy constraints by the relative entropy is a generalization of Ref.~\cite{Cheung:2018cwt}.
The key point of derivation of Eq.~\eqref{eq:tree1} is that the relative entropy must be evaluated around the local minimum of heavy degrees of freedom.
Otherwise, the saddle point approximation does not work well, and the perturbative corrections to the Euclidean effective action can be positive.

To see the loophole, let us consider following action in Minkowski space:
\begin{align}
    I^{(M)}=\int d^4 x \left(-\frac{1}{4}F_{\mu\nu}F^{\mu\nu}+ m_A^2 \phi_A^2+\frac{1}{M}\phi_A F_{\rho\sigma}F^{\rho\sigma} \right),\label{eq:ex1}
\end{align}
where $\phi_A$ is an auxiliary field.
The solution of the equation of motion of $\phi_A$ is calculated as
\begin{align}
    \widetilde{\phi}_A=-\frac{1}{2m_A^2 M}F_{\mu\nu}F^{\mu\nu}.
\end{align}
After integrating out $\phi_A$, Eq.~\eqref{eq:ex1} yields
\begin{align}
    I_{\rm eff}^{(M)}=\int d^4x \left(-\frac{1}{4}F_{\mu\nu}F^{\mu\nu}-\frac{1}{4m_A^2 M^2}(F_{\mu\nu}F^{\mu\nu})^2 \right).\label{eq:eff1}
\end{align}
In the Euclidean space, the second term in Eq.~\eqref{eq:eff1} increases the Euclidean effective action, and contradicts the entropy constraints.
This is because the solution of the equation of motion of $\phi_A$ is not a local minimum of $I$ in the Euclidean space.

Next, let us consider a doublet of real, shift-symmetric, massless scalar fields $\phi_i$, $i=1,2$ in Minkowski space:
\begin{align}
    I^{(M)}=\int d^4x \left(\frac{1}{2}(\partial_{\mu}\phi_i \partial^{\mu}\phi_i)+m_A^2 (X_{\mu\nu}X^{\mu\nu})-\frac{\epsilon^{il}}{M}(\partial_{\mu}\phi_i \partial_{\nu}\phi_l)X^{\mu\nu} \right),\label{eq:X1}
\end{align}
where $X_{\mu\nu}$ is an auxiliary field, and $\epsilon^{12}=-\epsilon^{21}=1$.
The equation of motion of $X_{\mu\nu}$ is calculated as
\begin{align}
    \widetilde{X}_{\mu\nu}=\frac{\epsilon^{il}}{2m_A^2 M} (\partial_{\mu}\phi_i \partial_{\nu}\phi_l).
\end{align}
After integrating out $X_{\mu\nu}$ Eq.~\eqref{eq:X1} yields
\begin{align}
    I_{\rm eff}
    =\int d^4x \left(\frac{1}{2}(\partial_{\mu}\phi_i \partial^{\mu}\phi_i)+\frac{1}{4 m_A^2 M^2}\epsilon^{il}\epsilon^{kj}(\partial_{\mu}\phi_i \partial^{\mu}\phi_j)(\partial_{\nu}\phi_k \partial^{\nu}\phi_l) \right).\label{eq:effX2}
\end{align}
Substituting a solution of the equation of motion of $\phi_i$: $
    \partial_{\mu}\bar{\phi}_1=\left(0,1,0,0\right),~\partial_{\mu}\bar{\phi}_2=\left(0,0,1,0\right)$ 
into Eq.~\eqref{eq:effX2}, we find that the second term of Eq.~\eqref{eq:effX2} is negative as follows:
\begin{align}
    \epsilon^{il}\epsilon^{kj}(\partial_{\mu}\bar{\phi}_i \partial^{\mu}\bar{\phi}_j)(\partial_{\nu}\bar{\phi}_k \partial^{\nu}\bar{\phi}_l)&=2\left((\partial_{\mu}\bar{\phi}_1 \partial^{\mu}\bar{\phi}_2)^2-(\partial_{\mu}\bar{\phi}_1 \partial^{\mu}\bar{\phi}_1)(\partial_{\nu}\bar{\phi}_2 \partial^{\nu}\bar{\phi}_2)  \right)=-2.
\end{align}
Therefore, the second term of Eq.~\eqref{eq:effX2} increases the Euclidean effective action, and a contradiction of the entropy constraint arises.
This is because $\tilde{X}_{\mu\nu}$ is not a local minimum of $I$ in the Euclidean space, and the saddle point approximation does not work well.
Note here that the negative shift of the Euclidean effective action arises when the sign of the second term of Eq.~\eqref{eq:X1} is flipped.
Consequently, the loophole of entropy constraints can arise from the classical solution of heavy degrees of freedom not being the local minimum, where the path integral method in the Euclidean space does not work. 
\\

\section{Conditions to apply entropy constraints}
\label{app:(V)}
We summarize the conditions to apply the entropy constraints.
For ease of understanding, we show the conditions as a flowchart in Fig.~\ref{fig:flow}.
In this Letter, the entropy constraints mainly denote three inequalities.
For each of the inequalities, we explain the conditions as follows:

\begin{itemize}
    \item $g\cdot {\langle I_I\rangle}_g \leq \Delta W_g^{(E)} \leq g\cdot {\langle I_I\rangle}_{g=0}$

    To derive the inequality~\eqref{eq:uplow}, we impose conditions:
    (a) the theories $A$ and $B$ are defined by $I_0$ and $I_g\equiv I_0+g\cdot I_I$, respectively, and the probability distribution functions $P_0$ and $P_g$ are defined by them, and
    (b) the tree level corrections from the heavy degrees of freedom to the Euclidean effective actions $W_0$ and $W_g$ arise from the local minimum of $I_g$ and $I_0$, respectively.
The condition (b) is relevant to the loophole discussed above.
Note here that, in general, Eq.~\eqref{eq:uplow} does not depend on whether $I_I$ represents the interactions between heavy and light degrees of freedom.
Since, however, in this Letter, we are interested in the constraints on higher derivative terms that arise from the interactions between heavy and light degrees of freedom, we suppose that $I_I$ represents the interactions between heavy and light degrees of freedom.
%


%
%

\item Positivity bounds on higher-derivative terms

To derive the positivity bounds on the Wilson coefficients of higher-derivative operators, in addition to the conditions (a) and (b), we use a condition: (c) quantum corrections to non-higher derivative terms can be  absorbed by redefinitions of light fields, and (d) $J[\phi]$ does not include the higher-derivative operators.
In general, the corrections from the interactions contribute to the non-higher derivative terms, but these conditions (c) and (d) allow us to remove such corrections.

\item $(\partial S/\partial \epsilon)_{M,\vec{Q}}\geq 0$

To derive the positive perturbative corrections from the higher-derivative terms to thermodynamic entropy at a fixed energy and charge, in addition to the conditions (a), (b), (c) and (d), we impose conditions: (e) thermodynamics relations hold in the system, and (f) the system is the weak-dynamics theory, where $\mathcal{O}(\epsilon^2)$ terms are negligible. 
\end{itemize}
%

\begin{figure}[t]
\begin{center}
\includegraphics[width=15cm]{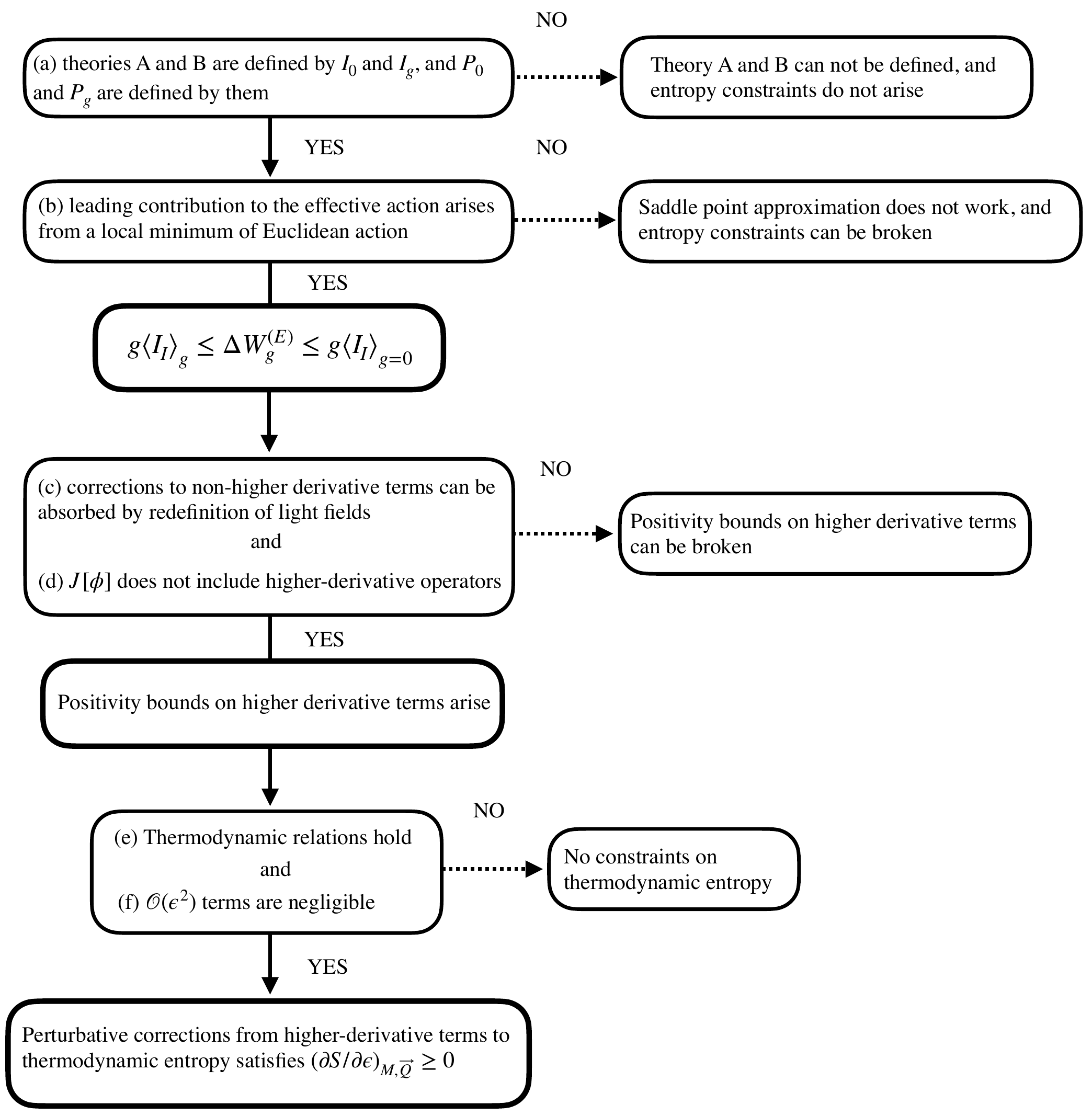}
\end{center}
\vspace{-0.4cm}
\caption{A flow chart for conditions of applicability of entropy constraints: Each step explain which conditions are necessary to use the entropy constraints.}
\label{fig:flow}
\end{figure}


\end{widetext}
\bibliography{ref}
\end{document}